\PassOptionsToPackage{hyphens}{url}
\PassOptionsToPackage{table}{xcolor}

\documentclass[sigconf, nonacm]{acmart}

\pdfoutput=1

\usepackage{amsmath,balance,multirow,multicol,tabularx}
\usepackage[inline]{enumitem}
\usepackage[binary-units]{siunitx}
\DeclareSIUnit{\nothing}{\relax}
\usepackage{amsthm}
\newtheorem{theorem}{Theorem}[section]
\newtheorem{proposition}[theorem]{Proposition}

\theoremstyle{definition}
\newtheorem{definition}[theorem]{Definition}
\newtheorem{example}[theorem]{Example}

\usepackage[binary-units]{siunitx}
\DeclareSIUnit{\txn}{txn}
\DeclareSIUnit{\batch}{batch}
\newcommand{\Replicas}[1]{\mathfrak{R}_{#1}}

\newcommand{\Shards}{\mathfrak{S}}
\newcommand{\Involved}{\mathfrak{I}}
\newcommand{\Shard}[1]{\textsc{#1}}
\newcommand{\Client}[1]{\textsc{c}_{#1}}
\newcommand{\Replica}[1]{\textsc{#1}}
\newcommand{\Primary}[1]{\textsc{p}_{#1}}
\newcommand{\Faulty}{\mathcal{F}}
\newcommand{\NonFaulty}{\mathcal{NF}}
\newcommand{\Transaction}[1]{\mathbb{T}_{#1}}
\newcommand{\Response}[1]{v_#1}

\newcommand{\z}{\mathbf{z}}
\newcommand{\ID}[1]{\operatorname{id}(#1)}
\newcommand{\n}{\mathbf{n}}
\newcommand{\s}{\mathbf{s}}
\newcommand{\f}{\mathbf{f}}
\newcommand{\nf}{\mathbf{nf}}
\newcommand{\kmax}{k_{\text{max}}}

\newcommand{\Message}[2]{{#1}(#2)}
\newcommand{\SignMessage}[2]{\langle#1\rangle_{#2}}

\newcommand{\Hash}[1]{H(#1)}
\newcommand{\Digest}{\Delta}
\newcommand{\Ledger}[1]{\mathfrak{L}_{#1}}
\newcommand{\Block}[1]{\mathfrak{B}_{#1}}

\newcommand{\Name}[1]{\textnormal{\textsc{#1}}}
\newcommand{\BName}[1]{\textnormal{\bf \textsc{#1}}}
\newcommand{\RingBFT}{\Name{RingBFT}}
\newcommand{\PoE}{\Name{PoE}}
\newcommand{\BFT}{\Name{Bft}}
\newcommand{\PBFT}{\Name{Pbft}}
\newcommand{\ZZ}{\Name{Zyzzyva}}
\newcommand{\SBFT}{\Name{Sbft}}
\newcommand{\HS}{\Name{HotStuff}}
\newcommand{\STW}{\Name{Steward}}
\newcommand{\AHL}{\Name{AHL}}
\newcommand{\ResilientDB}{\Name{Resilient\-DB}}

\newcommand{\Sharper}{\Name{Sharper}}
\newcommand{\twopc}{\Name{2pc}}
\newcommand{\CST}{\Name{cst}}

\newcommand{\abs}[1]{\lvert #1 \rvert}

\newcommand{\union}{\cup}
\newcommand{\intersect}{\cap}
\newcommand{\difference}{\setminus}

\usepackage{fontawesome}
\newcommand{\db}[3][purple]{\node[scale=2,color=black!25!#1] at (#2,#3) {\faDatabase};}
\newcommand{\cog}[2]{\node[scale=2,color=black!75] at (#1 + 0.35,#2 + 0.35) {\faCog};}
\newcommand{\dbcog}[3][purple]{\db[#1]{#2}{#3}\cog{#2}{#3}}

\newcommand{\dbuser}[3][blue]{\node[scale=2,color=black!25!#1] at (#2,#3) {\faUser};}

\newcommand{\RN}[1]{%
  \textup{\uppercase\expandafter{\romannumeral#1}}%
}

\definecolor{colO}{RGB}{230,159,0}
\definecolor{colB}{RGB}{86,180,233}
\definecolor{colG}{RGB}{0,158,115}
\definecolor{colY}{RGB}{240,228,66}
\definecolor{colN}{RGB}{0,114,178}
\definecolor{colR}{RGB}{213,94,0}
\definecolor{colP}{RGB}{204,121,167}

\usepackage[noend]{algorithmic}
\newenvironment{myprotocol}{
    \hrule
    \smallskip
    \scriptsize
    \algsetup{linenosize=\tiny}
    \begin{algorithmic}[1]
        \newcommand{\SPACE}{\item[]}	
	\newcommand{\GETS}{:=}
        \newcommand{\TITLE}[2]{\item[] \textbf{\underline{##1}} (##2) \textbf{:}\\[0.5pt]}
        \makeatletter
            \newcommand{\EVENT}[1]{\STATE \textbf{event} ##1 \textbf{do}\begin{ALC@g}}
            \newcommand{\ENDEVENT}{\end{ALC@g}}
        \makeatother
	\newcommand{\INITIAL}[2]{\item[] \textbf{\underline{##1}} ##2\\[0.5pt]}
	\newcommand{\MC}[1]{{\color{colN}// ##1}}
	
}{
    \end{algorithmic}
    \smallskip
    \hrule
}

\newcommand{\changed}[1]{{\color{black}#1}}

\usepackage{xcolor}
\usepackage{tikz,pgfplots,pgfplotstable}
\usetikzlibrary{arrows.meta}
\tikzset{
    >=Stealth,
    plot/.append style={baseline,scale=0.45},
    label/.append style={font=\strut\Large},
    dot/.style={circle,scale=0.35,draw=black,fill=black},
}
\pgfplotscreateplotcyclelist{mycyclelist}{
    violet   ,every mark/.append style={solid,fill=\pgfplotsmarklistfill},mark=*\\
    red     ,every mark/.append style={solid,fill=\pgfplotsmarklistfill},mark=square*\\
    blue    ,every mark/.append style={solid,fill=\pgfplotsmarklistfill},mark=triangle*\\
    black    ,every mark/.append style={solid,fill=\pgfplotsmarklistfill},mark=pentagon*\\
    brown   ,every mark/.append style={solid,fill=\pgfplotsmarklistfill},mark=halfsquare*\\
    orange  ,every mark/.append style={solid,fill=\pgfplotsmarklistfill},mark=halfcircle*\\
    teal  ,every mark/.append style={solid,fill=\pgfplotsmarklistfill,rotate=180},mark=halfdiamond*\\
    green  ,every mark/.append style={solid,fill=\pgfplotsmarklistfill,rotate=180},mark=diamond*\\
}
\pgfplotscreateplotcyclelist{mycyclelistex}{
    black   ,every mark/.append style={solid,fill=\pgfplotsmarklistfill},mark=*\\
    red     ,every mark/.append style={solid,fill=\pgfplotsmarklistfill},mark=square*\\
}
\pgfplotsset{
    tick label style={font=\Large},
    legend style={font=\Large,cells={anchor=west}},
    title style={font=\Large},
    label style={font=\Large},
    width=262.5pt,
    height=185pt,
    every axis/.append style={
            ylabel near ticks,
            xlabel near ticks,
            mark size=2.5pt,
            cycle list name=mycyclelist,
            font=\Large,
            y tick label style={
                    /pgf/number format/precision=1,
                    /pgf/number format/fixed,
                    /pgf/number format/fixed zerofill
                }
        },
}

\pgfplotstableread{
    Nodes RingBFT15 RingBFT0	PBFT	SBFT	HotStuff	Steward	RCC	POE	ZYZ
    4	 921713   1469213    194716	158571	9072	18815	300261	201691	230990
    16	 626374   960454     149677	132600	9072	18590	261271	159874	199895
    32	 417150   689390     95217	111412	9296	18388	221133	128835	154506

}\graphOtherProtocolsTp

\pgfplotstableread{
    Shards	RingBFT	Sharper	AHL
    3	77041	72736	35086
    5	82213	48484	19781
    7	80561	45779	13983
    9	80191	33745	10736
    11	85457	29690	8706
    15	80678	20063	5119
}\graphAftp
\pgfplotstableread{
    Shards	RingBFT	Sharper	AHL
    3	23014	21854	10582
    5	24741	14945	6399
    7	24614	14432	4346
    9	24425	10903	3705
    11	26102	9862	3162
    15	25199	7269	1969
}\graphActp
\pgfplotstableread{
    Shards	RingBFT	Sharper	AHL
    3	54027	50882	24504
    5	57472	33539	13382
    7	55947	31347	9637
    9	55766	22842	7031
    11	59355	19828	5544
    15	55479	12794	3150
}\graphAtp
\pgfplotstableread{
    Shards	RingBFT	Sharper	AHL
    3	1.167	2.17	4.501
    5	1.899	5.767	6.79
    7	2.552	8.321	14.343
    9	3.985	16.627	24.123
    11	4.553	22.534	36.072
    15	6.821	22.828	76.603
}\graphAlt
\pgfplotstableread{
    Nodes	RingBFT	Sharper	AHL
    10	194729	53551	16356
    16	136249	39995	10016
    22	100341	25924	3778
    28	80678	20063	5119
}\graphBftp
\pgfplotstableread{
    Nodes	RingBFT	Sharper	AHL
    10	59652	17523	5700
    16	42246	13257	3586
    22	30921	9095	1762
    28	25199	7269	1969
}\graphBctp
\pgfplotstableread{
    Nodes	RingBFT	Sharper	AHL
    10	135077	36028	10656
    16	94003	26738	6430
    22	69420	16829	2016
    28	55479	12794	3150
}\graphBtp
\pgfplotstableread{
    Nodes	RingBFT	Sharper	AHL
    10	5.685	17.546	30.662
    16	8.066	17.391	51.933
    22	5.342	18.196	64.578
    28	6.821	22.828	76.603
}\graphBlt
\pgfplotstableread{
    X-Precent	RingBFT	Sharper	AHL
    1	1197762	1222657	1146602
    5	254067	126326	39727
    10	196197	66935	20827
    15	137854	44584	13179
    30	80678	20063	5119
    60	41531	10479	2013
    100	26531	7342	1470
}\graphCftp
\pgfplotstableread{
    X-Precent	RingBFT	Sharper	AHL
    1	0	0	0
    5	14215	6603	2069
    10	21844	7239	2290
    15	22194	7470	2129
    30	25199	7269	1969
    60	25872	7822	795
    100	26531	7342	1470
}\graphCctp
\pgfplotstableread{
    X-Precent	RingBFT	Sharper	AHL
    1	1197762	1222657	1146602
    5	239852	119723	37658
    10	174353	59696	18537
    15	115660	37114	11050
    30	55479	12794	3150
    60	15659	2657	1218
    100	0	0	0
}\graphCtp
\pgfplotstableread{
    X-Precent	RingBFT	Sharper	AHL
    1	3.297	3.228	3.65
    5	6.405	7.843	11.54
    10	7.009	14.539	25.177
    15	5.105	21.042	38.767
    30	6.821	22.828	76.603
    60	9.897	36.202	107.164
    100	16.459	53.395	125.786
}\graphClt
\pgfplotstableread{
    BatchSize	RingBFT	Sharper	AHL
    10	    7437	2332	1610
    50	    40595	12739	4360
    100	    80678	20063	5119
    500	    138758	57538	5708
    1000	178309	87249	5392
    1500	189663	90395	5459
    3000	154129	81239	4793
}\graphDftp
\pgfplotstableread{
    BatchSize	RingBFT	Sharper	AHL
    10	    2461	726	441
    50	    13041	3958	1404
    100	    25199	7269	1969
    500	    43938	21771	3134
    1000	56061	28759	3064
    1500	60269	31336	1724
    3000	48874	29886	1408
}\graphDctp
\pgfplotstableread{
    BatchSize	RingBFT	Sharper	AHL
    10	    4976	1606	1169
    50	    27554	8781	2956
    100 	55479	12794	3150
    500	    94820	35767	2574
    1000	122248	58490	2328
    1500	129394	59059	3735
    3000	105255	51353	3385
}\graphDtp
\pgfplotstableread{
    BatchSize	RingBFT	Sharper	AHL
    10	    14.695	35.723	55.3
    50  	9.6	    24.987	46.62
    100	    6.821	22.828	76.603
    500	    7.233	22.017	70.142
    1000	8.953	17.015	98.771
    1500	11.024	15.918	123.365
    3000	14.062	17.919	137.791
}\graphDlt
\pgfplotstableread{
    InvShards	RingBFT	Sharper	AHL
    1	1197762	1222657	1146602
    3	377937	358660	22938
    6	216179	115285	14563
    9	136251	59203	9002
    15	80678	20063	5119
}\graphEftp
\pgfplotstableread{
    InvShards	RingBFT	Sharper	AHL
    1	0	0	0
    3	114172	107708	8755
    6	66130	35812	5093
    9	42198	19111	3665
    15	25199	7269	1969
}\graphEctp
\pgfplotstableread{
    InvShards	RingBFT	Sharper	AHL
    1	1197762	1222657	1146602
    3	263765	250952	14183
    6	150049	79473	9470
    9	94053	40092	5337
    15	55479	12794	3150
}\graphEtp
\pgfplotstableread{
    InvShards	RingBFT	Sharper	AHL
    1	3.297	3.228	3.65
    3	3.734	3.302	36.495
    6	7.104	12.011	37.822
    9	8.003	15.688	46.767
    15	6.821	22.828	76.603
}\graphElt
\pgfplotstableread{
    MaxInf	RingBFT	Sharper	AHL
    3000	69237	16166	1857
    5000	80678	20063	5616
    10000	78633	24244	8071
    15000	75686	19435	6417
    20000	70496	18482	7969
}\graphFftp
\pgfplotstableread{
    MaxInf	RingBFT	Sharper	AHL
    3000	21468	6737	1857
    5000	25199	7269	1853
    10000	25931	8481	1930
    15000	26074	7226	1227
    20000	24253	6638	646
}\graphFctp
\pgfplotstableread{
    MaxInf	RingBFT	Sharper	AHL
    3000	47769	9429	0
    5000	55479	12794	3763
    10000	52702	15763	6141
    15000	49639	12209	5190
    20000	46243	11844	7323
}\graphFtp
\pgfplotstableread{
    MaxInf	RingBFT	Sharper	AHL
    3000	4.533	15.383	46.244
    5000	6.821	22.828	73.557
    10000	14.244	58.211	81.194
    15000	25.48	61.062	96.997
    20000	37.275	73.591	105.89
}\graphFlt

\pgfplotstableread{
    Time	Throughput
    0       486227.1988
    5	    483810.0501
    10      461601.6865
    15      455925.0753
    20      438830.6552
    25      429254.5927
    30      422839.787
    35      421019.971
    40      429905.8006
    45      451377.6531
    50      458127.4441
    55      477158.4789
    60      478181.7192
    65      479625.4481
    70      479267.7121
    75      481760.5334
    80      481440.1456
    85      483303.8275
    90      484434.7415
    95      485660.3749
    100     486528.776
    105     487182.8757
    110     487779.3115
}\graphVCtp

\pgfplotstableread{
    Dep	RingBFT
    0	80678
    1	76537
    8	73442
    16	73015
    32	66450
    48	60017
    64	46853
}\graphDPftp
\pgfplotstableread{
    Dep	RingBFT
    0	25199
    1	25093
    8	24386
    16	24378
    32	22251
    48	20156
    64	16119
}\graphDPctp
\pgfplotstableread{
    Dep	RingBFT
    0	55479
    1	51444
    8	49056
    16	48637
    32	44199
    48	39861
    64	30734
}\graphDPtp
\pgfplotstableread{
    Dep	RingBFT
    0	7
    1	8.62
    8	8.908
    16	9.04
    32	9.965
    48	11.186
    64	15.715
}\graphDPlt

\newcommand{\axisShards}{Number of Shards ($\s$)}
\newcommand{\axisNodes}{Number of Nodes Per Shard ($\n$)}
\newcommand{\axisCross}{Cross-Shard Workload Rate}
\newcommand{\axisTTP}{Total Throughput (\si{\txn\per\second})}

\newcommand{\axisLat}{Latency (\si{\second})}

\newcommand{\graphOtherProtocols}{
    \begin{tikzpicture}[plot,scale=1.50]
        \begin{axis}[
                title={Scalability},
                scaled y ticks = false,
                xlabel={Number of Nodes},
                ylabel={\axisTTP},
                legend style={
                        at={(1.03,0.5)},
                        anchor=west
                    },
                ytick={0,200000, 400000, 600000, 800000, 1000000, 1200000, 1400000},
                yticklabels={0,200K,400K,600K,800K,1M,1.2M,1.4M},
                xtick={4,16,32,64},
            ]
            \addplot table[x={Nodes},y={RingBFT0}] {\graphOtherProtocolsTp};
            \addplot table[x={Nodes},y={RingBFT15}] {\graphOtherProtocolsTp};
            \addplot table[x={Nodes},y={PBFT}] {\graphOtherProtocolsTp};
            \addplot table[x={Nodes},y={SBFT}] {\graphOtherProtocolsTp};
            \addplot table[x={Nodes},y={HotStuff}] {\graphOtherProtocolsTp};
            \addplot table[x={Nodes},y={RCC}] {\graphOtherProtocolsTp};
            \addplot table[x={Nodes},y={POE}] {\graphOtherProtocolsTp};
            \addplot table[x={Nodes},y={ZYZ}] {\graphOtherProtocolsTp};
            \legend{\RingBFT{}, \RingBFT{}$_X$, \PBFT{}, \SBFT{}, \HS{}, \Name{Rcc}, \PoE{}, \ZZ{}}
        \end{axis}
    \end{tikzpicture}
}

\newcommand{\graphAFullTP}{
    \begin{tikzpicture}[plot]
        \begin{axis}[
                title={(\RN{1}) Impact of Shards (Throughput)},
                scaled y ticks = false,
                xlabel={\axisShards},
                ylabel={\axisTTP},
                ytick={0,20000,40000,60000,80000},
                yticklabels={0,20K,40K,60K,80K},
                xtick={3,5,7,9,11,15},
                legend to name={mainlegend},
                legend columns=-1,
                legend entries={\RingBFT{},\Sharper{},\AHL{}},
            ]
            \addplot table[x={Shards},y={RingBFT}] {\graphAftp};
            \addplot table[x={Shards},y={Sharper}] {\graphAftp};
            \addplot table[x={Shards},y={AHL}] {\graphAftp};
        \end{axis}
    \end{tikzpicture}
}

\newcommand{\graphALat}{
    \begin{tikzpicture}[plot]
        \begin{axis}[
                xlabel={\axisShards},
                ylabel={\axisLat},
                xtick={3,5,7,9,11,15},
                title={(\RN{2}) Impact of Shards (Latency)},
                legend to name={mainlegend},
                legend columns=-1
            ]
            \addplot table[x={Shards},y={RingBFT}] {\graphAlt};
            \addplot table[x={Shards},y={Sharper}] {\graphAlt};
            \addplot table[x={Shards},y={AHL}] {\graphAlt};
            \legend{\RingBFT{},\Sharper{},\AHL{}}
        \end{axis}
    \end{tikzpicture}
}
\newcommand{\graphBFullTP}{
    \begin{tikzpicture}[plot]
        \begin{axis}[
                scaled y ticks = false,
                xlabel={\axisNodes},
                ylabel={\axisTTP},
                ytick={0,50000,100000,150000,200000},
                yticklabels={0,50K,100K,150K,200K},
                xtick={10,16,22,28},
                title={(\RN{3}) Impact of Nodes per Shards (Throughput)},
                legend to name={mainlegend},
                legend columns=-1
            ]
            \addplot table[x={Nodes},y={RingBFT}] {\graphBftp};
            \addplot table[x={Nodes},y={Sharper}] {\graphBftp};
            \addplot table[x={Nodes},y={AHL}] {\graphBftp};
            \legend{\RingBFT{},\Sharper{},\AHL{}}
        \end{axis}
    \end{tikzpicture}
}

\newcommand{\graphBLat}{
    \begin{tikzpicture}[plot]
        \begin{axis}[
                xlabel={\axisNodes},
                ylabel={\axisLat},
                xtick={10,16,22,28},
                title={(\RN{4}) Impact of Nodes per Shards (Latency)},
                legend to name={mainlegend},
                legend columns=-1
            ]
            \addplot table[x={Nodes},y={RingBFT}] {\graphBlt};
            \addplot table[x={Nodes},y={Sharper}] {\graphBlt};
            \addplot table[x={Nodes},y={AHL}] {\graphBlt};
            \legend{\RingBFT{},\Sharper{},\AHL{}}
        \end{axis}
    \end{tikzpicture}
}
\newcommand{\graphCFullTP}{
    \begin{tikzpicture}[plot]
        \begin{axis}[
                xmode=log,
                ymode=log,
                xlabel={\axisCross},
                ylabel={\axisTTP},
                ytick={2000,10000,50000,200000,1000000},
                yticklabels={2K,10K,50K,200K,1M},
                xtick={1,5,10,15,30,60,100},
                xticklabels={0,5,10,15,30,60,100},
                title={(\RN{5}) Impact of X-Shard Workload Rate (Throughput)},
                legend to name={mainlegend},
                legend columns=-1
            ]
            \addplot table[x={X-Precent},y={RingBFT}] {\graphCftp};
            \addplot table[x={X-Precent},y={Sharper}] {\graphCftp};
            \addplot table[x={X-Precent},y={AHL}] {\graphCftp};
            \legend{\RingBFT{},\Sharper{},\AHL{}}
        \end{axis}
    \end{tikzpicture}
}

\newcommand{\graphCLat}{
    \begin{tikzpicture}[plot]
        \begin{axis}[
                xmode=log,
                xlabel={\axisCross},
                ylabel={\axisLat},
                xtick={1,5,10,15,30,60,100},
                xticklabels={0,5,10,15,30,60,100},
                title={(\RN{6}) Impact of X-Shard Workload Rate (Latency)},
                legend to name={mainlegend},
                legend columns=-1
            ]
            \addplot table[x={X-Precent},y={RingBFT}] {\graphClt};
            \addplot table[x={X-Precent},y={Sharper}] {\graphClt};
            \addplot table[x={X-Precent},y={AHL}] {\graphClt};
            \legend{\RingBFT{},\Sharper{},\AHL{}}
        \end{axis}
    \end{tikzpicture}
}

\newcommand{\graphDFullTP}{
    \begin{tikzpicture}[plot]
        \begin{axis}[
                title={(\RN{7}) Impact of Batch Size (Throughput)},
                scaled y ticks = false,
                xmode=log,
                xlabel={Batch Size},
                ylabel={\axisTTP},
                ytick={0,50000,100000,150000,200000},
                yticklabels={0,50K,100K,150K,200K},
                xtick={10,50,100,500,1000,2000,6000},
                xticklabels={10,50,100,500,1K,1.5K,3K},
                legend to name={mainlegend},
                legend columns=-1
            ]
            \addplot table[x={BatchSize},y={RingBFT}] {\graphDftp};
            \addplot table[x={BatchSize},y={Sharper}] {\graphDftp};
            \addplot table[x={BatchSize},y={AHL}] {\graphDftp};
            \legend{\RingBFT{},\Sharper{},\AHL{}}
        \end{axis}
    \end{tikzpicture}
}

\newcommand{\graphDLat}{
    \begin{tikzpicture}[plot]
        \begin{axis}[
                title={(\RN{8}) Impact of Batch Size (Latency)},
                xmode=log,
                xlabel={Batch Size},
                ylabel={\axisLat},
                xtick={10,50,100,500,1000,2000,6000},
                xticklabels={10,50,100,500,1K,1.5K,3K},
                legend to name={mainlegend},
                legend columns=-1
            ]
            \addplot table[x={BatchSize},y={RingBFT}] {\graphDlt};
            \addplot table[x={BatchSize},y={Sharper}] {\graphDlt};
            \addplot table[x={BatchSize},y={AHL}] {\graphDlt};
            \legend{\RingBFT{},\Sharper{},\AHL{}}
        \end{axis}
    \end{tikzpicture}
}

\newcommand{\graphEFullTP}{
    \begin{tikzpicture}[plot]
        \begin{axis}[
                title={(\RN{9})  Impact of Involved Shards (Throughput)},
                ymode=log,
                xlabel={Involved Shards},
                ylabel={\axisTTP},
                ytick={2000,10000,50000,200000,1000000},
                yticklabels={2K,10K,50K,200K,1M},
                xtick={1,3,6,9,15},
                xticklabels={1,3,6,9,15},
                legend to name={mainlegend},
                legend columns=-1
            ]
            \addplot table[x={InvShards},y={RingBFT}] {\graphEftp};
            \addplot table[x={InvShards},y={Sharper}] {\graphEftp};
            \addplot table[x={InvShards},y={AHL}] {\graphEftp};
            \legend{\RingBFT{},\Sharper{},\AHL{}}
        \end{axis}
    \end{tikzpicture}
}

\newcommand{\graphELat}{
    \begin{tikzpicture}[plot]
        \begin{axis}[
                title={(\RN{10}) Impact of Involved Shards (Latency)},
                xlabel={Involved Shards},
                ylabel={\axisLat},xtick={1,3,6,9,15},
                xticklabels={1,3,6,9,15},
                legend to name={mainlegend},
                legend columns=-1
            ]
            \addplot table[x={InvShards},y={RingBFT}] {\graphElt};
            \addplot table[x={InvShards},y={Sharper}] {\graphElt};
            \addplot table[x={InvShards},y={AHL}] {\graphElt};
            \legend{\RingBFT{},\Sharper{},\AHL{}}
        \end{axis}
    \end{tikzpicture}
}

\newcommand{\graphFFullTP}{
    \begin{tikzpicture}[plot]
        \begin{axis}[
                title={(\RN{11}) Impact of Inflight Transactions (Throughput)},
                scaled y ticks = false,
                scaled x ticks = false,
                xlabel={Clients},
                ylabel={\axisTTP},
                ytick={0,20000,40000,60000,80000},
                yticklabels={0,20K,40K,60K,80K},
                xtick={3000,5000,10000,15000,20000},
                xticklabels={3K,5K,10K,15K,20K},
                legend to name={mainlegend},
                legend columns=-1
            ]
            \addplot table[x={MaxInf},y={RingBFT}] {\graphFftp};
            \addplot table[x={MaxInf},y={Sharper}] {\graphFftp};
            \addplot table[x={MaxInf},y={AHL}] {\graphFftp};
            \legend{\RingBFT{},\Sharper{},\AHL{}}
        \end{axis}
    \end{tikzpicture}
}

\newcommand{\graphFLat}{
    \begin{tikzpicture}[plot]
        \begin{axis}[
                scaled y ticks = false,
                scaled x ticks = false,
                xlabel={Clients},
                ylabel={\axisLat},
                xtick={3000,5000,10000,15000,20000},
                xticklabels={3K,5K,10K,15K,20K},
                title={(\RN{12}) Impact of Inflight Transactions (Latency)},
                legend to name={mainlegend},
                legend columns=-1
            ]
            \addplot table[x={MaxInf},y={RingBFT}] {\graphFlt};
            \addplot table[x={MaxInf},y={Sharper}] {\graphFlt};
            \addplot table[x={MaxInf},y={AHL}] {\graphFlt};
            \legend{\RingBFT{},\Sharper{},\AHL{}}
        \end{axis}
    \end{tikzpicture}
}

\newcommand{\graphVC}{
    \definecolor{bar-green}{RGB}{105,200,76}
    \definecolor{bar-yellow}{RGB}{250,183,51}
    \definecolor{bar-orange}{RGB}{255,142,21}
    \definecolor{bar-red}{RGB}{255,13,13}
    \begin{tikzpicture}[plot, scale=1.3]
        \begin{axis}[
                title style={font=\LARGE},
                label style={font=\LARGE},
                tick label style={font=\LARGE},
                ybar,
                bar shift=0,
                bar width=6pt,
                scaled y ticks = false,
                xlabel={Time},
                ylabel={\axisTTP},
                ytick={100000,200000,300000,400000,500000},
                yticklabels={100K,200K,300K,400K,500K},
                xtick={0,10,20,30,40,50,60,70,80,90,100,110},
                title={Impact of Primary Failure in Three Shards}
            ]
            \addplot[fill=bar-green] coordinates    {(0,   486227.1988  )};
            \addplot[fill=bar-green] coordinates    {(5,   483810.0501  )};
            \addplot[fill=bar-yellow] coordinates   {(10,  461601.6865  )};
            \addplot[fill=bar-yellow] coordinates   {(15,  455925.0753  )};
            \addplot[fill=bar-orange] coordinates   {(20,  438830.6552  )};
            \addplot[fill=bar-orange] coordinates   {(25,  429254.5927  )};
            \addplot[fill=bar-red] coordinates      {(30,  422839.787  )};
            \addplot[fill=bar-orange] coordinates   {(35,  421019.971  )};
            \addplot[fill=bar-yellow] coordinates   {(40,  429905.8006  )};
            \addplot[fill=bar-yellow] coordinates   {(45,  451377.6531  )};
            \addplot[fill=bar-yellow] coordinates   {(50,  458127.4441  )};
            \addplot[fill=bar-green] coordinates    {(55,  477158.4789  )};
            \addplot[fill=bar-green] coordinates    {(60,  478181.7192  )};
            \addplot[fill=bar-green] coordinates    {(65,  479625.4481  )};
            \addplot[fill=bar-green] coordinates    {(70,  479267.7121  )};
            \addplot[fill=bar-green] coordinates    {(75,  481760.5334  )};
            \addplot[fill=bar-green] coordinates    {(80,  481440.1456  )};
            \addplot[fill=bar-green] coordinates    {(85,  483303.8275  )};
            \addplot[fill=bar-green] coordinates    {(90,  484434.7415  )};
            \addplot[fill=bar-green] coordinates    {(95,  485660.3749  )};
            \addplot[fill=bar-green] coordinates    {(100, 486528.776  )};
            \addplot[fill=bar-green] coordinates    {(105, 487182.8757  )};
            \addplot[fill=bar-green] coordinates    {(110, 487779.3115  )};
        \end{axis}
    \end{tikzpicture}
}

\newcommand{\graphDPFullTP}{
    \begin{tikzpicture}[plot, scale=.95]
        \begin{axis}[
                title style={font=\huge},
                label style={font=\huge},
                tick label style={font=\huge},
                xlabel style={align=center},
                title={(\RN{1}) Impact of Remote Reads (Throughput)},
                scaled y ticks = false,
                scaled x ticks = false,
                ymin=0,ymax=80000,
                xmin=0,xmax=64,
                extra y ticks={0},
                extra x ticks={0},
                xlabel={Number of Remote Reads\\Each Txn Requires},
                ylabel={\axisTTP},
                ytick={0,20000,40000,60000,80000},
                yticklabels={0,20K,40K,60K,80K},
                xtick={8,16,32,48,64},
                legend to name={mainlegend2},
                legend columns=-1
            ]
            \addplot table[x={Dep},y={RingBFT}] {\graphDPftp};
            \legend{\RingBFT{}}
        \end{axis}
    \end{tikzpicture}
}

\newcommand{\graphDPLat}{
    \begin{tikzpicture}[plot, scale=.95]
        \begin{axis}[
                title style={font=\huge},
                label style={font=\huge},
                tick label style={font=\huge},
                xlabel style={align=center},
                xlabel={Number of Remote Reads\\Each Txn Requires},
                ylabel={\axisLat},
                title={(\RN{2}) Impact of Remote Reads (Latency)},
                ymin=0,ymax=18,
                xmin=0,xmax=64,
                extra y ticks={0},
                extra x ticks={0},
                legend to name={mainlegend2},
                xtick={8,16,32,48,64},
                legend columns=-1
            ]
            \addplot table[x={Dep},y={RingBFT}] {\graphDPlt};
            \legend{\RingBFT{}}
        \end{axis}
    \end{tikzpicture}
}

\begin{document}
\title[RingBFT: Resilient Consensus over Sharded Ring Topology]{RingBFT: Resilient Consensus over Sharded Ring Topology}

\author{Sajjad Rahnama \quad Suyash Gupta \quad Rohan Sogani \quad Dhruv Krishnan \quad Mohammad Sadoghi}
\affiliation{%
  \institution{Exploratory Systems Lab \\University of California Davis}
}

\begin{abstract}
The recent surge in federated data management applications has brought forth concerns about the security 
of underlying data and the consistency of replicas in the presence of malicious attacks.
A prominent solution in this direction is to employ a permissioned blockchain framework that is modeled around
traditional Byzantine Fault-Tolerant (\BFT{}) consensus protocols.
Any federated application expects its data to be globally scattered to achieve faster access. 
But, prior works have shown that traditional \BFT{} protocols are slow. 

This has led to the rise of 
sharded-replicated blockchains.
Existing \BFT{} protocols for these sharded blockchains are efficient if client transactions 
require access to a single-shard, but face performance degradation if there is a cross-shard 
transaction that requires access to multiple shards.
As cross-shard transactions are common, to resolve this dilemma, we present \RingBFT{}, 
a novel meta-\BFT{} protocol for sharded blockchains.
\RingBFT{} requires shards to adhere to the ring order, and 
follow the principle of process, forward, and re-transmit while ensuring the communication between shards is linear. 
Our evaluation of \RingBFT{} against state-of-the-art sharding \BFT{} protocols illustrates that 
\RingBFT{} achieves up to $18\times$ higher throughput, gracefully scales to nearly $500$ 
globally distributed nodes, and achieves a peak throughput of $1.2$ million transactions per second.

\end{abstract}

\maketitle

\section{Introduction}
\label{s:intro}
A growing interest in {\em federated data management} illustrates 
an increased demand for multi-party database management~\cite{federated-databases,federated-geospatial,federated-query-optimization}.
In these multi-party systems, a common database is maintained by several parties. 
As all of these parties cannot be at the same location, so the system needs to be 
decentralized, which implies that the database is distributed.
There are two key ways in which a distributed database can be managed by 
multiple parties: {\em replication} and {\em sharding}~\cite{deneva,calvin,easyc,dapd,qstore,tpbook}.

In a replicated system, each party holds a {\em copy} of the database.
As a result, the effects of each client transaction are replicated across all the parties (replicas). 
In a sharded system, each party maintains a subset (shard) of the database.
Hence, each party can independently handle incoming client transactions that require access to its shard.

One of the factors that advocates the use of replicated databases is their ability to handle 
{\em failure} of one or more replicas.
This necessitates the need for keeping all the replicas at the same state.
To achieve this task, databases employ {\em crash-fault tolerant} protocols such as Paxos~\cite{paxos} 
and Raft~\cite{raft} 
to help all replicas reach a common order for each client transaction.
However, one or more replicas can get {\em compromised} due to a malicious attack. 
A compromised replica may wish: 
(i) to exclude transactions of some clients, 
(ii) to make the system unavailable to clients, and
(iii) to make replicas inconsistent.
These malicious attacks are so common that one estimate shows that cyberattacks 
alone cost the U.S. economy around \$57 billion dollars in 2016~\cite{ecodam}.
As a result, not all the replicas can be trusted.

A recent solution to guarantee secure federated data management is through the use of 
{\em permissioned} blockchain technology~\cite{bc-processing,blockchain-book}. 
These permissioned blockchains require their replicas to agree on the 
order for each transaction by participating in a {\em Byzantine-Fault Tolerant} ({\em \BFT{}}) consensus protocol. 
Post consensus, each replica logs the ordered transaction in a {\em block} 
that is part of an immutable append-only ledger--{\em blockchain}.
A blockchain is termed as immutable because each new block includes the hash of the previous block, 
and it allows verifying the state of the participating replicas.

\begin{figure}[t]
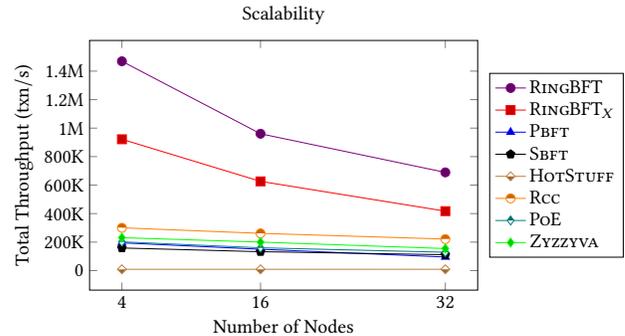

    \centering
    \graphOtherProtocols
    \caption{Comparing scalability of different \BFT{} protocols. 
	In this figure, we depict throughput of single-primary, multiple-primaries, geographically-scalable, and 
	sharding \BFT{} protocols.
	For \BName{RingBFT}, we require each shard to have number of replicas stated on x-axis.}
    \label{fig:all-protocols}
\end{figure}

In this paper, we present a novel meta-\BFT{} protocol \RingBFT{} that guards against Byzantine attacks, 
achieves high throughput, and incurs low latency.
Our \RingBFT{} protocol explores the landscape of sharded-replicated databases, and helps to scale 
{\em permissioned} blockchains, which in turn helps in designing efficient federated data management systems.
\RingBFT{} aims to make consensus inexpensive even when transactions require access to {\em multiple shards}.
In the rest of this section, we motivate our design choices.
To highlight the need for \RingBFT{}, we will be referring to Figure~\ref{fig:all-protocols}, which illustrates 
the throughput attained by the system when employing different \BFT{} consensus protocols.

\subsection{Challenges for Efficient BFT Consensus}
Existing permissioned blockchain applications employ traditional \BFT{} protocols 
to achieve consensus among their replicas~\cite{pbft,zyzzyva,caper,sharper}.
Over the past two decades, these \BFT{} protocols have undergone a series of evolutions to guarantee 
resilience against Byzantine attacks, while ensuring high throughput and low latency.
The seminal work by Castro and Liskov~\cite{pbft,pbftj} led to the design of the first practical \BFT{} protocol, 
\PBFT{}, which advocates a {\em primary-backup} paradigm where primary initiates the consensus and all the backups 
follow primary's lead.
\PBFT{} achieves consensus among the replicas in three phases, of which two require quadratic communication complexity.
Following this, several exciting primary-backup protocols, such as \ZZ{}~\cite{zyzzyvaj}, \SBFT{}~\cite{sbft}, 
and \PoE{}~\cite{poe}, have been proposed that try to yield higher throughputs from \BFT{} consensuses.
We use Figure~\ref{fig:all-protocols} to illustrate the benefits of these optimizations over \PBFT{}.
Prior works~\cite{steward,geobft} have illustrated that these {\em single} primary protocols are essentially {\em centralized} 
and prevent scaling the system to a large number of replicas.

An emerging solution to balance load among replicas is to employ {\em multi-primary} protocols like 
Honeybadger~\cite{honeybadger} and
\Name{Rcc}~\cite{bambft,multibft} that permit all replicas to act as primaries by running multiple consensuses concurrently.
However, multi-primary protocols also face scalability limitations as despite concurrent consensuses, 
each transaction requires communication between all the replicas.
Moreover, if the replicas are separated by geographically large distances, then these protocols incur low throughput and 
high latency due to low bandwidth and high round-trip time.
This led to the design of {\em topology-aware} protocols, such as \STW{}~\cite{steward} and \Name{Geobft}~\cite{geobft}, 
which cluster replicas based on their geographical distances. 
For instance, \Name{Geobft} expects each cluster to first locally order its client transaction by 
running the \PBFT{} protocol, and then exchange this ordered transaction with all the other clusters.
Although \Name{Geobft} is highly scalable, it necessitates total replication, which forces communicating 
large messages among geographically distant replicas.

\subsection{The Landscape for Sharding}
To mitigate the costs associated with replicated databases, a common strategy is to employ the {\em sharded-replicated} 
paradigm~\cite{distdb}. 
In a sharded-replicated database, the data is distributed across a set of {\em shards} where each shard manages a 
unique {\em partition} of the data. 
Further, each shard replicates its partition of data to ensure availability under failures.  
If each transaction accesses only one shard, then these sharded systems can fetch high throughput as 
consensus is restricted to a subset of replicas.

\AHL{}~\cite{ahl} was the first permissioned blockchain system to employ principles of sharding. 
\AHL's novel design helps to scale blockchain systems to hundreds of replicas across the globe and achieve high 
throughput for {\em single-shard} transactions. 
To tackle {\em cross-shard} transactions that require access to data in multiple shards, 
\AHL{} designates a set of replicas as a reference committee, which {\em globally orders} all such transactions.
Following \AHL's design, \Sharper{}~\cite{sharper} presents a sharding protocol that eliminates the barrier to rely on the reference committee for ordering cross-shard transactions, 
but necessitates {\em global and quadratic communication among all replicas of all the participating shards}.

{\bf Why} \BName{RingBFT}? 
Decades of research in database community has illustrated that {\em cross-shard} transactions are common~\cite{hekaton,spanner,calvin,deneva}.
In fact, heavy presence of these cross-shard transactions has led to development of several 
concurrency control~\cite{mvcc,deneva,qstore} and commit protocols~\cite{2pc,3pc,easyc}.
Hence, in this paper, we present our \RingBFT{} protocol that significantly reduces the costs associated with cross-shard transactions.

Akin to \AHL{} and \Sharper{}, \RingBFT{} assumes that the read-write sets of each transaction are 
known prior to the start of consensus. 
Given this, \RingBFT{} guarantees consensus for each cross-shard transaction in {\bf \em at most two rotations around the ring}.
In specific, \RingBFT{} envisions each shard participating in {\em multiple circular flows} or rings, simultaneously.
For each cross-shard transaction, \RingBFT{} follows the principle of {\bf \em process, forward, and re-transmit}.
This implies that each shard performs consensus on the transaction and forwards it to the next shard. 
This flow continues until each shard is aware of the fate of the transaction.
However, the real challenge with cross-shard transactions is to manage conflicts and to prevent deadlocks, 
which \RingBFT{} achieves by requiring cross-shard transactions to travel in {\bf \em ring order}.
Despite all of this, \RingBFT{} ensures communication between the shards is {\bf \em linear}, exhibiting a neighbor-to-neighbor communication.
This minimalistic design has allowed \RingBFT{} to achieve unprecedented gains in throughput and has allowed 
us to scale \BFT{} protocols to nearly $500$ nodes globally.
The benefits of our \RingBFT{} protocol are visible from Figure~\ref{fig:all-protocols} 
where we run \RingBFT{} in a system of $9$ shards with each shard having $4$, $16$ and $32$ replicas.
Further, we show the throughput with $0$ (\RingBFT{}) and $15\%$ (\RingBFT{}$_X$) cross-shard transactions.
We now list down our contributions.
\begin{enumerate}[wide]
\item We present a novel meta-\BFT{} protocol for sharded-replicated permissioned blockchain systems that requires 
participating shards to adhere to the ring order.
We term \RingBFT{} as ``meta'' because it can employ any single-primary protocols within each shard.

\item Our \RingBFT{} protocol presents a scalable consensus for cross-shard transactions that neither depends 
on any centralized committee nor requires all-to-all communication. 

\item We show that the cross-shard consensus provided by \RingBFT{} is safe, and live, despite
any Byzantine attacks.

\item We evaluate \RingBFT{} on our \ResilientDB%
\footnote{
\ResilientDB{} is open-sourced at {https://resilientdb.com/} and 
its source-code is available at {https://github.com/resilientdb/resilientdb}.
} 
framework~\cite{r-evalpaper,resilientdb-demo,phd-workshop,tut-vldb20,tut-middleware19,geobft,poe,multibft} 
against two state-of-the-art \BFT{} protocols for permissioned sharded systems, 
\AHL{}~\cite{ahl}, and \Sharper{}~\cite{sharper}. 
Our results show that \RingBFT{} easily scales to $428$ globally-distributed nodes, 
and achieves up to $18\times$ and $4\times$ times higher throughput than \AHL{} and \Sharper{}, respectively.

\end{enumerate}

\section{Cross-Shard Dilemma}
\label{s:back}
For any sharded system, ordering a single-shard transaction is trivial
as such a transaction requires access to only one shard. 
This implies that achieving consensus on a single-shard transaction just requires 
running a standard \BFT{} protocol. 
Further, single-shard transactions support {\em parallelism} as each shard can order its transaction in parallel, 
this without any communication between shards.

On the other hand, cross-shard transactions are {\em complex}. 
Not only do they require communication between shards but also their fate depends on the consent of each of the 
{\em involved shards}.
Further, two or more cross-shard transactions can {\em conflict} if they require access to same data. 
Such conflicts can cause one or more transactions to abort or worse, can create a {\em deadlock}.
Hence, we need an efficient protocol to order these cross-shard transactions, which ensures that the system 
is both {\em safe} and {\em live}.

{\bf \em Designated Committee (AHL).} 
One way to order cross-shard transactions is to designate a set of replicas with this task. 
AHL~\cite{ahl} defines a reference committee that assigns an order to each cross-shard transaction, which 
requires running \PBFT{} protocol among all the members of the reference committee.
Next, reference committee members run the Two-phase commit (\twopc{}) protocol with all the replicas of involved shards. 
Notice that the \twopc{} protocol requires: 
(1) each shard to send a vote to the reference committee, 
(2) reference committee collects these votes and takes a decision (abort or commit), and 
(3) each shard implements the decision.
Firstly, this solution requires each shard to run the \PBFT{} protocol to decide on the vote. 
Secondly, reference committee needs to again run \PBFT{} to reach a common decision.
Finally, these multiple phases of 2PC require {\em all-to-all communication} between
the replicas of each shard and the replicas of reference committee.

{\bf \em Initiator Shard (Sharper).}
Another way to process a cross-shard transaction is to designate one of the involved shards as the {\em initiator shard}. 
\Sharper~\cite{sharper} employs this approach by requiring each cross-shard transaction to be managed by the primary replica 
of one of the involved shards.
This {\em initiator primary} proposes the transaction to the primaries of other shards. 
Next, these primaries propose this transaction within their own shards. 
Following this there is an all-to-all communication between replicas of all the involved shards.

\section{System Model}
To explain our \RingBFT{} protocol in detail, we first lay down some notations and assumptions.
Our system comprises of a set $\Shards{}$ of shards where each shard $\Shard{S}$ provides 
a replicated service. 
In specific, each shard $\Shard{S}$ manages a {\em unique partition of the data}, which is 
replicated by a set $\Replicas{\Shard{S}}$ of \emph{replicas}.

In each shard $\Shard{S}$, there are $\Faulty \subseteq \Replicas{\Shard{S}}$ {\em Byzantine} replicas, of which
$\NonFaulty = \Replicas{\Shard{S}} \difference \Faulty$ are {\em non-faulty} replicas.
We expect non-faulty replicas to follow the protocol and act deterministic, that is, 
on identical inputs, all non-faulty replicas must produce identical outputs. 
We write $\z = \abs{\Shards}$ to denote the total number of shards and
$\n = \abs{\Replicas{\Shard{S}}}$, $\f = \abs{\Faulty}$, and $\nf = \abs{\NonFaulty}$ to denote the 
number of replicas, faulty replicas, and non-faulty replicas, respectively, in each shard.

{\bf \em Fault-Tolerance Requirement.} 
Traditional, \BFT{} protocols such as \PBFT{}, \ZZ{}, and \SBFT{} expect a total replicated system 
where the total number of Byzantine replicas are less than {\em one-third} of the total replicas in the system.
In our sharded-replicated model, we adopt a slightly weaker setting where at each shard 
the total number of Byzantine replicas are less than {\em one-third} of the total replicas in that shard. 
In specific, at each shard $\Shard{S}$, we expect $\n \ge 3\f+1$.
This does not imply that we want each shard to have an equal number of replicas. 
Each shard can have a different number of replicas till less than one-third are byzantine.
This requirement is in accordance with existing works in Byzantine sharding space~\cite{ahl,sharper}.

{\bf \em Cross-Shard Transactions.}
Each shard $\Shard{S} \in \Shards{}$ can receive a {\em single-shard} or {cross-shard} transaction. 
A single-shard transaction for $\Shard{S}$ leads to {\em intra-shard} communication, 
that is, all the messages necessary to order this transaction are exchanged among the replicas of $\Shard{S}$.
On the other hand, a cross-shard transaction requires access to data from a subset of shards (henceforth we use the abbreviation \CST{} to refer 
to a cross-shard transaction). 
We denote this subset of shards as $\Involved{}$ where $\Involved \subseteq \Shards$, and refer to it as {\em involved shards}.
Each \CST{} can be termed as {\em simple} or {\em complex}. 
A simple \CST{} is a collection of fragments where each shard can independently run consensus and execute its fragment.
On the other hand, a complex \CST{} includes dependencies, that is, an involved shard may require access to data from other 
involved shards to execute its fragment.

{\bf \em Deterministic Transactions.}
We define a deterministic transaction as the transaction for which the data-items it will read/write 
are known prior to the start of the consensus~\cite{calvin,qstore}.
Given a deterministic transaction, a replica can determine which data-items accessed by this transaction 
are present in its shard.

{\bf \em Ring Order.}
We assume shards in set $\Shards{}$ are {\em logically} arranged in a {\em ring topology}. 
In specific, each shard $\Shard{S} \in \Shards$ has a position in the ring, which we denote by $\ID{\Shard{S}}$, 
$1 \leq \ID{\Shard{S}} \leq \abs{\Shards}$.
\RingBFT{} employs these identifiers to specify the flow of a \CST{} or {\bf \em ring order}.
For instance, a simple ring policy can be that each \CST{} is processed by the involved shards in the increasing order 
of their identifiers.
\RingBFT{} can also adopt other complex permutations of these identifiers for determining the flow across the ring.

{\bf \em Authenticated Communication.}  
We assume that each message exchanged among clients and replicas is \emph{authenticated}.
Further, we assume that Byzantine replicas are unable to impersonate non-faulty replicas. 
Notice that authenticated communication is a minimal requirement to deal with Byzantine behavior.
For intra-shard communication, we employ cheap {\em message authentication codes} 
(\Name{MAC}s), while for cross-shard communication we employ digital signatures (\Name{DS}) 
to achieve authenticated communication.
\Name{MAC}s facilitate symmetric cryptography by requiring each pair of communicating nodes to share a  {\em secret key}.
We expect non-faulty replicas to keep their {\em secret keys} hidden. 
\Name{DS} follow asymmetric cryptography. 
In specific, prior to signing a message, each replica generates a pair of {\em public-key} and {\em private-key}. 
The signer keeps the private-key hidden and uses it to sign a message. 
Each receiver authenticates the message using the corresponding public-key. 
Although \Name{MAC}s are cheaper than \Name{DS}, they cannot guarantee {\em non-repudiation}.
We require non-repudiation property during cross-shard communication as it helps to prove that the message communicated 
was sent by the sender and the message's contents were not fabricated.

In the rest of this manuscript, if a message $m$ is signed by a replica $\Replica{r}$ using \Name{DS}, we represent it as 
$\SignMessage{m}{\Replica{r}}$ to explicitly identify replica $\Replica{r}$.
Otherwise, we assume that the message employs \Name{MAC}.

To ensure message integrity, we employ a \emph{collision-resistant cryptographic hash function} $\Hash{\cdot}$ that maps an arbitrary value $v$ to a constant-sized digest $\Hash{v}$~\cite{cryptobook}. We assume that there is a negligible probability to find another value $v'$, $v \neq v'$, such that $\Hash{v} = \Hash{v'}$. 
Further, we refer to a message as {\em well-formed} if a non-faulty receiver can validate the \Name{DS} or \Name{MAC}, 
verify the integrity of the message digest, and determine that the sender of the message is also the creator.

\section{RingBFT Consensus Protocol}
To achieve efficient consensus in sharded-replicated databases, we employ our \RingBFT{} protocol. 
While designing our \RingBFT{} protocol, we set following goals:

\begin{enumerate}[wide,nosep,label=(G\arabic*),ref={G\arabic*}]
\item Inexpensive consensus of single-shard transactions.
\item Flexibility of employing different existing consensus protocols for intra-shard consensus.
\item Deadlock-free two-ring consensus of deterministic cross-shard transactions.
\item Cheap communication between globally distributed shards.
\end{enumerate}

We define the {\em safety} and {\em liveness} guarantees provided by our \RingBFT{} protocol. 
\begin{definition}
Let $\Shards{}$ be a system of shards and $\Replicas{\Shard{S}}$ be a set of replicas in some shard 
$\Shard{S} \in \Shards{}$. 
Each run of a {\em consensus protocol} in this system should satisfy the following requirements:
\begin{description}[nosep]
\item[\bf Involvement] Each $\Shard{S} \in \Shards{}$ processes a transaction if $\Shard{S} \in \Involved{}$.
\item[\bf Termination] Each non-faulty replica in $\Replicas{\Shard{S}}$ executes a transaction.
\item[\bf Non-divergence] (intra-shard) All non-faulty replicas in $\Replicas{\Shard{S}}$ execute the same transaction.
\item[\bf Consistence] (cross-shard) Each non-faulty replica in $\Shards{}$ executes a conflicting transaction in same order.
\end{description}
\end{definition}
In traditional replicated systems, non-divergence implies {\em safety}, while termination implies {\em liveness}.
For a sharded-replicated system like \RingBFT{}, we need stronger guarantees. 
If a transaction requires access to only one shard, safety is provided by involvement and non-divergence, while termination sufficiently guarantees liveness.
For a cross-shard transaction, to guarantee safety, we also need consistency apart from involvement and non-divergence, 
while liveness is provided using involvement and termination.

\RingBFT{} guarantees safety in an asynchronous setting. 
In such a setting, messages may get lost, delayed, or duplicated, and up to $\f$ replicas in each shard may act Byzantine.
However, \RingBFT{} can only provide liveness during periods of synchrony. 
Notice that these assumptions are no harder than those required by existing protocols~\cite{pbft,ahl,sharper}.

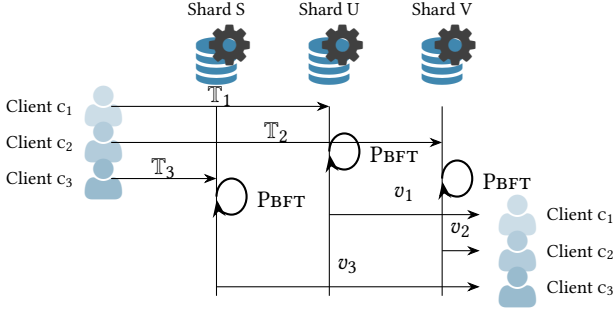
\begin{figure}[t]
    \centering
    \begin{tikzpicture}[yscale=0.6,xscale=0.5]
        \dbcog[colB]{3}{0}
        \dbcog[colB]{6}{0}
        \dbcog[colB]{9}{0}
        
        \node[above=11pt,label] at (3, 0)  {\footnotesize Shard \Shard{S}};
        \node[above=11pt,label] at (6, 0)  {\footnotesize Shard \Shard{U}};
        \node[above=11pt,label] at (9, 0)  {\footnotesize Shard \Shard{V}};

	\dbuser[colN!20]{0}{-1}
	\node[left=8pt,label] at (0, -1)  {\footnotesize Client $\Client{1}$};
	\path (0.2, -1) edge[->] node[above=-3pt] {$\Transaction{1}$} (6, -1);

	\dbuser[colN!30]{0}{-1.8}
	\node[left=8pt,label] at (0, -1.8)  {\footnotesize Client $\Client{2}$};
	\path (0.2, -1.8) edge[->] node[above=-3pt] {$\Transaction{2}$} (9, -1.8);

	\dbuser[colN!40]{0}{-2.6}
	\node[left=8pt,label] at (0, -2.6)  {\footnotesize Client $\Client{3}$};
	\path (0.2, -2.6) edge[->] node[above=-3pt] {$\Transaction{3}$} (3, -2.6);

	\draw[thick, ->] (6,-2) arc (180:-170:4mm) node[right=12pt] {$\PBFT{}$};
	\draw[thick, ->] (9,-2.6) arc (180:-170:4mm) node[right=12pt] {$\PBFT{}$};
	\draw[thick, ->] (3,-3) arc (180:-170:4mm) node[right=12pt] {$\PBFT{}$};

	\dbuser[colN!20]{11}{-3.4}
	\node[right=8pt,label] at (11, -3.4)  {\footnotesize Client $\Client{1}$};
	\path (6, -3.4) edge[->] node[above=2pt] {$\Response{1}$} (10, -3.4);

	\dbuser[colN!30]{11}{-4.2}
	\node[right=8pt,label] at (11, -4.2)  {\footnotesize Client $\Client{2}$};
	\path (9, -4.2) edge[->] node[above=2pt] {$\Response{2}$} (10, -4.2);

	\dbuser[colN!40]{11}{-5}
	\node[right=8pt,label] at (11, -5)  {\footnotesize Client $\Client{3}$};
	\path (3, -5) edge[->] node[above=2pt] {$\Response{3}$} (10, -5);

        \path (3, -1) edge ++(0, -4.2)
              (6, -1) edge ++(0, -4.2)
              (9, -1) edge ++(0, -4.2);

    \end{tikzpicture}
    \caption{\RingBFT{} consensus for single-shard transactions. Each of the three shards $\Shard{S}$, 
$\Shard{U}$, and $\Shard{V}$ receive transactions $\Transaction{1}$, $\Transaction{2}$, and $\Transaction{3}$ 
from their respective clients $\Client{1}$, $\Client{2}$, and $\Client{3}$ to execute. 
Each shard independently run $\PBFT{}$ consensus, and sends responses to respective clients.}
     \label{fig:ringbft-single-shard}
\end{figure}

\subsection{Single-Shard Consensus}
\label{ss:single-shard}
To order and execute single-shard transactions is trivial. 
For this task, \RingBFT{} employs one of the many available primary-backup consensus protocols and runs them at each shard.
In the rest of this section, without the loss of generality, we assume that \RingBFT{} employs the \PBFT{} consensus protocol 
to order single-shard transactions.
We use the following example to explain \RingBFT's single-shard consensus.

\begin{example}\label{ex:single-shard}
Assume a system that comprises of three shards $\Shard{S}$, $\Shard{U}$, and $\Shard{V}$. 
Say client $\Client{1}$ sends $\Transaction{1}$ to $\Shard{S}$, $\Client{2}$ sends $\Transaction{2}$ to $\Shard{U}$, and 
client $\Client{3}$ sends $\Transaction{3}$ to $\Shard{V}$. 
On receiving the client transaction, the primary of each shard initiates the \PBFT{} consensus protocol among its replicas. 
Once each replica successfully orders the transaction, it sends a response to the client.
Such a flow is depicted in Figure~\ref{fig:ringbft-single-shard}.
\end{example}

It is evident from Example~\ref{ex:single-shard} that there is no communication among the shards. 
This is the case because each transaction requires access to data available inside only one shard.
Hence, ordering single-shard transactions for shard $\Shard{S}$ requires running the \PBFT{} protocol among the 
replicas of $\Shard{S}$ without any synchronization with other shards.

\subsection{Cross-Shard Consensus: Process \& Forward}
\label{ss:cross-shard}
In this section, we illustrate how \RingBFT{} guarantees consensus of every {\em deterministic} cross-shard transaction (\CST{}) 
in {\em at most two rotations across the ring}.
To order a \CST{}, \RingBFT{} requires shards to adhere to the {\em ring order}, and follow the 
principle of process, forward, and re-transmit while ensuring the communication between shards is {\em linear}.
We use the following example to illustrate what we mean by following the ring order.

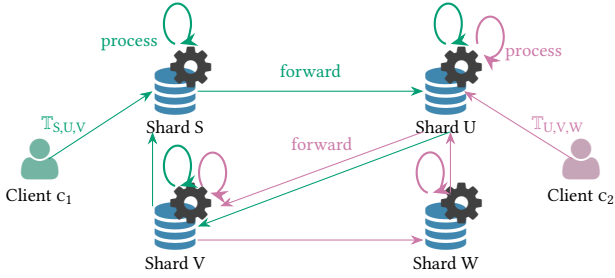
\begin{figure}[t]
    \centering
    \begin{tikzpicture}[xscale=0.6,yscale=0.9]
        \dbcog[colB]{3}{0}
        \dbcog[colB]{9}{0}
	\dbcog[colB]{3}{-2}
        \dbcog[colB]{9}{-2}
        
        \node[below=5pt,label] at (3, 0)  {\footnotesize Shard \Shard{S}};
        \node[below=5pt,label] at (9, 0)  {\footnotesize Shard \Shard{U}};
        \node[below=5pt,label] at (3, -2) {\footnotesize Shard \Shard{V}};
	\node[below=5pt,label] at (9, -2) {\footnotesize Shard \Shard{W}};

	\dbuser[colG!55]{0}{-1}
	\dbuser[colP!55]{12}{-1}
	
	\node[below=5pt,label] at (0, -1)  {\footnotesize Client $\Client{1}$};
	\node[below=5pt,label] at (12, -1) {\footnotesize Client $\Client{2}$};

	\path (0.2, -1) edge[colG,->] node[left=2pt] {\footnotesize $\Transaction{\Shard{S},\Shard{U},\Shard{V}}$} (2.5, 0);
	\draw[thick,colG, ->] (3,0.7) arc (260:-80:3mm) node[left=5pt] {\footnotesize process};
	\path (3.5, 0) edge[colG,->] node[above=2pt] {\footnotesize forward} (8.5, 0);
	\draw[thick,colG, ->] (9,0.7) arc (260:-80:3mm);
	\path (9.1, -0.6) edge[colG,->] (3.5, -2);
	\draw[thick,colG, ->] (3,-1.4) arc (260:-80:3mm);
	\path (2.5, -1.7) edge[colG,->] (2.5, -0.6);

	\path (11.8, -1) edge[colP,->] node[right=2pt] {\footnotesize $\Transaction{\Shard{U},\Shard{V},\Shard{W}}$} (9.4, 0);
	\draw[thick,colP, ->] (9.7,0.7) arc (200:-120:3mm) node[right=5pt] {\footnotesize process};
	\path (8.5, -0.6) edge[colP,->] node[above=4pt] {\footnotesize forward} (4, -1.7);
	\draw[thick,colP, ->] (3.6,-1.4) arc (220:-100:3mm);
	\path (3.5, -2.2) edge[colP,->] (8.5, -2.2);
	\draw[thick,colP, ->] (8.6,-1.5) arc (260:-80:3mm);
	\path (9.1, -1.5) edge[colP,->] (9.1, -0.6);

    \end{tikzpicture}
    \caption{\RingBFT's concurrent consensus of two cross-shard transactions 
	$\Transaction{\Shard{S},\Shard{U},\Shard{V}}$ and $\Transaction{\Shard{U},\Shard{V},\Shard{W}}$ across four shards. 
	The prescribed ring order is $\Shard{S} \rightarrow \Shard{U} \rightarrow \Shard{V} \rightarrow \Shard{W}$.}
     \label{fig:ringbft-cross-shard}
\end{figure}

\begin{example}\label{ex:cross-shard}
Assume a system that comprises of four shards $\Shard{S}$, $\Shard{U}$, $\Shard{V}$, and $\Shard{W}$ where the 
ring order has been defined as $\Shard{S} \rightarrow \Shard{U} \rightarrow \Shard{V} \rightarrow \Shard{W}$.
Say client $\Client{1}$ wants to process a transaction $\Transaction{\Shard{S},\Shard{U},\Shard{V}}$ that requires 
access to data from shards $\Shard{S}$, $\Shard{U}$, and $\Shard{V}$, and
client $\Client{2}$ wants to process a transaction $\Transaction{\Shard{U},\Shard{V}, \Shard{W}}$ that requires 
access to data from shards $\Shard{U}$, $\Shard{V}$, and $\Shard{W}$ (refer to Figure~\ref{fig:ringbft-cross-shard}).   
In this case, client $\Client{1}$ sends its transaction to the primary of shard $\Shard{S}$ while $\Client{2}$ sends 
its transaction to primary of $\Shard{U}$.  
On receiving $\Transaction{\Shard{S},\Shard{U},\Shard{V}}$, replicas of $\Shard{S}$ process the transaction and forward 
it to replicas of $\Shard{U}$. 
Next, replicas of $\Shard{U}$ process $\Transaction{\Shard{S},\Shard{U},\Shard{V}}$ and forward it to replicas of $\Shard{V}$. 
Finally, replicas of $\Shard{V}$ process $\Transaction{\Shard{S},\Shard{U},\Shard{V}}$ and send it back to replicas of $\Shard{S}$, 
which reply to client $\Client{1}$. 
Similar flow takes place while ordering transaction $\Transaction{\Shard{U},\Shard{V},\Shard{W}}$.
\end{example}

Although Example~\ref{ex:cross-shard} illustrates \RingBFT's design, it is unclear 
how multiple concurrent \CST{} are ordered in a deadlock-free manner. 
In specific, we wish to answer following questions regarding the design of our \RingBFT{} protocol.
\begin{enumerate}[wide,nosep,label=(Q\arabic*),ref={Q\arabic*}]
\item Can a shard concurrently order multiple \CST{}?
\item How does \RingBFT{} handle conflicting transactions?
\item Can shards running \RingBFT{} protocol deadlock?
\item How much communication is required between two shards?
\end{enumerate}
To answer these questions, we first present the transactional flow of a cross-shard transaction undergoing 
\RingBFT{} consensus, following which we lay down the steps of our \RingBFT{} consensus protocol.

\begin{figure*}
    \centering
    \begin{tikzpicture}[yscale=0.4,xscale=1.05]
        \draw[draw=colB,fill=colB!65] (3.5, 1.5) rectangle (4.5, 9);
	\draw[draw=colB,fill=colB!65] (7,   4.5) rectangle (8,   -7.5);
	\draw[draw=colB,fill=colB!65] (11,  9)   rectangle (12,  -7.5);
	\draw[draw=colB,fill=colB!65] (13.5, 1.5) rectangle (14.5, 9);
        
        \draw[thick,draw=black!75] (0.75, 10.5) edge[green!50!black!90] ++(15.5, 0)
                                   (0.75,  6) edge ++(15.5, 0)
                                   (0.75,  7) edge ++(15.5, 0)
                                   (0.75,  8) edge ++(15.5, 0)
                                   (0.75,  9) edge[blue!50!black!90] ++(15.5, 0)

				   (0.75, 1.5) edge ++(15.5, 0)
                                   (0.75, 2.5) edge ++(15.5, 0)
                                   (0.75, 3.5) edge ++(15.5, 0)
                                   (0.75, 4.5) edge[blue!50!black!90] ++(15.5, 0)

				   (0.75, -3) edge ++(15.5, 0)
                                   (0.75, -2) edge ++(15.5, 0)
                                   (0.75, -1) edge ++(15.5, 0)
                                   (0.75, 0) edge[blue!50!black!90] ++(15.5, 0)

				   (0.75, -7.5) edge ++(15.5, 0)
                                   (0.75, -6.5) edge ++(15.5, 0)
                                   (0.75, -5.5) edge ++(15.5, 0)
                                   (0.75, -4.5) edge[blue!50!black!90] ++(15.5, 0);

       \draw[thin,draw=black!75]  (1,   1.5) edge ++(0, 3)
                                  (2,   1.5) edge ++(0, 3)
                                  (3.5, 1.5) edge ++(0, 3)
                                  (4.5, 1.5) edge ++(0, 3)
                                  (5.5, 1.5) edge ++(0, 3)
				  (7,   1.5) edge ++(0, 3)
                                  (8,   1.5) edge ++(0, 3)
				  (9,   1.5) edge ++(0, 3)
                                  (10.5, 1.5) edge ++(0, 3)
				  (11, 1.5) edge ++(0, 3)
				  (12, 1.5) edge ++(0, 3)
				  (13, 1.5) edge ++(0, 3)
				  (13.5, 1.5) edge ++(0, 3)
				  (14.5, 1.5) edge ++(0, 3)
				  (15.5, 1.5) edge ++(0, 3)
				  (16, 1.5) edge ++(0, 3)
                                  
                                  (1,   6) edge ++(0, 4.5)
                                  (2,   6) edge ++(0, 4.5)
                                  (3.5, 6) edge ++(0, 4.5)
                                  (4.5, 6) edge ++(0, 4.5)
                                  (5.5, 6) edge ++(0, 4.5)
				  (7,   6) edge ++(0, 4.5)
                                  (8,   6) edge ++(0, 4.5)
				  (9,   6) edge ++(0, 4.5)
                                  (10.5, 6) edge ++(0, 4.5)
				  (11, 6) edge ++(0, 4.5)
				  (12, 6) edge ++(0, 4.5)
				  (13, 6) edge ++(0, 4.5)
				  (13.5, 6) edge ++(0, 4.5)
				  (14.5, 6) edge ++(0, 4.5)
				  (15.5, 6) edge ++(0, 4.5)
				  (16, 6) edge ++(0, 4.5)

				  (1,   -4.5) edge ++(0, -3)
                                  (2,   -4.5) edge ++(0, -3)
                                  (3.5, -4.5) edge ++(0, -3)
				  (4.5, -4.5) edge ++(0, -3)
                                  (5.5, -4.5) edge ++(0, -3)
                                  (7,   -4.5) edge ++(0, -3)
                                  (8,   -4.5) edge ++(0, -3)
				  (9,   -4.5) edge ++(0, -3)
                                  (10.5,-4.5) edge ++(0, -3)
				  (11,  -4.5) edge ++(0, -3)
				  (12,  -4.5) edge ++(0, -3)
				  (13,  -4.5) edge ++(0, -3)
				  (13.5,  -4.5) edge ++(0, -3)
				  (14.5,  -4.5) edge ++(0, -3)
				  (15.5,  -4.5) edge ++(0, -3)
				  (16,  -4.5) edge ++(0, -3);

	\node[left] at (0.8, 6) {\footnotesize $\Replica{r}_1$};
        \node[left] at (0.8, 7) {\footnotesize $\Replica{r}_2$};
        \node[left] at (0.8, 8) {\footnotesize $\Replica{r}_3$};
        \node[left] at (0.8, 9) {\footnotesize $\Primary{\Shard{S}}$};
        \node[left] at (0.8, 10.5) {\footnotesize $\Client{}$};

        \node[left] at (0.8, 1.5) {\footnotesize $\Replica{r}_1$};
        \node[left] at (0.8, 2.5) {\footnotesize $\Replica{r}_2$};
        \node[left] at (0.8, 3.5) {\footnotesize $\Replica{r}_3$};
        \node[left] at (0.8, 4.5) {\footnotesize $\Primary{\Shard{U}}$};

	\node[left] at (0.8, -3)  {\footnotesize $\Replica{r}_1$};
        \node[left] at (0.8, -2)  {\footnotesize $\Replica{r}_2$};
        \node[left] at (0.8, -1)  {\footnotesize $\Replica{r}_3$};
        \node[left] at (0.8, 0)   {\footnotesize $\Primary{\Shard{V}}$};

	\node[left] at (0.8, -7.5) {\footnotesize $\Replica{r}_1$};
        \node[left] at (0.8, -6.5) {\footnotesize $\Replica{r}_2$};
        \node[left] at (0.8, -5.5) {\footnotesize $\Replica{r}_3$};
        \node[left] at (0.8, -4.5) {\footnotesize $\Primary{\Shard{W}}$};

        \path[->] (1, 10.5) edge node[above=-4pt,xshift=4pt,label] {\footnotesize $\Transaction{\Shard{S},\Shard{U},\Shard{W}}$} (2, 9);
                  
	\path[->] (2, 9) edge (2.5, 8) edge (2.5, 7) edge (2.5, 6);

        \path[<-] (3.5, 9) edge (3, 8) edge (3, 7) edge (3, 6)
                  (3.5, 8) edge (3, 9) edge (3, 7) edge (3, 6)
                  (3.5, 7) edge (3, 9) edge (3, 8) edge (3, 6)
                  (3.5, 6) edge (3, 9) edge (3, 8) edge (3, 7);	

        \path[->] (3.5, 9) edge (4.5, 4.5)
		  (3.5, 8) edge (4.5, 3.5)
		  (3.5, 7) edge (4.5, 2.5)
		  (3.5, 6) edge (4.5, 1.5)
                  
                  (4.5, 3.5) edge (5.5, 2.5) edge (5.5, 1.5) edge (5.5, 4.5)
                  (4.5, 2.5) edge (5.5, 3.5) edge (5.5, 1.5) edge (5.5, 4.5)
		  (4.5, 1.5) edge (5.5, 4.5) edge (5.5, 3.5) edge (5.5, 2.5);

	\path[->] (5.5, 4.5) edge (6, 3.5) edge (6, 2.5) edge (6, 1.5);

        \path[<-] (7, 4.5) edge (6, 3.5) edge (6, 2.5) edge (6, 1.5)
                  (7, 3.5) edge (6, 2.5) edge (6, 1.5) edge (6, 4.5)
                  (7, 2.5) edge (6, 1.5) edge (6, 4.5) edge (6, 3.5)
                  (7, 1.5) edge (6, 4.5) edge (6, 3.5) edge (6, 2.5);

        \draw[draw=colP!60,fill=colP!40,rounded corners] (5.7, 4.7) rectangle (6.8, 1.3)
                                                             (2.2, 9.2) rectangle (3.3, 5.8);

	\node[align=center,label] at (6.25, 3)   {\footnotesize Local \PBFT{}\\ {\footnotesize Consensus} \\ {\footnotesize on $\Transaction{\Shard{S},\Shard{U},\Shard{W}}$}};
        \node[align=center,label] at (2.75, 7.5) {\footnotesize Local \PBFT{}\\ {\footnotesize Consensus} \\ {\footnotesize on $\Transaction{\Shard{S},\Shard{U},\Shard{W}}$}};

        \path[->] (7, 4.5) edge (8, -4.5)
		  (7, 3.5) edge (8, -5.5)
		  (7, 2.5) edge (8, -6.5);

	\path[->] (8, -4.5) edge (9, -5.5) edge (9, -6.5) edge (9, -7.5)
                  (8, -5.5) edge (9, -6.5) edge (9, -7.5) edge (9, -4.5)
		  (8, -6.5) edge (9, -7.5) edge (9, -4.5) edge (9, -5.5);

	\path[->] (9, -4.5) edge (9.5, -5.5) edge (9.5, -6.5) edge (9.5, -7.5);

        \path[<-] (10.5, -4.5) edge (9.5, -5.5) edge (9.5, -6.5) edge (9.5, -7.5)
                  (10.5, -5.5) edge (9.5, -6.5) edge (9.5, -7.5) edge (9.5, -4.5)
                  (10.5, -6.5) edge (9.5, -7.5) edge (9.5, -4.5) edge (9.5, -5.5)
                  (10.5, -7.5) edge (9.5, -4.5) edge (9.5, -5.5) edge (9.5, -6.5);

	\draw[draw=colP!60,fill=colP!40,rounded corners] (9.2, -4.3) rectangle (10.3, -7.7);
	\node[align=center,label] at (9.75, -6) {\footnotesize Local \PBFT{}\\ {\footnotesize Consensus} \\ {\footnotesize on $\Transaction{\Shard{S},\Shard{U},\Shard{W}}$}};

        \path[->] (11, -4.5) edge (12, 9)
		  (11, -5.5) edge (12, 8)
		  (11, -6.5) edge (12, 7);

	\path[->] (12, 9) edge (13, 8) edge (13, 7) edge (13, 6)
                  (12, 8) edge (13, 9) edge (13, 7) edge (13, 6)
		  (12, 7) edge (13, 9) edge (13, 8) edge (13, 6);

        \path[->] (13.5, 9) edge (14.5, 3.5)
		  (13.5, 8) edge (14.5, 2.5)
		  (13.5, 7) edge (14.5, 1.5)
                  
                  (14.5, 3.5) edge (15.5, 2.5) edge (15.5, 1.5) edge (15.5, 4.5)
                  (14.5, 2.5) edge (15.5, 3.5) edge (15.5, 1.5) edge (15.5, 4.5)
		  (14.5, 1.5) edge (15.5, 4.5) edge (15.5, 3.5) edge (15.5, 2.5);

        \draw[draw=colN!20,fill=colN!10,rounded corners]  (10.55, -4.4) rectangle (10.95, -7.6) (13.05, 5.4) rectangle (13.45, 9.6) (15.55, 4.6) rectangle (15.95, 1.3);
        \node[align=center,label,rotate=270] at (10.75, -6)  {\footnotesize Execute};
        \node[align=center,label,rotate=270] at (13.25, 7.5) {\footnotesize Execute};
	\node[align=center,label,rotate=270] at (15.75, 3)   {\footnotesize Execute};

        \path[<-] (14.5, 10.5) edge (13.5, 9) edge (13.5, 8) edge (13.5, 7) edge (13.5, 6);

        \node[align=center,below,label] at (1.5, -7.5)  {{\footnotesize Local} \\ {\footnotesize Request}};
        \node[align=center,below,label] at (2.7, -7.5)  {{\footnotesize Local} \\ {\footnotesize Replication}};
        \node[align=center,below,label] at (4, -7.5)    {{\footnotesize Forward}};
        \node[align=center,below,label] at (5, -7.5)    {{\footnotesize Local}  \\ {\footnotesize Sharing}};
        \node[align=center,below,label] at (6.3, -7.5)  {{\footnotesize Local}  \\ {\footnotesize Replication}};
	\node[align=center,below,label] at (7.5, -7.5)  {{\footnotesize Global} \\ {\footnotesize Sharing}};
        \node[align=center,below,label] at (8.5, -7.5)  {{\footnotesize Local}  \\ {\footnotesize Sharing}};
        \node[align=center,below,label] at (9.7, -7.5)  {{\footnotesize Local}  \\ {\footnotesize Replication}};
	\node[align=center,below,label] at (11.5, -7.5) {{\footnotesize Global} \\ {\footnotesize Sharing}};
        \node[align=center,below,label] at (12.5, -7.5) {{\footnotesize Local}  \\ {\footnotesize Sharing}};
	\node[align=center,below,label] at (14, -7.5)   {{\footnotesize Client} \\ {\footnotesize Response}};
        
        \draw[decoration={brace},decorate] (0.25, 6) -- (0.25, 10.5);
        \node[left=2pt] at (0.25, 8) {\small $\Shard{S}$};
        
        \draw[decoration={brace},decorate] (0.25, 1.5) -- (0.25, 4.5);
        \node[left=2pt] at (0.25, 3) {\small $\Shard{U}$};

	\draw[decoration={brace},decorate] (0.25, -3) -- (0.25, 0);
        \node[left=2pt] at (0.25, -2) {\small $\Shard{V}$};

	\draw[decoration={brace},decorate] (0.25, -7.5) -- (0.25, -4.5);
        \node[left=2pt] at (0.25, -6) {\small $\Shard{W}$};

	\draw[decoration={brace},decorate] (10.9, -9.35) -- (1, -9.35);
        \node[below=2pt] at (6.05, -9.25) {\small Round 1};
	
	\draw[decoration={brace},decorate] (16, -9.35) -- (11.2, -9.35);
        \node[below=2pt] at (13.7, -9.25) {\small Round 2};

    \end{tikzpicture}%
    \caption{Representation of the normal-case flow of \RingBFT{} in a system of four shards where client sends a cross-shard transaction 
	$\Transaction{\Shard{S},\Shard{U},\Shard{W}}$ that requires access to data in {\em three} shards: $\Shard{S}$, $\Shard{U}$, and 
	$\Shard{W}$.}
    \label{fig:ringbft_sketch}
\end{figure*}
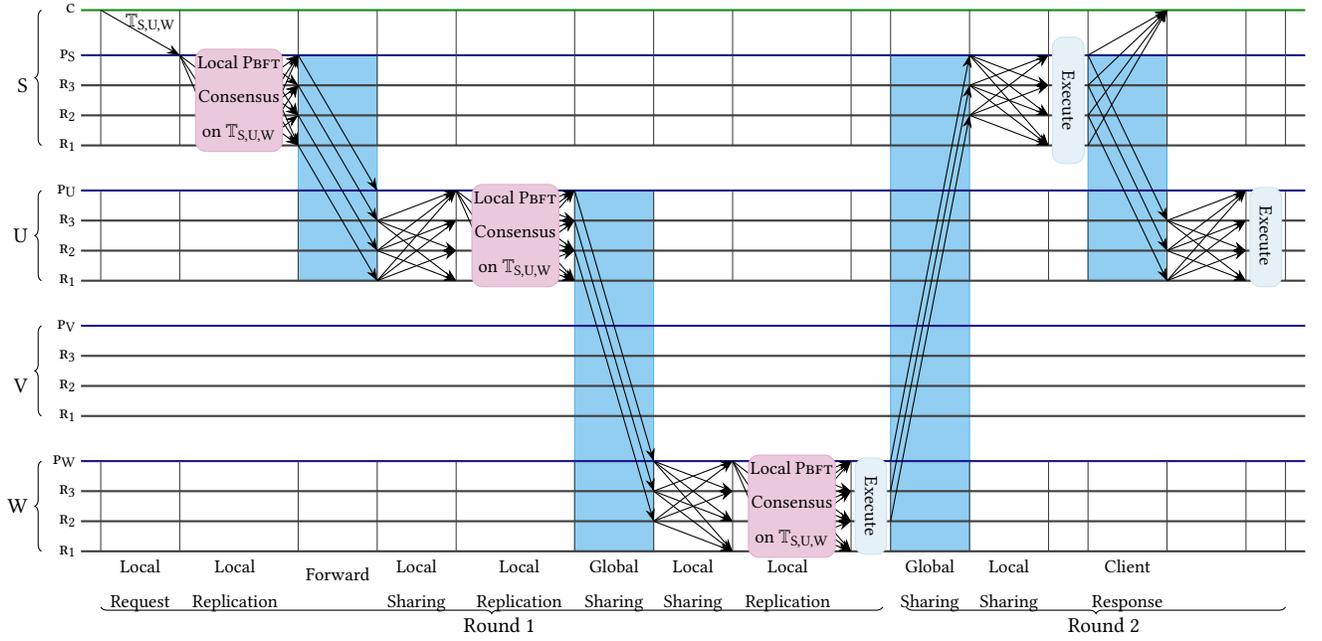

\begin{figure}[ht]
    \begin{myprotocol}
	\INITIAL{Initialization:}{\newline
	{\color{colN}
	// $\kmax$ \GETS 0  (maximum sequence number in shard $\Shard{S}$)\newline
	// $\Sigma^{\Involved{}} \GETS \emptyset$ (set of data-fragments of each shard)\newline
	// $\pi \GETS \emptyset$ (list of pending transactions at a replica)
	}}
	\SPACE

        \TITLE{Client-role}{used by client $\Client{}$ to request transaction $\Transaction{\Involved{}}$}
        \STATE Sends $\SignMessage{\Transaction{\Involved{}}}{\Client{}}$ to the primary $\Primary{\Shard{S}}$ of shard $\Shard{S}$.
        \STATE Awaits receipt of messages $\Message{\Name{Response}}{\SignMessage{\Transaction{\Involved{}}}{\Client{}}, k, r}$ from 
		$\f+1$ replicas of $\Shard{S}$.
        \STATE Considers $\Transaction{\Involved{}}$ executed, with result $r$, as the $k$-th transaction.\label{fig:pa:cc}
        \SPACE \SPACE

	\MC{This event is only triggered at the primary replica of each shard.}
        \TITLE{Primary-role}{running at the primary $\Primary{\Shard{S}}$ of shard $\Shard{S}$}
	 \EVENT{$\Primary{\Shard{S}}$ receives $\SignMessage{\Transaction{\Involved{}}}{\Client{}}$}
	 	\IF{$\Shard{S} \in \Involved{} ~\wedge~ \ID{\Shard{S}} = {\tt FirstInRingOrder}(\Involved{})$}
		 \STATE Calculate digest $\Digest := \Hash{\SignMessage{\Transaction{}}{\Client{}}}$.
		 \STATE Broadcast $\Message{\Name{Preprepare}}{\SignMessage{\Transaction{\Involved{}}}{\Client{}},\Digest,k}$ in shard $\Shard{S}$ (order at sequence $k$).
		\ELSE
		 \STATE Send to primary $\Primary{\Shard{U}}$ of shard $\Shard{U}, \Shard{U} \in \Shards{} \wedge \ID{\Shard{U}} = {\tt FirstInRingOrder}(\Involved{})$
	 	\ENDIF
	 \ENDEVENT
	 \SPACE \SPACE

	\MC{This event is only triggered at a non-primary replica.}
	\TITLE{Non-Primary Replica-role}{running at the replica $\Replica{r}$ of shard $\Shard{S}$}
	 \EVENT{$\Replica{r}$ receives $\Message{\Name{Preprepare}}{\SignMessage{\Transaction{\Involved{}}}{\Client{}},\Digest,k}$ from $\Primary{\Shard{S}}$ such that:\\
		\qquad message is well-formed, and $\Replica{r}$ did not accept a $k$-th proposal from $\Primary{\Shard{S}}$.
	 }
		\STATE Broadcast $\Message{\Name{Prepare}}{\Digest,k}$ to replicas in $\Replicas{\Shard{S}}$.	
	 \ENDEVENT
	 \SPACE \SPACE

	\MC{Following events are triggered at every replica irrespective of whether it is the primary or a non-primary replica.}
	\TITLE{Replica-role}{running at \changed{any} replica $\Replica{r}$ of shard $\Shard{S}$}
	 \EVENT{$\Replica{r}$ receives well-formed $\Message{\Name{Prepare}}{\Digest,k}$ messages from $\nf$ replicas 
	 	in $\Shard{S}$
 	 }
	 
		\STATE Broadcast $\SignMessage{\Message{\Name{Commit}}{\Digest,k}}{\Replica{r}}$ to replicas in 
				$\Replicas{\Shard{S}}$.
	 \ENDEVENT
	 \SPACE

	 \EVENT{$\Replica{r}$ receives $\nf$ $m := \SignMessage{\Message{\Name{Commit}}{\Digest,k}}{\Replica{q}}$ messages such that:\\
                	\qquad each message $m$ is well-formed and is sent by a distinct replica $\Replica{q} \in \Replicas{\Shard{S}}$.
	 }
		\STATE $\Shard{U}$ be the shard to forward such that $\ID{\Shard{U}} = {\tt NextInRingOrder}({\Involved{}})$.
		\STATE $A$ := set of \Name{DS} of these $\nf$ messages.\label{alg:ds-commit}

		\IF{$k = \kmax + 1$ \MC{Forward to next shard}}
			\STATE Lock data-fragment corresponding to $\SignMessage{\Transaction{\Involved{}}}{\Client{}}$.\label{alg:lock}
			\STATE Send $\SignMessage{\Message{\Name{Forward}}{\SignMessage{\Transaction{\Involved{}}}{\Client{}},A,m,\Delta,}}{\Replica{r}}$ 
			to replica $\Replica{o}$,  
			where $\Replica{o} \in \Replicas{\Shard{U}} ~\wedge~ \ID{\Replica{r}} = \ID{\Replica{o}}$\label{alg:forward}

		\ELSE
			\STATE Store $\SignMessage{\Message{\Name{Forward}}{\SignMessage{\Transaction{\Involved{}}}{\Client{}},A,m,\Delta,}}{\Replica{r}}$ in $\pi$. 
		\ENDIF

		\WHILE{$\pi != \emptyset$ \MC{Pop out waiting transaction.}}\label{alg:loop}
			\STATE Extract transaction at $\kmax + 1$ from $\pi$ (if any).\label{alg:pi-push}
			\IF{Corresponding data-fragment is not locked}
				\STATE $\kmax = \kmax + 1$
				\STATE Follow lines~\ref{alg:lock} and~\ref{alg:forward}.
			\ELSE	
				\STATE Store transaction at $\kmax$ in $\pi$ and {\em exit} the loop.
			\ENDIF
		\ENDWHILE
         \ENDEVENT
	 \SPACE
	 \SPACE

	 \MC{Locally share any message from previous shard.}
	 \EVENT{$\Replica{r}$ receives message $m \GETS \SignMessage{\Name{message-type}}{\Replica{q}}$ such that:\\
			\qquad $m$ is well-formed and sent by replica $\Replica{q}$, where  \\ \qquad \qquad 
				$\ID{\Shard{U}} = {\tt PrevInRingOrder}({\Involved{}})$, 
				$\Replica{q} \in \Replicas{\Shard{U}} ~\wedge~ \ID{\Replica{r}} = \ID{\Replica{q}}$
	 }
		\STATE Broadcast $m$ to all replicas in $\Shard{S}$. 
	 \ENDEVENT
	\SPACE
	\SPACE

	 \MC{\Name{Forward} message from previous shard.}
	 \EVENT{$\Replica{r}$ receives $\f+1$ $m' := \SignMessage{\Message{\Name{Forward}}{\SignMessage{\Transaction{\Involved{}}}{\Client{}},A,m,\Delta}}{\Replica{q}}$ such that:\\
			\qquad each $m'$ is well-formed; and set $A$ includes valid \Name{DS} from $\nf$ replicas for $m$.
	 }
		\IF{Data-fragment corresponding to $\SignMessage{\Transaction{\Involved{}}}{\Client{}}$ is locked \MC{Second Rotation}}
			\STATE Execute data-fragment of $\SignMessage{\Transaction{\Involved{}}}{\Client{}}$ and add to log.\label{alg:exelock}
			\STATE Push result to set $\Sigma^{\Involved{}}$.
			\STATE Release the locks from corresponding data-fragment.
			\STATE $\Shard{V}$ be the shard to forward such that $\ID{\Shard{V}} = {\tt NextInRingOrder}({\Involved{}})$.
			\STATE Send $\SignMessage{\Message{\Name{Execute}}{\Delta,\Sigma^{\Involved{}}}}{\Replica{r}}$ 
			to replica $\Replica{o}$,  
			where $\Replica{o} \in \Replicas{\Shard{V}} ~\wedge~ \ID{\Replica{r}} = \ID{\Replica{o}}$.\label{alg:exesend}
		\ELSIF{$\Replica{r} = \Primary{\Shard{S}}$ \MC{Primary initiates consensus}}
 			\STATE Broadcast $\Message{\Name{Preprepare}}{\SignMessage{\Transaction{\Involved{}}}{\Client{}},\Digest,k'}$ in shard $\Shard{S}$ (order at sequence $k'$).
		\ENDIF
	 \ENDEVENT
	\SPACE

	 \EVENT{$\Replica{r}$ receives $m' \GETS \SignMessage{\Message{\Name{Execute}}{\Delta,\Sigma^{\Involved{}}}}{\Replica{q}}$ such that:\\
			\qquad $m'$ is sent by replica $\Replica{q}$, where
				$\Replica{q} \in \Replicas{\Shard{U}} ~\wedge~ \ID{\Replica{r}} = \ID{\Replica{q}}$
	 }
		\IF{Already executed $\SignMessage{\Transaction{\Involved{}}}{\Client{}}$ \MC{Reply to client}}
			\STATE Send client $\Client{}$ the result $r$.
		\ELSE
			\STATE Follow lines~\ref{alg:exelock} to~\ref{alg:exesend}.
		\ENDIF
	 \ENDEVENT
    \end{myprotocol}
    \caption{The \changed{event-based} normal-case algorithm of \RingBFT{}.
	\changed{Depending on the type of message the primary replica or a non-primary replica receives, specific events are triggered.}
}
    \label{alg:ringbft}
\end{figure}

\subsubsection{Cross-shard Transactional Flow}
\RingBFT{} assumes shards are arranged in a logical ring. 
For the sake of explanation, we assume the ring order of {\em lowest to highest identifier}.
For each \CST{}, we denote one shard as the {\bf \em initiator shard}, which is responsible for starting consensus on 
the client transaction.
How do we select the initiator shard?
Of all the involved shards a \CST{} accesses, the shard with the {\em lowest identifier in ring order} 
is denoted as the initiator shard.

\RingBFT{} also guarantees consensus for each deterministic \CST{} in at most two rotations across the ring.
This implies that for achieving consensus on a deterministic \CST{}, each involved shard
$\Shard{S} \in \Involved$ needs to process it at most two times.
Notice that if a \CST{} is simple, then a single rotation around the ring is sufficient to ensure that 
each involved shard $\Shard{S}$ safely executes its fragment.

Prior to presenting our \RingBFT's consensus protocol that safely orders each \CST{}, 
we sketch the flow of a \CST{} in Figure~\ref{fig:ringbft_sketch}.
In this figure, we assume a system of four shards: $\Shard{S}$, $\Shard{U}$, $\Shard{V}$, and $\Shard{W}$ where 
$\ID{\Shard{S}} < \ID{\Shard{U}} < \ID{\Shard{V}} < \ID{\Shard{W}}$.
The client creates a transaction $\Transaction{\Shard{S},\Shard{U},\Shard{W}}$ that requires access to data in shards 
$\Shard{S}$, $\Shard{U}$, and $\Shard{W}$ and sends this transaction to the primary $\Primary{S}$ of $\Shard{S}$. 
On receiving this transaction, $\Primary{S}$ initiates the \PBFT{} consensus protocol ({\em local replication}) among its replicas. 
If the local replication is successful, then all the replicas of $\Shard{S}$ {\bf \em lock} the corresponding data.
This locking of data-items in the ring-order helps in preventing deadlocks.
Next, replicas of $\Shard{S}$ forward the transaction to replicas of shard $\Shard{U}$. 
Notice that only {\em linear communication} takes place between replicas of $\Shard{S}$ and $\Shard{U}$. 
Hence, to handle any failures, replicas of $\Shard{U}$ share this message among themselves. 
Next, replicas of $\Shard{U}$ also follow similar steps and forward transaction to $\Shard{W}$. 
As $\Shard{W}$ is the last shard in the ring of involved shards, it goes ahead and executes the \CST{} if all the dependencies 
are met.
Finally, replicas of shards $\Shard{S}$ and $\Shard{U}$ also execute the transaction and replicas of $\Shard{S}$ send 
the result of execution to the client.

\subsection{Cross-Shard Consensus Algorithm}
We use Figure~\ref{alg:ringbft} to present \RingBFT's algorithm for ordering cross-shard transactions. 
Next, we discuss these steps in detail.

\subsubsection{Client Request}
When a client $\Client{}$ wants to process a cross-shard transaction $\Transaction{\Involved{}}$, it creates a 
$\SignMessage{\Transaction{\Involved{}}}{\Client{}}$ message and sends it to the primary of the {\em first shard in ring order}.
As part of this transaction, the client $\Client{}$ specifies the information regarding all the involved shards ($\Involved{}$),
such as their identifiers and the necessary {\em read-write} sets of each shard.
\changed{Notice that the client signs this message using \Name{DS} to prevent repudiation attacks.}

\subsubsection{Client Request Reception} 
When the primary $\Primary{\Shard{S}}$ of shard $\Shard{S}$
receives a client request $\Transaction{\Involved{}}$, it first checks if the message is well-formed. 
If this is the case, then $\Primary{\Shard{S}}$ checks if among the set of involved shards $\Involved{}$, 
$\Shard{S}$ is the first shard in ring order. 
If this condition is met, then $\Primary{\Shard{S}}$ assigns this request a linearly increasing sequence number $k$, 
calculates the digest $\Digest{}$, and broadcasts a $\Name{Preprepare}$ message to all the replicas $\Replicas{\Shard{S}}$ of its shard.
In the case when $\Shard{S}$ is not the first shard in the ring order, $\Primary{\Shard{S}}$ forwards the transaction to the primary 
of the appropriate shard.

\subsubsection{Pre-prepare Phase}
When a replica $\Replica{r} \in \Replicas{\Shard{S}}$ receives the $\Name{Preprepare}$ message from $\Primary{\Shard{S}}$, 
it checks if the request is well-formed. 
If this is the case and if $\Replica{r}$ has not agreed to support any other request from $\Primary{\Shard{S}}$ as the 
$k$-th request, then it broadcasts a $\Name{Prepare}$ message in its shard $\Shard{S}$.

\subsubsection{Prepare Phase}
When a replica $\Replica{r} \in \Replicas{\Shard{S}}$ receives identical $\Name{Prepare}$ messages from $\nf$ distinct replicas, it gets an assurance 
that a majority of non-faulty replicas are supporting this request. 
At this point, each replica $\Replica{r}$ broadcasts a $\Name{Commit}$ message to all the replicas in $\Shard{S}$.
Once a transaction passes this phase, the replica $\Replica{r}$ marks it {\em prepared}.

\subsubsection{Commit and Data Locking}
When a replica $\Replica{r}$ receives well-formed identical $\Name{Commit}$ messages from $\nf$ distinct replicas in $\Shard{S}$, 
it checks if it also prepared this transaction at same sequence number.
If this is the case, \RingBFT{} requires each replica $\Replica{r}$ to {\em lock} all the read-write sets that transaction 
$\Transaction{\Involved{}}$ needs to access in shard $\Shard{S}$.
In \RingBFT{}, we allow replicas to process and broadcast $\Name{Prepare}$ and $\Name{Commit}$ messages {\em out-of-order}, 
but require each replica to acquire locks on data in transactional sequence order. 
This out-of-ordering helps replicas to continuously perform useful work by concurrently participating in consensus of 
several transactions.
To achieve these tasks, each replica $\Replica{r}$ tracks the maximum sequence number ($\kmax$), 
which indicates the sequence number of the last transaction to lock data. 
If sequence number $k$ for a transaction $\Transaction{\Involved{}}$ is greater than $\kmax+1$, we store the transaction 
in a list $\pi$ until transaction at $\kmax+1$ has acquired the locks.
Once the $\kmax+1$-th transaction has acquired locks, we gradually release transactions in $\pi$ until there is a 
transaction that wishes to lock already locked data-fragments. 
We illustrate this through the following example.

\begin{example}
Assume the use of following notations for four transactions and the data-fragments they access at shard $\Shard{S}$: 
$\Transaction{1,a}$, $\Transaction{2,b}$, $\Transaction{3,a}$, and $\Transaction{4,c}$. 
For instance, $\Transaction{1,a}$ implies that transaction at sequence $1$ requires access to data-item $a$.
Next, due to out-of-order message processing, assume a replica $\Replica{r}$ in $\Shard{S}$ 
receives $\nf$ $\Name{Commit}$ messages for $\Transaction{2,b}$, $\Transaction{3,a}$, and $\Transaction{4,c}$ 
before $\Transaction{1,a}$.
Hence, $\pi = \{\Transaction{2,b},\Transaction{3,a},\Transaction{4,c}\}$.
Once $\Replica{r}$ receiving $\nf$ $\Name{Commit}$ messages for $\Transaction{1,a}$, 
it locks data-item $a$ and extracts $\Transaction{2,b}$ from $\pi$. 
As $\Transaction{2,b}$ wishes to lock a distinct data-item, so $\Replica{r}$ continues processing $\Transaction{2,b}$. 
Next, $\Replica{r}$ moves to $\Transaction{3,a}$ but it cannot process $\Transaction{3,a}$ due to lock-conflicts. 
Hence, it places back $\Transaction{3,a}$ in $\pi$ and stops processing transactions in $\pi$ until lock is available for 
$\Transaction{3,a}$.
\end{example}

Notice that if the client transaction $\Transaction{\Involved{}}$ is a {\em single-shard} transaction, 
it requires access to data-items in only this shard. 
In such a case, this commit phase is the final phase of consensus and 
each replica executes $\Transaction{\Involved{}}$ and replies to the client
when the lock for the corresponding data-item is available.

\subsubsection{Forward to next Shard via Linear Communication}
\label{ss:forward}
Once a replica $\Replica{r}$ in $\Shard{S}$ locks the data corresponding to \CST{} $\Transaction{\Involved{}}$,  
it sends a $\Name{Forward}$ message to {\em only one} replica $\Replica{q}$ of the next shard in ring order. 
As one of the key goals of \RingBFT{} is to ensure communication between two shards is linear, 
so we design a communication primitive that builds on top of the optimal bound for communication between two shards~\cite{csp,geobft}.
We define \RingBFT's cross-shard communication primitive as follows:

\begin{description}[wide]
\item[\bf Linear Communication Primitive.]
In a system $\Shards{}$ of shards, where each shard $\Shard{S}, \Shard{U} \in \Shards{}$ 
has at most $\f$ Byzantine replicas, 
if each replica in shard $\Shard{S}$ communicates with a distinct replica in shard $\Shard{U}$, 
then at least $\f+1$ non-faulty replicas from $\Shard{S}$ will communicate with 
$\f+1$ non-faulty replicas in $\Shard{U}$.
\end{description}

Our linear communication primitive guarantees that 
to reliably communicate a message $m$ between two shards requires 
only sending a linear number of messages in comparison 
to protocols like \AHL{} and \Sharper{} which require {\em quadratic} 
communication.
Using this communication primitive, to communicate message $m$ from shard
$\Shard{S}$ to shard $\Shard{U}$, we need to exchange only $\n$ messages.

So, \changed{how does \RingBFT{} achieve} this task?
We require each replica of $\Shard{S}$ to initiate communication with the replica of 
$\Shard{U}$ having the same identifier.
Hence, replica $\Replica{r}$ of shard $\Shard{S}$ sends a $\Name{Forward}$ message to replica $\Replica{q}$ in shard $\Shard{U}$ 
such that $\ID{\Replica{r}} = \ID{\Replica{q}}$.
By sending a $\Name{Forward}$ message, $\Replica{r}$ is requesting $\Replica{q}$ to initiate 
consensus on $\SignMessage{\Transaction{\Involved{}}}{\Client{}}$. 
For $\Replica{q}$ to support such a request, it needs a proof that $\SignMessage{\Transaction{\Involved{}}}{\Client{}}$ 
was successfully ordered in shard $\Shard{S}$. 
Hence, $\Replica{r}$ includes the \Name{DS} on $\Name{Commit}$ messages from $\nf$ distinct replicas 
(Line~\ref{alg:ds-commit}).

\changed{
Until now, we assumed that each shard has an equal number of replicas. 
If we forgo this assumption, it will not affect the intra-shard consensus, 
that is, the \BFT{} consensus protocol running at each shard remains unchanged. 
Further, the transaction execution explained in the next section also remains unaffected.
The only visible change occurs in our linear communication primitive. 
However, even this change does not impact the correctness of our \RingBFT{} protocol 
as our linear communication primitive builds on the optimal bound for 
communication between two shards, which permits shards to have a different  
number of replicas while guaranteeing linear communication complexity ~\cite{csp,geobft}.

Finally, in Section~\ref{sss:remote-view-change}, we illustrate how our 
linear communication primitive can handle
attacks by byzantine replicas.
}

\subsubsection{Execution and Final Rotation}
Once a client request has been ordered on all the involved shards, we call it {\em one complete rotation} around the ring. 
This is a significant event because it implies that all the necessary data-fragments have been locked by each of the involved shards. 
If a \CST{} is simple, then each shard can independently execute its fragment without any further communication
between the shards. 
In the case a \CST{} is complex, at the end of the first rotation, 
the replicas of the first shard in ring order ($\Shard{S}$) will receive a $\Name{Forward}$ message 
from the replicas of the last shard in ring order.

Next, the replicas of $\Shard{S}$ will attempt to execute parts of transaction, which are their responsibility.
Post execution, replicas of $\Shard{S}$ send $\Name{Execute}$ messages to the replicas in next shard using our
communication primitive.
Notice that the $\Name{Execute}$ message includes updated write sets 
($\Sigma^{\Involved{}}$), which help in resolving any dependencies during execution.
Finally, when the execution is completed across all the shards, 
the first shard in ring order replies to the client.

\section{Uncivil Executions}
In previous sections, we discussed transactional flows under the assumption that the network is stable and 
replicas will follow the stated protocol.
However, any Byzantine-Fault Tolerant protocol should provide safety under asynchronous settings and
liveness in the period of synchrony even if up to $\f$ replicas are Byzantine.

\RingBFT{} offers safety in an asynchronous environment. 
To guarantee liveness during periods of synchrony, \RingBFT{} offers several recovery protocols, such as 
{\em checkpoint}, {\em retransmission}, and {\em view-change}, to counter malicious attacks.
The first step in recovery against any attack is {\em detection}. 
To do so, we require each replica $\Replica{r}$ to employ a set of {\em timers}. 
When a timer at a replica $\Replica{r}$ {\em timeouts}, then $\Replica{r}$ initiates an appropriate recovery mechanism.
In specific, each replica $\Replica{r}$ sets following timers:
\begin{itemize}[nosep]
\item {\bf Local Timer} -- To track successful replication of a transaction in its shard.
\item {\bf Transmit Timer} -- To re-transmit a successfully replicated cross-shard transaction to next shard.
\item {\bf Remote Timer} -- To track replication of a cross-shard transaction in the previous shard.
\end{itemize}
 
Each of these timers is initiated at the occurrence of a distinct event and its timeout leads to running a 
specific recovery mechanism. 
When a local timer expires, then the corresponding replica initiates replacement of the primary of its shard (view-change), 
while a remote timer timeout requires the replica to inform the previous shard in ring order about the insufficient communication.
This brings us to following observation regarding the consensus offered by \RingBFT{}:
\begin{proposition}\label{prop:non-faulty-primary}
If the network is reliable and the primary of each shard is non-faulty, then the Byzantine replicas in the system cannot affect the 
consensus protocol.
\end{proposition}

Notice that Proposition~\ref{prop:non-faulty-primary} holds implicitly as no step in Figure~\ref{alg:ringbft} 
depends on the correct working of non-primary Byzantine replicas; 
in each shard $\Shard{S}$, local replication of each transaction is managed by the primary of $\Shard{S}$ 
and communication between any two shards $\Shard{S}$ and $\Shard{U}$ involves all the replicas.
This implies that we need only consider cases when the network is unreliable and/or primary is Byzantine. 
We know that \RingBFT{} guarantees safety even in unreliable communication and requires a reliable network for 
assuring liveness. 
Hence, we will illustrate mechanisms to tackle attacks by Byzantine primaries. 
Next, we illustrate how \RingBFT{} resolves all the possible attacks it encounters.
\begin{enumerate}[wide,label={\bf (A\arabic*)},ref={A\arabic*}]
\item\label{A:mal-client} {\bf Client Behavior and Attacks.} 
In the case, the primary is Byzantine and/or network is unreliable, client is the key entity at loss.
Client requested the primary to process its transaction, but due to an ongoing Byzantine attack, 
client did not receive sufficient responses.
Clearly, client cannot wait indefinitely to receive valid responses. 
Hence, we require each client $\Client{}$ to start a timer when it sends its transaction $\Transaction{\Involved{}}$ 
to the primary $\Primary{\Shard{S}}$ of shard $\Shard{S}$.
If the timer timeouts prior to $\Client{}$ receiving at least $\f+1$ identical responses, 
$\Client{}$ broadcasts $\Transaction{\Involved{}}$ to all the replicas $\Replica{r} \in \Replicas{\Shard{S}}$ of shard $\Shard{S}$. 

When a non-primary replica $\Replica{r}$ receives a transaction from $\Client{}$, 
it forwards that transaction to $\Primary{\Shard{S}}$ and waits on a timer for 
$\Primary{\Shard{S}}$ to initiate consensus on $\Transaction{\Involved{}}$.
During this time, $\Replica{r}$ expects $\Primary{\Shard{S}}$ to start consensus on at least one transaction from $\Client{}$, 
otherwise it initiates view-change protocol.
Notice that a {\em Byzantine client} can always forward its request to all the replicas of some shard to {\em blame} a non-faulty primary. 
Such an attack will not succeed as if $\Client{}$ sends to $\Replica{r}$ an already executed request, 
$\Replica{r}$ simply replies with the stored response.
Moreover, if $\Replica{r}$ belongs to some shard $\Shard{S}$, which is not the first shard
in ring order, then $\Replica{r}$ ignores the client transaction.

\item\label{A:faulty-primary} {\bf Faulty Primary and/or Unreliable network.} 
A faulty primary can prevent successful consensus of a client transaction. 
Such a primary can be trivially detected as at most $\f$ non-faulty replicas would have 
successfully {\em committed} the transaction (received at least $\n-\f$ $\Name{Commit}$ messages).

An unreliable network can cause messages to get lost or indefinitely delayed. 
Such an attack is difficult to detect and non-faulty replicas may blame the primary.

Each primary represents a {\em view} of a shard. 
Hence, the term view-change is often used to imply primary replacement. 
Notice that each shard in \RingBFT{} is a replicated system.
Further, \RingBFT{} is a meta-protocol, which employs existing \BFT{} protocols, 
such as \PBFT{}, to run consensus. 
These properties allow \RingBFT{} to use the accompanying {\em view-change protocol}. 
Specifically, in this paper, we use \PBFT's view change protocol (for \Name{MAC}-based authentication) 
to detect and replace a faulty primary~\cite{pbftj}.

A replica $\Replica{r} \in \Replicas{\Shard{S}}$ initiates the view-change protocol to replace 
its primary $\Primary{\Shard{S}}$ in response to a timeout.
As discussed earlier in this section, there are two main causes for such timeouts:
(i) $\Replica{r}$ does not receive $\nf$ identical $\Name{Commit}$ 
messages from distinct replicas, and 
(ii) $\Primary{\Shard{S}}$ fails to propose a request from client $\Client{}$.

\item\label{A:mal-primary} {\bf Malicious Primary.} 
A malicious primary $\Primary{}$ can ensure that up to $\f$ non-faulty replicas 
in its shard $\Shard{S}$ are unable to make progress (in {\em dark}).
Under such conditions, the affected non-faulty replicas will request a view-change, but they will not be successful 
as the next primary may not receive sufficient $\Name{ViewChange}$ messages (from at least $\nf$ replicas) 
to initiate a new view. 
Further, the remaining $\f+1$ non-faulty replicas will not support such $\Name{ViewChange}$ requests as it is impossible 
for them to distinguish between this set of $\f$ non-faulty replicas and the actual $\f$ Byzantine replicas.

To ensure these replicas in dark make progress, traditional protocols periodically send checkpoint messages. 
These checkpoint messages include all client transactions and the corresponding $\nf$ $\Name{Commit}$ messages 
since the last checkpoint.

\end{enumerate}

\vspace{-1mm}
\subsection{Cross-Shard Attacks}
Until now, we have discussed attacks that can be resolved by replicas of any shard independent of the functioning 
of other shards.
However, the existence of cross-shard transactions unravels new attacks, which may span multiple shards. 
We use the term {\em cross-shard attacks} to denote attacks that thwart successful consensus of a \CST{}, 
First, we describe such attacks, and then we present solutions to recover from these attacks.

In \RingBFT{}, we know that the consensus of each \CST{} follows a ring order. 
In specific, for a cross-shard transaction $\Transaction{\Involved{}}$, each of its involved shards
$\Shard{S}, \Shard{U} \in \Involved{}$ first run a local consensus and then communicate the data  
to the next shard in ring order.
Earlier in this section, we observed that if at least $\f+1$ non-faulty replicas of any shard are unable to reach 
consensus on $\Transaction{\Involved{}}$, then that shard will undergo local view-change.
Hence, we are interested in those cross-shard attacks where {\em neither the involved shards are able to trigger local 
view change by themselves, nor are they able to execute the transaction and reply to the client}.
This can only occur when all the involved shards of a cross-shard transaction $\Transaction{\Involved{}}$,
either successfully completed consensus on $\Transaction{\Involved{}}$, or are unable to initiate the consensus on 
$\Transaction{\Involved{}}$.
Next, we describe these attacks. 

Assume $\Replicas{S}$ and $\Replicas{U}$ represent the sets of replicas in shards $\Shard{S}$ and $\Shard{U}$, respectively.
\begin{enumerate}[wide,nosep,label=(C\arabic*),ref={C\arabic*}]
\item\label{C:no-communication} {\bf No Communication.} 
Under a no communication attack, we expect that the replicas in $\Replicas{S}$ are unable to 
send any messages to replicas of $\Replicas{U}$.

\item\label{C:partial-communication} {\bf Partial Communication.} 
Under a partial communication attack, we expect that at least $\f+1$ replicas in $\Replicas{U}$ receive less than 
$\f+1$ $\Name{Forward}$ messages from replicas in $\Replicas{S}$.
\end{enumerate}

Both of these attacks could occur solely due to an unreliable network that causes message loss or indefinite message delays.
Further, a malicious primary can collude with an adversarial network to accelerate the frequency of such attacks.
In either of the cases, to recover from such cross-shard attacks, all the involved shards may need to communicate among themselves.

\subsubsection{\bf \em Message Retransmission}
In \RingBFT{}, to handle a {\em no communication} attack, affected replicas of the preceding shard retransmit their original 
message to the next shard in ring order. 
Specifically, when a replica $\Replica{r}$ of shard $\Shard{S}$ successfully completes the consensus on transaction 
$\Transaction{\Involved{}}$, it sets the {\em transmit timer} for this request prior to sending the $\Name{Forward}$ 
message to replica $\Replica{Q}$ of shard $\Shard{U}$ (next shard in ring order).
When the transmit timer of $\Replica{r}$ timeouts, it again sends the $\Name{Forward}$ message to $\Replica{q}$.

\begin{figure}[t]
    \begin{myprotocol}
	\TITLE{Replica-role}{running at the replica $\Replica{q}$ of shard $\Shard{U}$}
	 \EVENT{Remote timer of $\Replica{q}$ timeouts such that:\\
                \qquad $\Replica{q}$ has received at most $\f$ $\SignMessage{\Message{\Name{Forward}}{\SignMessage{\Transaction{\Involved{}}}{\Client{}},A,m,\Delta,}}{\Replica{r}}$ messages, where  \\ \qquad \qquad 
				$\ID{\Shard{S}} = {\tt PrevInRingOrder}({\Involved{}})$, 
				$\Replica{r} \in \Replicas{\Shard{S}}$
	 }
		\STATE Send $\SignMessage{\Message{\Name{RemoteView}}{\SignMessage{\Transaction{\Involved{}}}{\Client{}},\Delta}}{\Replica{q}}$
			to replica $\Replica{o}$,  
			where $\Replica{o} \in \Replicas{\Shard{S}} ~\wedge~ \ID{\Replica{q}} = \ID{\Replica{o}}$ 
	 \ENDEVENT
	 \SPACE

	 \EVENT{$\Replica{r}$ receives message $m \GETS \SignMessage{\Message{\Name{RemoteView}}{\SignMessage{\Transaction{\Involved{}}}{\Client{}},\Delta}}{\Replica{q}}$ such that:\\
			\qquad $m$ is well-formed and sent by replica $\Replica{q}$, where  \\ \qquad \qquad 
				$\ID{\Shard{U}} = {\tt NextInRingOrder}({\Involved{}})$, 
				$\Replica{q} \in \Replicas{\Shard{U}} ~\wedge~ \ID{\Replica{r}} = \ID{\Replica{q}}$
	 }
		\STATE Broadcast $m$ to all replicas in $\Shard{S}$. 
	 \ENDEVENT
	 \SPACE

	 \EVENT{$\Replica{r}$ receives $\f+1$ $\SignMessage{\Message{\Name{RemoteView}}{\SignMessage{\Transaction{\Involved{}}}{\Client{}},\Delta}}{\Replica{q}}$ messages
	 }
		\STATE Initiate Local view-change protocol. 
	 \ENDEVENT
	
    \end{myprotocol}
    \caption{\small The remote view-change algorithm of \RingBFT{}.}
    \label{alg:remote-view}
\end{figure}

\subsubsection{\bf \em Remote View Change}
\label{sss:remote-view-change}
A {\em partial communication} attack could be either due to a Byzantine primary or unreliable network. 
If the primary $\Primary{\Shard{S}}$ of shard $\Shard{S}$ is Byzantine, then it will ensure that at most $\f$ non-faulty replicas 
replicate a cross-shard transaction $\Transaction{\Involved{}}$ ($\Shard{S}, \Shard{U} \in \Involved{}$), locally.
As a result, replicas of next shard $\Shard{U}$ will receive at most $\f$ $\Name{Forward}$ messages.
Another case is where the network is unreliable, and under such conditions, replicas of $\Shard{U}$ 
may again receive at most $\f$ $\Name{Forward}$ messages.  

From Figure~\ref{alg:ringbft}, we know that when replica $\Replica{q}$ of shard $\Shard{U}$ receives a $\Name{Forward}$ 
message from replica $\Replica{r}$ of shard $\Shard{S}$ such that $\ID{\Replica{r}} = \ID{\Replica{q}}$, 
then $\Replica{q}$ broadcasts this $\Name{Forward}$ message to all the replicas in $\Shard{U}$. 
At this point, \RingBFT{} also requires replica $\Replica{q}$ to start the {\em remote timer}. 
If any replica $\Replica{q}$ in shard $\Shard{U}$ does not receive identical $\Name{Forward}$ messages from $\f+1$ distinct replicas 
of shard $\Shard{S}$, prior to the timeout of its remote timer, then $\Replica{q}$ detects a cross-shard attack and
 sends a $\Name{RemoteView}$ 
message to the replica $\Replica{r}$ of shard $\Shard{S}$, where $\ID{\Replica{r}} = \ID{\Replica{q}}$.
Following this, $\Replica{r}$ broadcasts the received $\Name{RemoteView}$ message to all the replicas in $\Shard{S}$. 
Finally, when any replica $\Replica{r}$ of shard $\Shard{S}$ receives $\Name{RemoteView}$ messages from $\f+1$ replicas of $\Shard{U}$, 
it supports the view change request and initiates the view-change protocol.
We illustrate this process in Figure~\ref{alg:remote-view}.

\subsubsection*{Triggering of Timers.} 
In \RingBFT{}, we know that for each cross-shard transaction, each replica $\Replica{r}$ of $\Shard{S}$ 
sets three distinct timers. 
Although each timer helps in recovering against a specific attack, there needs to be an order in which they timeout. 
As local timers lead to detecting a local malicious primary, we expect a local timer to have the shortest duration. 
Further, a remote timer helps to detect a lack of communication due to which it has a longer duration than local timers. 
Similarly, we require the duration of retransmit timer to be the longest.

\section{RingBFT Guarantees}
We now state the safety, liveness, and no deadlock guarantees provided by our \RingBFT{} protocol. 
\begin{proposition}\label{prop:non_divergent}
Let $\Replica{R}_i$, $i \in \{1, 2\}$,  be two non-faulty replicas in shard $\Shard{S}$ that committed to 
$\SignMessage{\Transaction{i}}{\Client{i}}$ as the $k$-th transaction sent by $\Primary{}$. 
If $\n > 3\f$, then $\SignMessage{\Transaction{1}}{\Client{1}} = \SignMessage{\Transaction{2}}{\Client{2}}$.
\end{proposition}
\begin{proof}
Replica $\Replica{r}_i$ only committed to $\SignMessage{\Transaction{}}{\Client{i}}$ after $\Replica{r}_i$ 
received identical $\Message{\Name{Commit}}{\Digest,k}$ messages from $\nf$ distinct replicas in $\Shard{S}$.
Let $X_i$ be the set of such $\nf$ replicas and 
$Y_i = X_i \difference \Faulty$ be the non-faulty replicas in $X_i$.
As $\abs{\Faulty} = \f$, so $\abs{Y_i} \geq \nf - \f$.
We know that each non-faulty replica only supports one transaction from primary $\Primary{}$ as the 
$k$-th transaction, and it will send only one $\Name{Prepare}$ message.
This implies that sets $Y_1$ and $Y_2$ must not overlap.
Hence, $\abs{X_1 \union X_2} \geq 2(\nf - \f)$.
As $\abs{X_1 \union X_2} = \nf$, the above inequality simplifies to $3\f \geq \n$, which contradicts $\n > 3\f$. 
Thus, we conclude $\SignMessage{\Transaction{1}}{\Client{1}} = \SignMessage{\Transaction{2}}{\Client{2}}$.
\end{proof}

\begin{theorem}
{\bf \em No Deadlock:} 
In a system $\Shards{}$ of shards, where $\Shard{S}, \Shard{U} \in \Shards{}$ and $\Shard{S} \neq \Shard{U}$,
no two replicas $\Replica{r} \in \Shard{S}$ and $\Replica{q} \in \Shard{U}$  
that order two conflicting transactions $\Transaction{\Involved_1}$ and $\Transaction{\Involved_2}$ such that 
$\Shard{S}, \Shard{U} \in \Involved_1 \intersect \Involved_2$ will execute $\Transaction{\Involved_1}$ and $\Transaction{\Involved_2}$ 
in different orders.
\end{theorem}

\begin{proof}
We know that \RingBFT{} associates an identifier with each shard and uses this identifier 
to define a ring order.
Let $\ID{\Shard{S}} < \ID{\Shard{U}}$, and 
the ring order be defined as lowest to highest identifier.
Assume that the conflicting transactions $\Transaction{\Involved_1}$ and $\Transaction{\Involved_2}$ 
are in a deadlock at shards $\Shard{S}$ and $\Shard{U}$, where
 $\Shard{S}, \Shard{U} \in \Involved_1 \intersect \Involved_2$. 
This implies that each non-faulty replica $\Replica{r} \in \Shard{S}$ has locked some data-item 
for $\Transaction{\Involved_1}$ that is required by $\Transaction{\Involved_2}$ 
while each non-faulty replica $\Replica{q} \in \Shard{U}$ has locked some data-item 
for $\Transaction{\Involved_2}$ that is required by $\Transaction{\Involved_1}$ 
or vice versa.

As each transaction $\Transaction{\Involved_i}$, $i \in [1,2]$ 
accesses $\Shard{S}$ and $\Shard{U}$ in ring order, 
so each transaction $\Transaction{\Involved_i}$ was initiated by $\Shard{S}$. 
This implies that the primary of $\Shard{S}$ would have assigned 
these transactions distinct sequence numbers $k_i, i \in [1,2]$, such that 
$k_1 < k_2$ or $k_1 > k_2$ ($k_1 = k_2$ is not possible as it will be 
detected as a Byzantine attack). 
During the commit phase, each replica $\Replica{r}$ will
put the transaction with larger sequence number $k_i$ in the $\pi$ list 
and lock the corresponding data-item
(Figure~\ref{alg:ringbft}, Line~\ref{alg:pi-push}), 
while the transaction with smaller $k_i$ is forwarded to the next shard $\Shard{U}$.
The transaction present in the $\pi$ list is only extracted once the 
data-item is unlocked.
Hence, there is a contradiction, that is, shards $\Shard{S}$ and $\Shard{U}$ 
will not suffer deadlock.
\end{proof}

\begin{theorem}
{\bf \em Safety:}
In a system $\Shards{}$ of shards, where each shard $\Shard{S} \in \Shards{}$ has at most $\f$ Byzantine replicas, 
each replica $\Replica{r}$ follows the Involvement, Non-divergence, and Consistence properties.
Specifically, all the replicas of $\Shard{S}$ execute each transaction in the same order, and every 
conflicting cross-shard transaction is executed by all the replicas of all the involved shards in the same order.
\end{theorem}

\begin{proof}
Using Proposition~\ref{prop:non_divergent} we have already illustrated that \RingBFT{} safely replicates a single-shard 
transaction, despite a malicious primary and/or unreliable network. 
In specific, any non-faulty replica $\Replica{R} \in \Replicas{\Shard{S}}$ will only commit a single-shard transaction if 
it receives $\Name{Commit}$ messages from $\nf$ distinct replicas in $\Replicas{\Shard{S}}$. 
When a non-faulty replica receives less than $\nf$ $\Name{Commit}$ messages, then eventually its local timer will timeout 
and it will participate in the view-change protocol.
Post the view-change protocol, any request that was committed by at least one non-faulty replica will persist across views.

Similarly, we can show that each cross-shard transaction is also safely replicated across all replicas of all the involved shards. 
In \RingBFT{}, each cross-shard transaction $\Transaction{\Involved{}}$ is processed in ring order by all the 
involved shards $\Involved{}$.
Let shards $\Shard{S}, \Shard{U} \in \Involved{}$ and $\ID{\Shard{S}} < \ID{\Shard{U}}$ such that 
ring order is based on lowest to highest identifier.
Hence, replicas of shard $\Shard{U}$ will only start consensus on $\Transaction{\Involved{}}$ if they receive $\Name{Forward}$ 
messages from $\f+1$ distinct replicas of $\Shard{S}$. 
Further, each of these $\Name{Forward}$ messages includes \Name{DS} from $\nf$ distinct replicas of $\Shard{S}$ on identical $\Name{Commit}$
messages corresponding to $\Transaction{\Involved{}}$, which guarantees that $\Transaction{\Involved{}}$ 
was replicated in $\Shard{S}$.
If the network is unreliable and/or primary of shard $\Shard{S}$ is Byzantine, 
then replicas of $\Shard{U}$ will receive less than $\f+1$ $\Name{Forward}$ messages. 
In such a case, either the remote timer at replicas of $\Shard{U}$ will timeout, or 
one of the two timers (local timer or transmit timer) of replicas of $\Shard{S}$ will timeout. 
In any case, following the specific recovery procedure, replicas of $\Shard{U}$ will receive a sufficient number of $\Name{Forward}$ messages. 
\end{proof}

\begin{theorem}
{\bf \em Liveness:}
In a system $\Shards{}$ of shards, where each shard $\Shard{S} \in \Shards{}$ has at most $\f$ Byzantine replica, 
if the network is reliable, then each replica $\Replica{r}$ follows the Involvement and Termination properties. 
Specifically, all the replicas continue making progress, and good clients continue receiving responses for their 
transactions.
\end{theorem}

\begin{proof}
In the case of a single-shard transaction, 
if the primary is non-faulty, then each replica will continue processing client transactions. 
If the primary is faulty, and prevents a request from replicating by allowing at most $\f$ replicas 
to receive $\Name{Commit}$ messages, then such a primary will be replaced through view-change protocol, 
following which a new primary will ensure that the replicas continue processing subsequent transactions. 
Notice that there can be at most $\f$ such faulty primaries, and the system will eventually make progress. 
If the primary is malicious, then it can keep up to $\f$ non-faulty replicas in dark, which will continue 
making progress through periodic checkpoints.
In the case of a cross-shard transaction, 
there is nothing extra that a faulty primary $\Primary{\Shard{S}}$ 
can do than preventing local replication of the transaction. 
If $\Primary{\Shard{S}}$ does that, then as discussed above, $\Primary{\Shard{S}}$ will be replaced. 
Further, during any communication between two shards, primary has no extra advantage over other replicas 
in the system. 
Further, the existence of transmit and remote timers help replicas of all the involved shards to keep track 
of any malicious act by primaries.
\end{proof}

\section{Design \& Implementation}
\label{sec:impl}
\RingBFT{} aims to scale permissioned blockchains to hundreds of replicas through efficient sharding.
To argue the benefits of our \RingBFT{} protocol, we need to first implement it in a permissioned blockchain fabric. 
For this purpose, we employed a state-of-the-art permissioned blockchain fabric,  
\ResilientDB{}~\cite{r-evalpaper,resilientdb-demo,phd-workshop,tut-vldb20,tut-middleware19,geobft,poe,multibft}.
In our prior works, we illustrated how \ResilientDB{} offers an optimal system-centric design that eases 
implementing novel \BFT{} consensus protocols. 
Further, \ResilientDB{} presents an architecture that allows even classical protocols like \PBFT{} to achieve 
high throughputs and low latencies.

In this section, we describe in brief \ResilientDB's architecture and explain the design decisions we took to 
implement \RingBFT{}.

{\bf \em Network Layer.} 
\ResilientDB{} provides a network layer to manage communication among clients and replicas.
The network layer provides TCP/IP capabilities through {\em Nanomsg-NG} to communicate messages.
To facilitate uninterrupted processing of millions of messages, at each replica, 
\ResilientDB{} offers multiple input and output threads to communicate with the network.

\begin{figure}[t]
    \centering
    \includegraphics[width=0.44\textwidth]{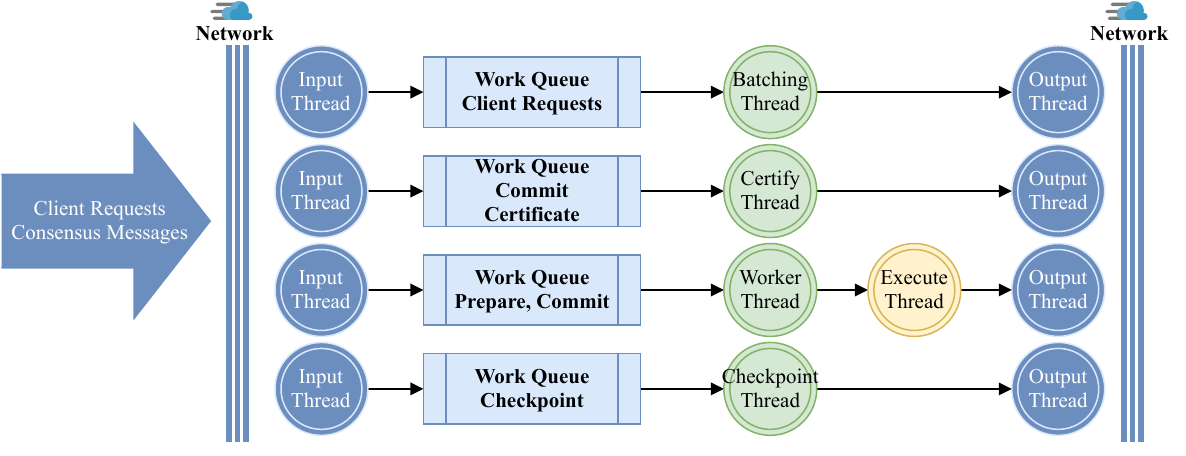}
    \caption{The parallel-pipelined architecture provided by \ResilientDB{} fabric for efficiently implementing \RingBFT{}.}
    \label{fig:ringbft_pipe}
\end{figure}

{\bf \em Pipelined Consensus.} 
Once a message is received from the network, the key challenge is to process it efficiently. 
If all the ensuing consensus tasks are performed sequentially, the resulting system output would be abysmally low.
Moreover, such a system would be unable to utilize the available computational and network capabilities.
Hence, \ResilientDB{} associates with each replica a {\em parallel pipelined} architecture, 
which we illustrate in Figure~\ref{fig:ringbft_pipe}.

When an  input thread receives a message from the network, it places them in a specific work queues based on 
the type of the message.
As depicted in Figure~\ref{fig:ringbft_pipe}, \ResilientDB{} provides dedicated threads for processing each type of message.

{\bf \em Blockchain.} 
To securely record each successfully replicated transaction, we also implement an immutable ledger--blockchain. 
For systems running fully-replicated \BFT{} consensus protocols like \PBFT{} and \ZZ{}, 
blockchain is maintained as a single linked-list of all transactions where each replica 
stores a copy of the blockchain. 
However, in the case of sharding protocols like \RingBFT, 
each shard maintains its own blockchain. 
As a result, no single shard can provide a complete state of all the transactions. 
Hence, we refer to the ledger maintained at each shard as a {\em partial-blockchain}.

Let, $\Shards{}$ be the system of $\z = \abs{\Shards{}}$ shards.
Say, we use the representation $\Shard{S}_1, \Shard{S}_2,...,\Shard{S}_i \in \Shards{}$,
to denote the shards in $\Shards{}$ where  $1 \le i \le z$.
In this sharded system, we represent the blockchain ledger maintained by replicas of $\Shard{S}_i$ as 
$\Ledger{\Shard{S}_i}$. 
Hence the complete state of the system  can be expressed as:
\begin{equation}
\Ledger{\Shard{S}_1} \union \Ledger{\Shard{S}_2} \union ... \union \Ledger{\Shard{S}_i} \union ... \union \Ledger{\Shard{S}_z}
\end{equation}
Further, we know that each ledger $\Ledger{\Shard{S}_i}$ is a linked list of blocks:
\begin{equation}
\Ledger{\Shard{S}_i} = \{\Block{1},\Block{2},...,\Block{k}\}
\end{equation}
where chaining is guaranteed by requiring each block to include the hash of the previous block:  
\begin{equation}
\Block{k} = \{k, \Digest, \Primary{\Shard{S}_i}, \Hash{\Block{k-1}} \}
\end{equation}

In \ResilientDB{}, for efficient processing, we follow existing literature and require the primary $\Primary{\Shard{S}_i}$
of shard $\Shard{S}_i$ to aggregate transactions in a batch and perform consensus on this batch.
Hence, each $k$-th block $\Block{k}$ in $\Ledger{\Shard{S}_i}$ represents a batch of transactions that replicas of 
$\Shard{S}_i$ successfully committed at sequence $k$.
Note: we expect each block to include all the transactions that access the same shards. 

If a block includes cross-shard transactions, then such a block is appended to the ledger of all the involved shards $\Involved{}$.
In specific, if a block $\Block{}$ includes a transaction $\Transaction{\Involved{}}$, 
such that $\Shard{S}_i, \Shard{S}_j \in \Involved{}$, then $\Block{} \in \Ledger{\Shard{S}_i}$ and 
$\Block{} \in \Ledger{\Shard{S}_j}$.
Notice that the order in which these blocks appear in each individual chain can be different.
However, if two blocks $\Block{x}$ and $\Block{y}$ include conflicting transactions that access intersecting 
set of shards, and consensus on $\Block{x}$ {\em happens before} $\Block{y}$, then 
in each ledger $\Block{x}$ is appended before $\Block{y}$.

Depending on the choice of storage, each block can include either all the transactional information or 
the Merkle Root~\cite{merkle} of all transactions in the block. 
A Merkle Root ($\Digest{}$) helps to optimize the size of each block, and is generated by assuming all the transactions 
in a batch as leaf nodes, followed by a pair-wise hashing up till the root.
To initialize each blockchain, every replica adds an agreed upon dummy block termed as the 
{\em genesis block}~\cite{bc-processing}.

\section{Evaluation}\label{sec:eval}
In this section, we evaluate our \RingBFT{} protocol.
To do so, we implement \RingBFT{} on our high throughput yielding permissioned blockchain 
fabric, \ResilientDB.

For experimentation, we deploy \ResilientDB{} on Google \changed{Cloud} Platform (GCP) in {\bf \em fifteen regions}
across {\bf \em five continents}, namely:
Oregon, Iowa, Montreal, Netherlands, Taiwan, Sydney, Singapore, South Carolina,
North Virginia, Los Angeles, Las Vegas, London, Belgium, Tokyo, and Hong Kong. 
In any experiment involving less than $15$ shards, the choice of the shards is in the order 
we have mentioned above.
We deploy each replica on a $16$-core N1 machine having Intel Broadwell CPUs with a $2.2$GHz clock and $32$GB RAM.
For deploying clients, we use the $4$-core variants having $16$GB RAM.
For each experiment, we equally distribute the clients in all regions.

{\bf \em Benchmark.} 
To provide workload for our experiments, we use the {\em Yahoo Cloud Serving Benchmark} 
\changed{from the BlockBench suite} (YCSB)~\cite{ycsb,blockbench} .
Each client transaction queries a YCSB table with an active set of $\SI{600}{\kilo\nothing}$ records.
For our evaluation, we adopt transactions that read 
and modify existing records. Prior to each experiment, each replica initializes an identical copy of 
the YCSB table. 
\changed{
YCSB workloads help us to create cross-shard client transactions with varying degrees of conflict, 
while other workloads aim to evaluate the cost of executing a transaction, which is orthogonal to our RingBFT consensus.
}

{\bf \em Existing Protocols.}
In all our experiments, we compare the performance of \RingBFT{} against two other state-of-the-art sharding \BFT{} 
protocols, \AHL~\cite{ahl} and \Sharper~\cite{sharper}.
In Section~\ref{s:back}, we highlighted key properties of these protocols. 
Like \RingBFT{}, both \AHL{} and \Sharper{} employ \PBFT{} to achieve consensus on single-shard transactions. 
Hence, all three protocols have identical implementations for replicating single-shard transactions.
For achieving consensus on cross-shard transactions, we follow the respective algorithms and 
modify \ResilientDB{} appropriately.

{\bf \em WAN Bandwidth and Round-Trip Costs.}
As the majority of experiments take place in a geo-scaled WAN environment spanning multiple continents, 
available bandwidth and round-trip costs between two regions play a crucial role.
Prior works~\cite{steward,geobft} have illustrated that if the available bandwidth is low and round-trip costs 
are high, then the protocols dependent on a subset of replicas face performance degradation.
In the case of \AHL{}, the reference committee is responsible for managing cross-shard consensus, while 
for \Sharper{}, the primary of coordinating shard leads the cross-shard consensus. 
Hence, both of these protocols observe low throughput and high latency in proportion to available bandwidth 
and round-trip costs.
Although \RingBFT{} requires cross-shard communication in the form of $\Name{Forward}$ and $\Name{Execute}$ messages, 
the system is comparably less burdened as all the replicas participate equally in this communication process.

{\bf \em Standard Settings.} 
Unless {\em explicitly} stated, we use the following settings for all our experiments.
We run with a mixture of single-shard and cross-shard transactions, of which $30\%$ are cross-shard transactions.
Each cross-shard \changed{transaction} accesses all the $15$ regions, and in each shard we deploy $28$ replicas, 
that is, a total of $420$ globally distributed replicas.
\changed{The number of key-value pairs accessed by each transaction varies in accordance with the number 
of regions accessed. 
For example, if a transaction accesses three regions, then it accesses three key-value pairs.}
In these experiments, we allow up to $50$K clients to send transactions.
Further, we require clients and replicas to employ batching and create batches of transactions of size $100$.

The sizes of messages communicated during \RingBFT{} consensus are:
$\Name{Preprepare}$  ($5408$B),
$\Name{Prepare}$ ($216$B),
$\Name{Commit}$ ($269$B),
$\Name{Forward}$ ($6147$B),
$\Name{Checkpoint}$ ($164$B), and
$\Name{Execute}$ ($1732$B).

\changed{
Note: Our \RingBFT{} protocol provides support for standard multi-statement transactions 
that are widely adopted by deterministic databases~\cite{qstore,calvin,deneva,easyc,tpbook}. 
Hence, the complexity of designing \RingBFT{} is similar to running an application on 
top of a deterministic database. 
Hence, we believe a developer would not face any new challenges.
}

Through our experiments, we want to answer the following:
\begin{enumerate}[wide,nosep,label=(Q\arabic*),ref={Q\arabic*}]
    \item \label{Q:scale-shards} What is the effect of increasing the number of shards on consensus provided by \RingBFT{}?
    \item \label{Q:scale-replicas} How does varying the number of replicas per shard affects the performance of \RingBFT{}?
    \item \label{Q:cross-shard} What is the impact of increasing the percentage of cross-shard transactions on \RingBFT{}?
    \item \label{Q:batch-size} How does batching affect the system performance?
    \item \label{Q:involved-shards} What is the effect of varying the number of involved shards in a cross-shard transaction on \RingBFT{}?
    \item \label{Q:clients} What is the impact of varying number of clients on consensus provided by \RingBFT{}?
    \item \label{Q:view-change} How do faulty primary and view change affect the performance of \RingBFT{}?
\end{enumerate}

\begin{figure*}
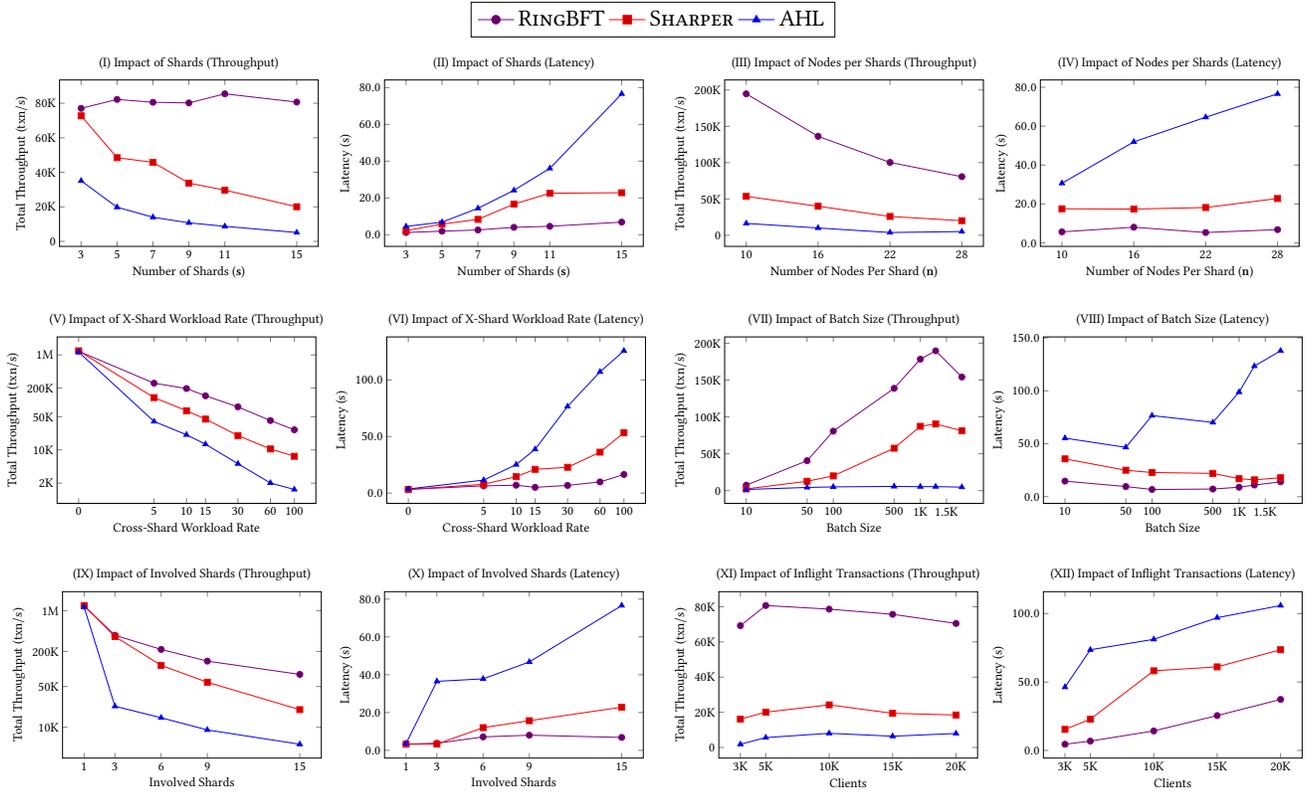

    \centering
    \setlength{\tabcolsep}{1pt}
    \scalebox{0.8}{\ref{mainlegend}}\\[5pt]
    \begin{tabular}{cc@{\quad}cc}
    \graphAFullTP & \graphALat & \graphBFullTP & \graphBLat \\ \\
    \graphCFullTP & \graphCLat & \graphDFullTP & \graphDLat \\ \\
    \graphEFullTP & \graphELat & \graphFFullTP & \graphFLat  \\ \\

    \end{tabular}
    \caption{Measuring system throughput and average latency on running different \BFT{} sharding consensus protocols.}
    \label{fig:plots}
\end{figure*}

\subsection{Scaling Number of Shards.}
For our first set of experiments, we study the effect of scaling the number of shards. 
In specific, we require clients to send cross-shard transactions that can access from $3$, $5$, $7$, $9$, $11$, and $15$ shards, while
keeping other parameters at the standard setting.
We use Figures~\ref{fig:plots} (\RN{1}) and (\RN{2}) to illustrate the throughput and latency metrics. 

\RingBFT{} achieves $16\times$ and $4\times$ higher throughput than \AHL{} and \Sharper{} 
in the 15 shard setting, respectively. 
An increase in the number of shards only increases the length of the ring while keeping the amount of communication between 
two shards at constant.
As a result, for \RingBFT{}, we observe an increase in latency as there is an increase in time to go around the ring, namely, a linear neighbor-to-neighbor communication.
From three shards to 15 shards, the latency increases from $1.17\s$ to $6.82\s$.
Notice that the throughput for \RingBFT{} is nearly constant since the size of shards and the amount of communication among shards are constant. 
This is a consequence of an increase in the number of shards that can perform consensus on single-shard transactions in parallel. 
Although on increasing the number of shards, there is a proportional increase in the number of involved shards per transaction, 
the linear communication pattern of \RingBFT{} prevents throughput degradation.

In the case of \AHL{}, the consensus on cross-shard transactions is led by the reference committee, which essentially 
centralizes the communication in the global setting and affects the system performance.
In contrast, \Sharper{} scales better because there is no single reference committee leading all cross-shard consensuses. 
However, even \Sharper{} sees a fall in throughput due to two rounds of communication 
between all replicas of all the involved shards.
For a system where all the shards are globally scattered, quadratic communication complexity and communication between 
all the shards impacts the scalability of the system.

\subsection{Scaling Number of Replicas per Shard.}
We now study the effects of varying different parameters
within a single shard.
Our next set of experiments aim to increase the amount of replication within a single shard. 
In specific, we allow each shard to have $10$, $16$, $22$, and $28$ replicas. 
We use Figures~\ref{fig:plots} (\RN{3}) and (\RN{4}) to illustrate the throughput and latency metrics.

These plots reaffirm our theory that \RingBFT{} ensures up to $16\times$ higher throughput and $11\times$ lower latency than the other two protocols. 
As the number of replicas in each shard increases, there is a corresponding decrease in throughput for \RingBFT{}. 
This decrease is not surprising because \RingBFT{} employs the \PBFT{} protocol for local replication, 
which necessitates two phases of quadratic communication complexity.
This, in turn increases the size (and as a result cost) of $\Name{Forward}$ messages communicated between shards.

In the case of \AHL{}, the existence of a reference committee acts as a performance bottleneck to an extent that 
$30\%$ cross-shard transactions involving all the $15$ shards subsides the benefits due to reduced replication 
($10$ or $16$ replicas).
\Sharper{} also observes a drop in its performance as it relies on \PBFT{}, and is unable to scale at smaller 
configurations due to expensive communication that requires an all-to-all communication between the replicas of 
involved shards.
{\bf \em To summarize:} \RingBFT{} achieves up to $4\times$ and $16\times$ higher throughput than \Sharper{} 
and \AHL{}, respectively.

\subsection{Varying percentage of Cross-shard Transactions.}
For our next study, we 
allow client workloads to have $0$, $5\%$, $10\%$, $15\%$, $30\%$, $60\%$, and $100\%$ cross-shard transactions. 
We use Figures~\ref{fig:plots} (\RN{5}) and (\RN{6}) to illustrate the throughput and latency metrics. 

When the workload contains no cross-shard transactions, it simply indicates a system where all the transactions 
access only one shard. 
In this case, all the three protocols attain the same throughput and latency as all of them employ \PBFT{} for 
reaching consensus on single-shard transactions.
They achieve \textbf{1.2 Million \si{\txn\per\second}} throughput among 500 nodes in 15 globally distributed regions.
With a small ($5\%$) introduction of cross-shard transactions in the workload, there is a significant decrease 
for all the protocols. 
The amount of decrease is in accordance to the reasons we discussed in previous sections.
However, \RingBFT{} continues to outperform other protocols. 
In the extreme case of $100\%$ cross-shard workload, \RingBFT achieve $4\times$ and $18\times$ higher throughput and $3.3\times$ and $7.8\times$ lower latency than \Sharper{} and \AHL{}, respectively.

\subsection{Varying the Batch Size.}
Next, we study the impact of batching transactions on system performance.
We require the three protocols to run consensus on batches of client transactions with sizes
$10$, $50$, $100$, $500$, $1$K, and $5$K.
We use Figures~\ref{fig:plots} (\RN{7}) and (\RN{8}) to illustrate the throughput and latency metrics.

As the number of transactions in a batch increases, there is a proportional decrease in the number of consensuses.
For example, with a batch size of 10 and 100 for 5000 transactions, we need 500 and 50 instances of consensus.
However, larger batches also cause an increase in latency due to the increased cost of communication and time for processing all the transactions in the batch.
Hence, we observe an increase in throughput on moving from small batches of $10$ 
transactions to large batches of $1$K transactions.
On further increase (after $1.5$K), the system throughput hits saturation and eventually decreases as benefits of batching 
are over-shadowed by increased communication costs.

Starting from the batch size of 10, on increasing the batch size, the throughput increases up to $27\times$ in \RingBFT{} because, with less communication and fewer messages, we are processing more transactions.
This trend lasts until the system reaches its saturation point in terms of communication and computation, which is the batch size of $1.5$K for \RingBFT{}.
Once the system is at filling its network bandwidth, adding more transactions to the batch will not increase the throughput
because it cannot process more, and sending those batches will be a bottleneck for the system.
Ideally, it should get constant after some point but because of implementation details and queuing, it drops slightly after some time. 

Ideally, we expect the latency to also decrease with an increase in batch size. 
However, for \RingBFT{}, more transactions in a batch implies more time spent processing the transactions around the ring. 
This causes an increase in latency for the client.
{\bf \em To summarize:} Using the optimal batch size improve the throughput of \RingBFT{}, \Sharper{} and \AHL{}, $27\times$, $45\times$, and $3\times$ respectively.

\subsection{Varying Number of Involved Shards.}
We now keep the number of shards fixed at $15$ and require all clients to create transactions that access a subset of these shards. 
In specific, clients send transactions that access $1$, $3$, $6$, $9$, and $15$ shards.
As our selected order for shards gives no preference to their proximity to each other (to prevent any bias), 
our clients select consecutive shards in order to generate the workload.

We use Figures~\ref{fig:plots} (\RN{9}) and (\RN{10}) to illustrate the throughput and latency metrics.
As expected, all three protocols observe a drop in performance on the increase in the number of involved shards.
However, \RingBFT{} still outperforms the other two protocols. As we increase the number of involved shards, the performance gap between \RingBFT{} and the other two protocols increases.
As shown in the graph, with three shards involved, \RingBFT{} has a $4\%$  performance gap, increasing to $4\times$ with 15 shards involved.

\subsection{Varying Number of Clients.}
Each system can reach optimal latency only if it is not overwhelmed by incoming client requests. 
In this section, we study the impact of the same by varying the number of incoming client transactions 
through a gradual increase in the number of clients from $5$K to $20$K.
We use Figures~\ref{fig:plots} (\RN{11}) and (\RN{12}) to illustrate resulting throughput and latency metrics.
As we increase the number of clients transmitting transactions, we observe a $15-20\%$ increase in throughput, reaching the saturation point.
Having more clients causes a decrease between $7\%$ and $9\%$,  which is a result of various queues being full with incoming requests, which in turn causes a replica to perform extensive memory management. 
Due to similar reasons, there is a significant increase in latency as the time to process each request has increased proportionally.
We observed $32.75\s$, $58.21\s$, and $59.64\s$ increase in \RingBFT{}, \Sharper{}, and \AHL{} respectively.
Despite this, \RingBFT{} scales better than other protocols even when the system is overwhelmed by clients.

\subsection{Impact of Primary Failure.}
\begin{figure}[t]
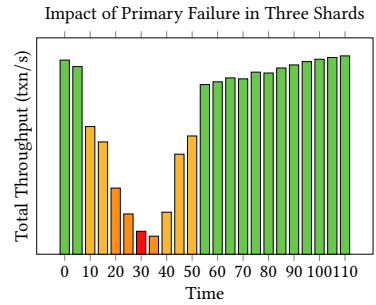

    \centering
    \graphVC
    \caption{\RingBFT's  throughput under the primary failure of three shards out of nine. ($\s=10$) primary fails;
    ($\s=20$) replicas timeout and send view-change messages; ($\s=30$) new primary starts the new view;
    ($\s=35$) system's throughput start increasing and returns back to normal at $\s=55$.}
    \label{fig:view_change}
\end{figure}

Next, we evaluate the effect of replacing a faulty primary in different shards.
For this experiment, we run experiments with $9$ shards and allow workload to consist of $30\%$ cross-shard transactions.
We use Figure~\ref{fig:view_change} to show the throughput attained by \RingBFT{} when the primary of 
the first three shards fail, and the replicas run the view change protocol to replace the faulty primary.
The primaries of these shards fail at $10s$, and the system's average throughput starts decreasing 
while other shards are processing their clients' requests.
\RingBFT{} observes a $15\%$ decrease in throughput and post view change; it again observes an increase in throughput.

\subsection{Impact of Complex Cross-Shard Transactions.}
Until now, we have experimented with simple \CST{} where for a given \CST{} each shard 
could independently execute its data-fragment.
However, a sharded system may encounter a complex \CST{} where 
each shard may require access to data (and needs to check constraints) present in other shards while executing its 
data-fragment.
These data-access dependencies require each shard to read the data from remote shards.

\begin{figure}
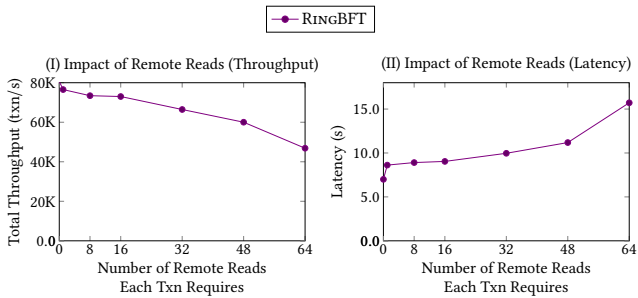

    \centering
    \setlength{\tabcolsep}{1pt}
    \scalebox{0.6}{\ref{mainlegend2}}\\[7pt]
    \begin{tabular}{cc}
    \graphDPFullTP & \graphDPLat 
    \end{tabular}
    \caption{\RingBFT's throughput and latency on encountering complex cross-shard transactions with dependencies varying from $0$ to $64$.}
    \label{fig:DPplots}
\end{figure}

Our \RingBFT{} protocol performs this task by requiring each shard to send its read-write sets 
along with the $\Name{Forward}$ message.
In this section, we study the cost of communicating the read-write sets 
of a complex \CST{} on our \RingBFT{} protocol.
We use Figure~\ref{fig:DPplots} to illustrate the throughput and latency metrics 
on varying the number of data-access dependencies from $0$ to $64$ distributed 
randomly across 15 shards. These figures illustrate that our \RingBFT{} protocol provides
reasonable throughput and latency even for a \CST{} with extensive dependencies.

Note that we have not included \Sharper{} and \AHL{} in Figure~\ref{fig:DPplots} 
as supporting complex \CST{} is not covered in \cite{sharper,ahl} and remains as 
an open problem. For example, to support remote reads, first, there must be a 
consensus on the remote shard to agree on the requested operations and their values. 
Second, on the receiving end, there must be another local consensus on the values 
received. If the remote values are not received, then a consensus is needed to detect 
failures in order to invoke remote recovery to restore liveness. 
Now \Sharper{}
has a single global consensus that coordinates among all shards and their replicas. 
Thus, extending \Sharper{} is nontrivial because it is unclear as to when and how 
the additional remote consensuses and recoveries could be invoked. 
In the case of \AHL{}, due to 
its 2PC design, invoking remote consensus on each shard to process remote read is 
simple, but it is challenging to invoke remote view change when the 
network is unreliable or the primary of the remote shard behaves maliciously. 
Moreover, keeping the question of feasibility aside, we observe that in Figure~\ref{fig:plots}(\RN{1}), at 15 shards, 
the throughputs of both \Sharper{} and \AHL{} are under $20$K while \RingBFT{} 
sustains $80$K transactions/second. However, in Figure~\ref{fig:DPplots}, when we 
scale up to $64$ remote operations across $15$ shards, \RingBFT{} yields a throughput of at least 
$45$K transactions/second, surpassing both baselines with no remote operations.

\section{Related Work}
In Section~\ref{s:intro}, we presented an overview of different types of \BFT{} protocols. 
Further, we have extensively studied the architecture of state-of-the-art permissioned sharding \BFT{} protocols, \AHL{} and \Sharper{}.
We now summarize other works in the space of Byzantine Fault-Tolerance consensus.

{\em Traditional \BFT{} consensus.} 
The consensus problems such as Byzantine Agreement and Interactive Consistency have been studied in literature in great detail~\cite{byzgen,dolevbound,interbound,flp,possibleasync,distalgo}. 
With the introduction of \PBFT{}-powered \emph{BFS}---a fault-tolerant version of the networked file system~\cite{nfs}---by Castro et al.~\cite{pbft,pbftj} there has been an unprecedented interest in the design of high-performance \BFT{} consensus protocols.
This has led to the design of several consensus protocols that have optimized different aspects of \PBFT{}, e.g, 
\ZZ{}, \SBFT{}, and \textsc{PoE}, as discussed in the Introduction. 
To further improve on the performance of \PBFT{}, some consensus protocols consider providing less failure resilience~\cite{qubft,byzq,phalanx}, focused on a theoretical framework to support weaker consistency and isolation semantics such as dirty reads and committed reads~\cite{byshard}, or rely on trusted components~\cite{hybster,less-replica-2,sgx-trust}.  

Guerraoui et al.~\cite{stretchingbft} introduced the StretchingBFT protocol that aims to improve on \PBFT{} 
by arranging replicas in a ring-like topology where each replica communicates with its two neighbors.
Our \RingBFT{} is a meta-protocol that can utilize any of these \BFT{} protocol to achieve optimal intra-shard consensus.
Hence, these protocols complement our design.
Further, these protocols cannot scale to hundreds of replicas scattered across the globe, and this is where our vision 
of \RingBFT{} acts as a resolve.

{\em Permissionless Sharded Blockchains.}
Permissionless space includes several sharding \BFT{} consensus protocols, such as {Conflux}~\cite{conflux}, {Elastico}~\cite{elastico}, 
{MeshCash}~\cite{meshcash}, {OmniLedger}~\cite{omniledger}, and {Spectre}~\cite{spectre}.
All of these protocols require each of their shards to run either the Proof-of-Work or Proof-of-Stake 
protocol during some phase of the consensus. 
As a result these protocols offer a magnitude lower throughput than both \AHL{} and \Sharper{}, which are included in our evaluation. 

In our recent sharding work, we have developed a comprehensive theoretical framework to study a wide range of consistency models and isolation semantics (e.g., dirty reads, committed reads, serializability) and communication patterns (e.g., centralized vs. distributed) ~\cite{byshard}. We have further developed a hybrid sharding protocol intended for the permissionless setting optimized for the widely used unspent transaction model~\cite{cerberus}.

\section{Conclusions}
In this paper, we present \RingBFT{}--a novel meta-\BFT{} protocol for permissioned sharded blockchains.
For a single-shard transaction, \RingBFT{} performs as efficient as any state-of-the-art 
sharding \BFT{} consensus protocol. 
However, existing sharding \BFT{} protocols face severe fall in throughput when they 
have to achieve consensus on a cross-shard transaction.
\RingBFT{} resolves this situation by requiring each shard to participate in at most two rotations around the ring. 
In specific, \RingBFT{} expects each shard to adhere to the prescribed ring order, 
and follow the \textit{principle of process, forward, and re-transmit}, while ensuring the communication between shards is linear.
We implement \RingBFT{} on our efficient \ResilientDB{} fabric, and evaluate 
it against state-of-the-art sharding \BFT{} protocols. 
Our results illustrates that \RingBFT{} achieves up to $18\times$ higher 
throughput than the most recent sharding protocols and easily scales to nearly $500$ globally-distributed nodes.

\bibliographystyle{ACM-Reference-Format}
\bibliography{reference}


\begin{thebibliography}{65}


\ifx \showCODEN    \undefined \def \showCODEN     #1{\unskip}     \fi
\ifx \showDOI      \undefined \def \showDOI       #1{#1}\fi
\ifx \showISBNx    \undefined \def \showISBNx     #1{\unskip}     \fi
\ifx \showISBNxiii \undefined \def \showISBNxiii  #1{\unskip}     \fi
\ifx \showISSN     \undefined \def \showISSN      #1{\unskip}     \fi
\ifx \showLCCN     \undefined \def \showLCCN      #1{\unskip}     \fi
\ifx \shownote     \undefined \def \shownote      #1{#1}          \fi
\ifx \showarticletitle \undefined \def \showarticletitle #1{#1}   \fi
\ifx \showURL      \undefined \def \showURL       {\relax}        \fi
\providecommand\bibfield[2]{#2}
\providecommand\bibinfo[2]{#2}
\providecommand\natexlab[1]{#1}
\providecommand\showeprint[2][]{arXiv:#2}

\bibitem[\protect\citeauthoryear{Abd-El-Malek, Ganger, Goodson, Reiter, and
  Wylie}{Abd-El-Malek et~al\mbox{.}}{2005}]%
        {qubft}
\bibfield{author}{\bibinfo{person}{Michael Abd-El-Malek},
  \bibinfo{person}{Gregory~R. Ganger}, \bibinfo{person}{Garth~R. Goodson},
  \bibinfo{person}{Michael~K. Reiter}, {and} \bibinfo{person}{Jay~J. Wylie}.}
  \bibinfo{year}{2005}\natexlab{}.
\newblock \showarticletitle{Fault-scalable Byzantine Fault-tolerant Services}.
  In \bibinfo{booktitle}{\emph{Proceedings of the Twentieth ACM Symposium on
  Operating Systems Principles}}. \bibinfo{publisher}{ACM},
  \bibinfo{pages}{59--74}.
\newblock
\urldef\tempurl%
\url{https://doi.org/10.1145/1095810.1095817}
\showDOI{\tempurl}


\bibitem[\protect\citeauthoryear{Amir, Danilov, Kirsch, Lane, Dolev,
  Nita-Rotaru, Olsen, and Zage}{Amir et~al\mbox{.}}{2006}]%
        {steward}
\bibfield{author}{\bibinfo{person}{Yair Amir}, \bibinfo{person}{Claudiu
  Danilov}, \bibinfo{person}{Jonathan Kirsch}, \bibinfo{person}{John Lane},
  \bibinfo{person}{Danny Dolev}, \bibinfo{person}{Cristina Nita-Rotaru},
  \bibinfo{person}{Josh Olsen}, {and} \bibinfo{person}{David Zage}.}
  \bibinfo{year}{2006}\natexlab{}.
\newblock \showarticletitle{Scaling Byzantine Fault-Tolerant Replication to
  Wide Area Networks}. In \bibinfo{booktitle}{\emph{International Conference on
  Dependable Systems and Networks (DSN'06)}}. \bibinfo{pages}{105--114}.
\newblock
\urldef\tempurl%
\url{https://doi.org/10.1109/DSN.2006.63}
\showDOI{\tempurl}


\bibitem[\protect\citeauthoryear{Amiri, Agrawal, and Abbadi}{Amiri
  et~al\mbox{.}}{2019a}]%
        {caper}
\bibfield{author}{\bibinfo{person}{Mohammad~Javad Amiri},
  \bibinfo{person}{Divyakant Agrawal}, {and} \bibinfo{person}{Amr~El Abbadi}.}
  \bibinfo{year}{2019}\natexlab{a}.
\newblock \showarticletitle{{CAPER}: A Cross-application Permissioned
  Blockchain}.
\newblock \bibinfo{journal}{\emph{Proc. VLDB Endow.}} \bibinfo{volume}{12},
  \bibinfo{number}{11} (\bibinfo{year}{2019}), \bibinfo{pages}{1385--1398}.
\newblock
\showISSN{2150-8097}
\urldef\tempurl%
\url{https://doi.org/10.14778/3342263.3342275}
\showDOI{\tempurl}


\bibitem[\protect\citeauthoryear{Amiri, Agrawal, and El~Abbadi}{Amiri
  et~al\mbox{.}}{2019b}]%
        {sharper}
\bibfield{author}{\bibinfo{person}{Mohammad~Javad Amiri},
  \bibinfo{person}{Divyakant Agrawal}, {and} \bibinfo{person}{Amr El~Abbadi}.}
  \bibinfo{year}{2019}\natexlab{b}.
\newblock \bibinfo{title}{{SharPer}: Sharding Permissioned Blockchains Over
  Network Clusters}.
\newblock
\newblock
\urldef\tempurl%
\url{https://arxiv.org/abs/1910.00765v1}
\showURL{%
\tempurl}


\bibitem[\protect\citeauthoryear{Behl, Distler, and Kapitza}{Behl
  et~al\mbox{.}}{2017}]%
        {hybster}
\bibfield{author}{\bibinfo{person}{Johannes Behl}, \bibinfo{person}{Tobias
  Distler}, {and} \bibinfo{person}{R\"{u}diger Kapitza}.}
  \bibinfo{year}{2017}\natexlab{}.
\newblock \showarticletitle{Hybrids on Steroids: {SGX}-Based High Performance
  {BFT}}. In \bibinfo{booktitle}{\emph{Proceedings of the Twelfth European
  Conference on Computer Systems}}. \bibinfo{publisher}{ACM},
  \bibinfo{pages}{222--237}.
\newblock
\urldef\tempurl%
\url{https://doi.org/10.1145/3064176.3064213}
\showDOI{\tempurl}


\bibitem[\protect\citeauthoryear{Bentov, Hub\'{a}\v{c}ek, Moran, and
  Nadler}{Bentov et~al\mbox{.}}{2017}]%
        {meshcash}
\bibfield{author}{\bibinfo{person}{Iddo Bentov}, \bibinfo{person}{Pavel
  Hub\'{a}\v{c}ek}, \bibinfo{person}{Tal Moran}, {and} \bibinfo{person}{Asaf
  Nadler}.} \bibinfo{year}{2017}\natexlab{}.
\newblock \bibinfo{title}{Tortoise and Hares Consensus: the Meshcash Framework
  for Incentive-Compatible, Scalable Cryptocurrencies}.
\newblock
\newblock
\urldef\tempurl%
\url{https://eprint.iacr.org/2017/300}
\showURL{%
\tempurl}


\bibitem[\protect\citeauthoryear{Bernstein and Goodman}{Bernstein and
  Goodman}{1983}]%
        {mvcc}
\bibfield{author}{\bibinfo{person}{P.~A. Bernstein} {and} \bibinfo{person}{N.
  Goodman}.} \bibinfo{year}{1983}\natexlab{}.
\newblock \showarticletitle{{Multiversion Concurrency Control - Theory and
  Algorithms}}.
\newblock \bibinfo{journal}{\emph{ACM TODS}} \bibinfo{volume}{8},
  \bibinfo{number}{4} (\bibinfo{year}{1983}), \bibinfo{pages}{465--483}.
\newblock


\bibitem[\protect\citeauthoryear{Butenuth, v.~Gösseln, Tiedge, Heipke, Lipeck,
  and Sester}{Butenuth et~al\mbox{.}}{2007}]%
        {federated-geospatial}
\bibfield{author}{\bibinfo{person}{Matthias Butenuth}, \bibinfo{person}{Guido
  v. Gösseln}, \bibinfo{person}{Michael Tiedge}, \bibinfo{person}{Christian
  Heipke}, \bibinfo{person}{Udo Lipeck}, {and} \bibinfo{person}{Monika
  Sester}.} \bibinfo{year}{2007}\natexlab{}.
\newblock \showarticletitle{Integration of heterogeneous geospatial data in a
  federated database}.
\newblock \bibinfo{journal}{\emph{ISPRS Journal of Photogrammetry and Remote
  Sensing}} \bibinfo{volume}{62}, \bibinfo{number}{5} (\bibinfo{year}{2007}),
  \bibinfo{pages}{328 -- 346}.
\newblock
\urldef\tempurl%
\url{https://doi.org/10.1016/j.isprsjprs.2007.04.003}
\showDOI{\tempurl}
\newblock
\shownote{Theme Issue: Distributed Geoinformatics.}


\bibitem[\protect\citeauthoryear{Castro and Liskov}{Castro and Liskov}{1999}]%
        {pbft}
\bibfield{author}{\bibinfo{person}{Miguel Castro} {and}
  \bibinfo{person}{Barbara Liskov}.} \bibinfo{year}{1999}\natexlab{}.
\newblock \showarticletitle{Practical Byzantine Fault Tolerance}. In
  \bibinfo{booktitle}{\emph{Proceedings of the Third Symposium on Operating
  Systems Design and Implementation}}. \bibinfo{publisher}{USENIX},
  \bibinfo{address}{USA}, \bibinfo{pages}{173--186}.
\newblock
\showISBNx{1880446391}


\bibitem[\protect\citeauthoryear{Castro and Liskov}{Castro and Liskov}{2002}]%
        {pbftj}
\bibfield{author}{\bibinfo{person}{Miguel Castro} {and}
  \bibinfo{person}{Barbara Liskov}.} \bibinfo{year}{2002}\natexlab{}.
\newblock \showarticletitle{Practical Byzantine Fault Tolerance and Proactive
  Recovery}.
\newblock \bibinfo{journal}{\emph{ACM Trans. Comput. Syst.}}
  \bibinfo{volume}{20}, \bibinfo{number}{4} (\bibinfo{year}{2002}),
  \bibinfo{pages}{398--461}.
\newblock
\urldef\tempurl%
\url{https://doi.org/10.1145/571637.571640}
\showDOI{\tempurl}


\bibitem[\protect\citeauthoryear{Chun, Maniatis, Shenker, and Kubiatowicz}{Chun
  et~al\mbox{.}}{2007}]%
        {less-replica-2}
\bibfield{author}{\bibinfo{person}{Byung-Gon Chun}, \bibinfo{person}{Petros
  Maniatis}, \bibinfo{person}{Scott Shenker}, {and} \bibinfo{person}{John
  Kubiatowicz}.} \bibinfo{year}{2007}\natexlab{}.
\newblock \showarticletitle{Attested Append-only Memory: Making Adversaries
  Stick to Their Word}. In \bibinfo{booktitle}{\emph{Proceedings of
  Twenty-first ACM SIGOPS Symposium on Operating Systems Principles}}.
  \bibinfo{publisher}{ACM}, \bibinfo{pages}{189--204}.
\newblock
\urldef\tempurl%
\url{https://doi.org/10.1145/1294261.1294280}
\showDOI{\tempurl}


\bibitem[\protect\citeauthoryear{Cooper, Silberstein, Tam, Ramakrishnan, and
  Sears}{Cooper et~al\mbox{.}}{2010}]%
        {ycsb}
\bibfield{author}{\bibinfo{person}{Brian~F. Cooper}, \bibinfo{person}{Adam
  Silberstein}, \bibinfo{person}{Erwin Tam}, \bibinfo{person}{Raghu
  Ramakrishnan}, {and} \bibinfo{person}{Russell Sears}.}
  \bibinfo{year}{2010}\natexlab{}.
\newblock \showarticletitle{Benchmarking Cloud Serving Systems with {YCSB}}. In
  \bibinfo{booktitle}{\emph{Proceedings of the 1st ACM Symposium on Cloud
  Computing}}. \bibinfo{publisher}{ACM}, \bibinfo{pages}{143--154}.
\newblock
\urldef\tempurl%
\url{https://doi.org/10.1145/1807128.1807152}
\showDOI{\tempurl}


\bibitem[\protect\citeauthoryear{Corbett, Dean, Epstein, Fikes, Frost, Furman,
  Ghemawat, Gubarev, Heiser, Hochschild, Hsieh, Kanthak, Kogan, Li, Lloyd,
  Melnik, Mwaura, Nagle, Quinlan, Rao, Rolig, Saito, Szymaniak, Taylor, Wang,
  and Woodford}{Corbett et~al\mbox{.}}{2012}]%
        {spanner}
\bibfield{author}{\bibinfo{person}{J.~C. Corbett}, \bibinfo{person}{Jeffrey
  Dean}, \bibinfo{person}{Michael Epstein}, \bibinfo{person}{Andrew Fikes},
  \bibinfo{person}{Christopher Frost}, \bibinfo{person}{JJ Furman},
  \bibinfo{person}{Sanjay Ghemawat}, \bibinfo{person}{Andrey Gubarev},
  \bibinfo{person}{Christopher Heiser}, \bibinfo{person}{Peter Hochschild},
  \bibinfo{person}{Wilson Hsieh}, \bibinfo{person}{Sebastian Kanthak},
  \bibinfo{person}{Eugene Kogan}, \bibinfo{person}{Hongyi Li},
  \bibinfo{person}{Alexander Lloyd}, \bibinfo{person}{Sergey Melnik},
  \bibinfo{person}{David Mwaura}, \bibinfo{person}{David Nagle},
  \bibinfo{person}{Sean Quinlan}, \bibinfo{person}{Rajesh Rao},
  \bibinfo{person}{Lindsay Rolig}, \bibinfo{person}{Yasushi Saito},
  \bibinfo{person}{Michal Szymaniak}, \bibinfo{person}{Christopher Taylor},
  \bibinfo{person}{Ruth Wang}, {and} \bibinfo{person}{Dale Woodford}.}
  \bibinfo{year}{2012}\natexlab{}.
\newblock \showarticletitle{{Spanner: Google{\textquoteright}s
  Globally-Distributed Database}}. In \bibinfo{booktitle}{\emph{10th {USENIX}
  Symposium on Operating Systems Design and Implementation ({OSDI} 12)}}.
  \bibinfo{publisher}{{USENIX} Association}, \bibinfo{pages}{261--264}.
\newblock
\showISBNx{978-1-931971-96-6}


\bibitem[\protect\citeauthoryear{Dang, Dinh, Loghin, Chang, Lin, and Ooi}{Dang
  et~al\mbox{.}}{2019}]%
        {ahl}
\bibfield{author}{\bibinfo{person}{Hung Dang}, \bibinfo{person}{Tien Tuan~Anh
  Dinh}, \bibinfo{person}{Dumitrel Loghin}, \bibinfo{person}{Ee-Chien Chang},
  \bibinfo{person}{Qian Lin}, {and} \bibinfo{person}{Beng~Chin Ooi}.}
  \bibinfo{year}{2019}\natexlab{}.
\newblock \showarticletitle{Towards Scaling Blockchain Systems via Sharding}.
  In \bibinfo{booktitle}{\emph{Proceedings of the 2019 International Conference
  on Management of Data}}. \bibinfo{publisher}{ACM}, \bibinfo{pages}{123--140}.
\newblock
\urldef\tempurl%
\url{https://doi.org/10.1145/3299869.3319889}
\showDOI{\tempurl}


\bibitem[\protect\citeauthoryear{{Deshpande} and {Hellerstein}}{{Deshpande} and
  {Hellerstein}}{2002}]%
        {federated-query-optimization}
\bibfield{author}{\bibinfo{person}{A. {Deshpande}} {and} \bibinfo{person}{J.~M.
  {Hellerstein}}.} \bibinfo{year}{2002}\natexlab{}.
\newblock \showarticletitle{{Decoupled query optimization for federated
  database systems}}. In \bibinfo{booktitle}{\emph{Proceedings 18th
  International Conference on Data Engineering}}. \bibinfo{pages}{716--727}.
\newblock
\urldef\tempurl%
\url{https://doi.org/10.1109/ICDE.2002.994788}
\showDOI{\tempurl}


\bibitem[\protect\citeauthoryear{Diaconu, Freedman, Ismert, Larson, Mittal,
  Stonecipher, Verma, and Zwilling}{Diaconu et~al\mbox{.}}{2013}]%
        {hekaton}
\bibfield{author}{\bibinfo{person}{C. Diaconu}, \bibinfo{person}{C. Freedman},
  \bibinfo{person}{E. Ismert}, \bibinfo{person}{P.-A. Larson},
  \bibinfo{person}{P. Mittal}, \bibinfo{person}{R. Stonecipher},
  \bibinfo{person}{N. Verma}, {and} \bibinfo{person}{M. Zwilling}.}
  \bibinfo{year}{2013}\natexlab{}.
\newblock \showarticletitle{{Hekaton: SQL Server's Memory-optimized OLTP
  Engine}}. \bibinfo{publisher}{ACM}, \bibinfo{pages}{1243--1254}.
\newblock
\urldef\tempurl%
\url{https://doi.org/10.1145/2463676.2463710}
\showDOI{\tempurl}


\bibitem[\protect\citeauthoryear{Dinh, Wang, Chen, Liu, Ooi, and Tan}{Dinh
  et~al\mbox{.}}{2017}]%
        {blockbench}
\bibfield{author}{\bibinfo{person}{Tien Tuan~Anh Dinh}, \bibinfo{person}{Ji
  Wang}, \bibinfo{person}{Gang Chen}, \bibinfo{person}{Rui Liu},
  \bibinfo{person}{Beng~Chin Ooi}, {and} \bibinfo{person}{Kian-Lee Tan}.}
  \bibinfo{year}{2017}\natexlab{}.
\newblock \showarticletitle{{BLOCKBENCH}: A Framework for Analyzing Private
  Blockchains}. In \bibinfo{booktitle}{\emph{Proceedings of the 2017 ACM
  International Conference on Management of Data}}. \bibinfo{publisher}{ACM},
  \bibinfo{pages}{1085--1100}.
\newblock
\urldef\tempurl%
\url{https://doi.org/10.1145/3035918.3064033}
\showDOI{\tempurl}


\bibitem[\protect\citeauthoryear{Dolev}{Dolev}{1982}]%
        {byzgen}
\bibfield{author}{\bibinfo{person}{Danny Dolev}.}
  \bibinfo{year}{1982}\natexlab{}.
\newblock \showarticletitle{The Byzantine generals strike again}.
\newblock \bibinfo{journal}{\emph{Journal of Algorithms}} \bibinfo{volume}{3},
  \bibinfo{number}{1} (\bibinfo{year}{1982}), \bibinfo{pages}{14--30}.
\newblock
\urldef\tempurl%
\url{https://doi.org/10.1016/0196-6774(82)90004-9}
\showDOI{\tempurl}


\bibitem[\protect\citeauthoryear{Dolev and Reischuk}{Dolev and
  Reischuk}{1985}]%
        {dolevbound}
\bibfield{author}{\bibinfo{person}{Danny Dolev} {and}
  \bibinfo{person}{R\"{u}diger Reischuk}.} \bibinfo{year}{1985}\natexlab{}.
\newblock \showarticletitle{Bounds on Information Exchange for Byzantine
  Agreement}.
\newblock \bibinfo{journal}{\emph{J. ACM}} \bibinfo{volume}{32},
  \bibinfo{number}{1} (\bibinfo{year}{1985}), \bibinfo{pages}{191--204}.
\newblock
\urldef\tempurl%
\url{https://doi.org/10.1145/2455.214112}
\showDOI{\tempurl}


\bibitem[\protect\citeauthoryear{Fischer and Lynch}{Fischer and Lynch}{1982}]%
        {interbound}
\bibfield{author}{\bibinfo{person}{Michael~J. Fischer} {and}
  \bibinfo{person}{Nancy~A. Lynch}.} \bibinfo{year}{1982}\natexlab{}.
\newblock \showarticletitle{A lower bound for the time to assure interactive
  consistency}.
\newblock \bibinfo{journal}{\emph{Inform. Process. Lett.}}
  \bibinfo{volume}{14}, \bibinfo{number}{4} (\bibinfo{year}{1982}),
  \bibinfo{pages}{183--186}.
\newblock
\urldef\tempurl%
\url{https://doi.org/10.1016/0020-0190(82)90033-3}
\showDOI{\tempurl}


\bibitem[\protect\citeauthoryear{Fischer, Lynch, and Paterson}{Fischer
  et~al\mbox{.}}{1985}]%
        {flp}
\bibfield{author}{\bibinfo{person}{Michael~J. Fischer},
  \bibinfo{person}{Nancy~A. Lynch}, {and} \bibinfo{person}{Michael~S.
  Paterson}.} \bibinfo{year}{1985}\natexlab{}.
\newblock \showarticletitle{Impossibility of Distributed Consensus with One
  Faulty Process}.
\newblock \bibinfo{journal}{\emph{J. ACM}} \bibinfo{volume}{32},
  \bibinfo{number}{2} (\bibinfo{year}{1985}), \bibinfo{pages}{374--382}.
\newblock
\urldef\tempurl%
\url{https://doi.org/10.1145/3149.214121}
\showDOI{\tempurl}


\bibitem[\protect\citeauthoryear{Golan~Gueta, Abraham, Grossman, Malkhi,
  Pinkas, Reiter, Seredinschi, Tamir, and Tomescu}{Golan~Gueta
  et~al\mbox{.}}{2019}]%
        {sbft}
\bibfield{author}{\bibinfo{person}{Guy Golan~Gueta}, \bibinfo{person}{Ittai
  Abraham}, \bibinfo{person}{Shelly Grossman}, \bibinfo{person}{Dahlia Malkhi},
  \bibinfo{person}{Benny Pinkas}, \bibinfo{person}{Michael Reiter},
  \bibinfo{person}{Dragos-Adrian Seredinschi}, \bibinfo{person}{Orr Tamir},
  {and} \bibinfo{person}{Alin Tomescu}.} \bibinfo{year}{2019}\natexlab{}.
\newblock \showarticletitle{{SBFT}: A Scalable and Decentralized Trust
  Infrastructure}. In \bibinfo{booktitle}{\emph{2019 49th Annual IEEE/IFIP
  International Conference on Dependable Systems and Networks (DSN)}}.
  \bibinfo{publisher}{IEEE}, \bibinfo{pages}{568--580}.
\newblock
\urldef\tempurl%
\url{https://doi.org/10.1109/DSN.2019.00063}
\showDOI{\tempurl}


\bibitem[\protect\citeauthoryear{Gray}{Gray}{1978}]%
        {2pc}
\bibfield{author}{\bibinfo{person}{Jim Gray}.} \bibinfo{year}{1978}\natexlab{}.
\newblock \showarticletitle{Notes on Data Base Operating Systems}.
\newblock


\bibitem[\protect\citeauthoryear{Guerraoui, Knezevic, Quema, and
  Vukolic}{Guerraoui et~al\mbox{.}}{2010}]%
        {stretchingbft}
\bibfield{author}{\bibinfo{person}{Rachid Guerraoui}, \bibinfo{person}{Nikola
  Knezevic}, \bibinfo{person}{Vivien Quema}, {and} \bibinfo{person}{Marko
  Vukolic}.} \bibinfo{year}{2010}\natexlab{}.
\newblock \showarticletitle{{Stretching BFT}}. \bibinfo{publisher}{Infoscience
  EPFL}.
\newblock


\bibitem[\protect\citeauthoryear{Gupta}{Gupta}{2020}]%
        {phd-workshop}
\bibfield{author}{\bibinfo{person}{Suyash Gupta}.}
  \bibinfo{year}{2020}\natexlab{}.
\newblock \showarticletitle{Resilient and Scalable Architecture for
  Permissioned Blockchain Fabrics}. In \bibinfo{booktitle}{\emph{Proceedings of
  the {VLDB} 2020 PhD Workshop co-located with the 46th International
  Conference on Very Large Databases}} \emph{(\bibinfo{series}{{CEUR} Workshop
  Proceedings})}, Vol.~\bibinfo{volume}{2652}.
  \bibinfo{publisher}{CEUR-WS.org}.
\newblock


\bibitem[\protect\citeauthoryear{Gupta, Hellings, Rahnama, and Sadoghi}{Gupta
  et~al\mbox{.}}{2019b}]%
        {tut-middleware19}
\bibfield{author}{\bibinfo{person}{Suyash Gupta}, \bibinfo{person}{Jelle
  Hellings}, \bibinfo{person}{Sajjad Rahnama}, {and} \bibinfo{person}{Mohammad
  Sadoghi}.} \bibinfo{year}{2019}\natexlab{b}.
\newblock \showarticletitle{An In-Depth Look of {BFT} Consensus in Blockchain:
  Challenges and Opportunities}. In \bibinfo{booktitle}{\emph{Proceedings of
  the 20th International Middleware Conference Tutorials, Middleware}}.
  \bibinfo{publisher}{ACM}, \bibinfo{pages}{6--10}.
\newblock
\urldef\tempurl%
\url{https://doi.org/10.1145/3366625.3369437}
\showDOI{\tempurl}


\bibitem[\protect\citeauthoryear{Gupta, Hellings, Rahnama, and Sadoghi}{Gupta
  et~al\mbox{.}}{2020a}]%
        {tut-vldb20}
\bibfield{author}{\bibinfo{person}{Suyash Gupta}, \bibinfo{person}{Jelle
  Hellings}, \bibinfo{person}{Sajjad Rahnama}, {and} \bibinfo{person}{Mohammad
  Sadoghi}.} \bibinfo{year}{2020}\natexlab{a}.
\newblock \showarticletitle{Building High Throughput Permissioned Blockchain
  Fabrics: Challenges and Opportunities}.
\newblock \bibinfo{journal}{\emph{Proc. VLDB Endow.}} \bibinfo{volume}{13},
  \bibinfo{number}{12} (\bibinfo{year}{2020}), \bibinfo{pages}{3441--3444}.
\newblock
\urldef\tempurl%
\url{https://doi.org/10.14778/3415478.3415565}
\showDOI{\tempurl}


\bibitem[\protect\citeauthoryear{Gupta, Hellings, Rahnama, and Sadoghi}{Gupta
  et~al\mbox{.}}{2021c}]%
        {poe}
\bibfield{author}{\bibinfo{person}{Suyash Gupta}, \bibinfo{person}{Jelle
  Hellings}, \bibinfo{person}{Sajjad Rahnama}, {and} \bibinfo{person}{Mohammad
  Sadoghi}.} \bibinfo{year}{2021}\natexlab{c}.
\newblock \showarticletitle{{Proof-of-Execution: Reaching Consensus through
  Fault-Tolerant Speculation}}. In \bibinfo{booktitle}{\emph{Proceedings of the
  24th International Conference on Extending Database Technology, {EDBT}}}.
  \bibinfo{publisher}{OpenProceedings.org}, \bibinfo{pages}{301--312}.
\newblock
\urldef\tempurl%
\url{https://doi.org/10.5441/002/edbt.2021.27}
\showDOI{\tempurl}


\bibitem[\protect\citeauthoryear{Gupta, Hellings, and Sadoghi}{Gupta
  et~al\mbox{.}}{2019a}]%
        {bambft}
\bibfield{author}{\bibinfo{person}{Suyash Gupta}, \bibinfo{person}{Jelle
  Hellings}, {and} \bibinfo{person}{Mohammad Sadoghi}.}
  \bibinfo{year}{2019}\natexlab{a}.
\newblock \showarticletitle{{Brief Announcement: Revisiting Consensus Protocols
  through Wait-Free Parallelization}}. In \bibinfo{booktitle}{\emph{33rd
  International Symposium on Distributed Computing, {DISC}}}
  \emph{(\bibinfo{series}{LIPIcs})}, Vol.~\bibinfo{volume}{146}.
  \bibinfo{publisher}{Schloss Dagstuhl - Leibniz-Zentrum f{\"{u}}r Informatik},
  \bibinfo{pages}{44:1--44:3}.
\newblock
\urldef\tempurl%
\url{https://doi.org/10.4230/LIPIcs.DISC.2019.44}
\showDOI{\tempurl}


\bibitem[\protect\citeauthoryear{Gupta, Hellings, and Sadoghi}{Gupta
  et~al\mbox{.}}{2021a}]%
        {blockchain-book}
\bibfield{author}{\bibinfo{person}{Suyash Gupta}, \bibinfo{person}{Jelle
  Hellings}, {and} \bibinfo{person}{Mohammad Sadoghi}.}
  \bibinfo{year}{2021}\natexlab{a}.
\newblock \bibinfo{booktitle}{\emph{{Fault-Tolerant Distributed Transactions on
  Blockchain}}}.
\newblock \bibinfo{publisher}{Morgan {\&} Claypool Publishers}.
\newblock
\urldef\tempurl%
\url{https://doi.org/10.2200/S01068ED1V01Y202012DTM065}
\showDOI{\tempurl}


\bibitem[\protect\citeauthoryear{Gupta, Hellings, and Sadoghi}{Gupta
  et~al\mbox{.}}{2021b}]%
        {multibft}
\bibfield{author}{\bibinfo{person}{Suyash Gupta}, \bibinfo{person}{Jelle
  Hellings}, {and} \bibinfo{person}{Mohammad Sadoghi}.}
  \bibinfo{year}{2021}\natexlab{b}.
\newblock \showarticletitle{{RCC: Resilient Concurrent Consensus for
  High-Throughput Secure Transaction Processing}}. In
  \bibinfo{booktitle}{\emph{37th {IEEE} International Conference on Data
  Engineering, {ICDE}}}. \bibinfo{pages}{1392--1403}.
\newblock
\urldef\tempurl%
\url{https://doi.org/10.1109/ICDE51399.2021.00124}
\showDOI{\tempurl}


\bibitem[\protect\citeauthoryear{Gupta, Rahnama, Hellings, and Sadoghi}{Gupta
  et~al\mbox{.}}{2020c}]%
        {geobft}
\bibfield{author}{\bibinfo{person}{Suyash Gupta}, \bibinfo{person}{Sajjad
  Rahnama}, \bibinfo{person}{Jelle Hellings}, {and} \bibinfo{person}{Mohammad
  Sadoghi}.} \bibinfo{year}{2020}\natexlab{c}.
\newblock \showarticletitle{{ResilientDB: Global Scale Resilient Blockchain
  Fabric}}.
\newblock \bibinfo{journal}{\emph{Proc. {VLDB} Endow.}} \bibinfo{volume}{13},
  \bibinfo{number}{6} (\bibinfo{year}{2020}), \bibinfo{pages}{868--883}.
\newblock
\urldef\tempurl%
\url{https://doi.org/10.14778/3380750.3380757}
\showDOI{\tempurl}


\bibitem[\protect\citeauthoryear{Gupta, Rahnama, Pandey, Crooks, and
  Sadoghi}{Gupta et~al\mbox{.}}{2022}]%
        {sgx-trust}
\bibfield{author}{\bibinfo{person}{Suyash Gupta}, \bibinfo{person}{Sajjad
  Rahnama}, \bibinfo{person}{Shubham Pandey}, \bibinfo{person}{Natacha Crooks},
  {and} \bibinfo{person}{Mohammad Sadoghi}.} \bibinfo{year}{2022}\natexlab{}.
\newblock \showarticletitle{{Dissecting {BFT} Consensus: In Trusted Components
  we Trust!}}
\newblock \bibinfo{journal}{\emph{CoRR}}  \bibinfo{volume}{abs/2202.01354}
  (\bibinfo{year}{2022}).
\newblock
\showeprint[arXiv]{2202.01354}


\bibitem[\protect\citeauthoryear{Gupta, Rahnama, and Sadoghi}{Gupta
  et~al\mbox{.}}{2020b}]%
        {r-evalpaper}
\bibfield{author}{\bibinfo{person}{Suyash Gupta}, \bibinfo{person}{Sajjad
  Rahnama}, {and} \bibinfo{person}{Mohammad Sadoghi}.}
  \bibinfo{year}{2020}\natexlab{b}.
\newblock \showarticletitle{{Permissioned Blockchain Through the Looking Glass:
  Architectural and Implementation Lessons Learned}}. In
  \bibinfo{booktitle}{\emph{40th {IEEE} International Conference on Distributed
  Computing Systems, {ICDCS}}}. \bibinfo{pages}{754--764}.
\newblock
\urldef\tempurl%
\url{https://doi.org/10.1109/ICDCS47774.2020.00012}
\showDOI{\tempurl}


\bibitem[\protect\citeauthoryear{Gupta and Sadoghi}{Gupta and Sadoghi}{2018}]%
        {easyc}
\bibfield{author}{\bibinfo{person}{Suyash Gupta} {and}
  \bibinfo{person}{Mohammad Sadoghi}.} \bibinfo{year}{2018}\natexlab{}.
\newblock \showarticletitle{{EasyCommit: A Non-blocking Two-phase Commit
  Protocol}}. In \bibinfo{booktitle}{\emph{Proceedings of the 21st
  International Conference on Extending Database Technology, {EDBT}}}.
  \bibinfo{publisher}{OpenProceedings.org}, \bibinfo{pages}{157--168}.
\newblock
\urldef\tempurl%
\url{https://doi.org/10.5441/002/edbt.2018.15}
\showDOI{\tempurl}


\bibitem[\protect\citeauthoryear{Gupta and Sadoghi}{Gupta and Sadoghi}{2019}]%
        {bc-processing}
\bibfield{author}{\bibinfo{person}{Suyash Gupta} {and}
  \bibinfo{person}{Mohammad Sadoghi}.} \bibinfo{year}{2019}\natexlab{}.
\newblock \showarticletitle{{Blockchain Transaction Processing}}.
\newblock In \bibinfo{booktitle}{\emph{Encyclopedia of Big Data Technologies}}.
  \bibinfo{publisher}{Springer}, \bibinfo{pages}{1--11}.
\newblock
\urldef\tempurl%
\url{https://doi.org/10.1007/978-3-319-63962-8_333-1}
\showDOI{\tempurl}


\bibitem[\protect\citeauthoryear{Gupta and Sadoghi}{Gupta and Sadoghi}{2020}]%
        {dapd}
\bibfield{author}{\bibinfo{person}{Suyash Gupta} {and}
  \bibinfo{person}{Mohammad Sadoghi}.} \bibinfo{year}{2020}\natexlab{}.
\newblock \showarticletitle{{Efficient and non-blocking agreement protocols}}.
\newblock \bibinfo{journal}{\emph{Distributed Parallel Databases}}
  \bibinfo{volume}{38}, \bibinfo{number}{2} (\bibinfo{year}{2020}),
  \bibinfo{pages}{287--333}.
\newblock
\urldef\tempurl%
\url{https://doi.org/10.1007/s10619-019-07267-w}
\showDOI{\tempurl}


\bibitem[\protect\citeauthoryear{Harding, Van~Aken, Pavlo, and
  Stonebraker}{Harding et~al\mbox{.}}{2017}]%
        {deneva}
\bibfield{author}{\bibinfo{person}{R. Harding}, \bibinfo{person}{D. Van~Aken},
  \bibinfo{person}{A. Pavlo}, {and} \bibinfo{person}{M. Stonebraker}.}
  \bibinfo{year}{2017}\natexlab{}.
\newblock \showarticletitle{{An Evaluation of Distributed Concurrency
  Control}}.
\newblock \bibinfo{journal}{\emph{Proc. VLDB Endow.}} \bibinfo{volume}{10},
  \bibinfo{number}{5} (\bibinfo{year}{2017}), \bibinfo{pages}{553--564}.
\newblock
\urldef\tempurl%
\url{https://doi.org/10.14778/3055540.3055548}
\showDOI{\tempurl}


\bibitem[\protect\citeauthoryear{Haynes and Noveck}{Haynes and Noveck}{2015}]%
        {nfs}
\bibfield{author}{\bibinfo{person}{Thomas Haynes} {and} \bibinfo{person}{David
  Noveck}.} \bibinfo{year}{2015}\natexlab{}.
\newblock \bibinfo{title}{{RFC} 7530: Network File System ({NFS}) Version 4
  Protocol}.
\newblock
\newblock
\urldef\tempurl%
\url{https://tools.ietf.org/html/rfc7530}
\showURL{%
\tempurl}


\bibitem[\protect\citeauthoryear{Hellings, Hughes, Primero, and
  Sadoghi}{Hellings et~al\mbox{.}}{2020}]%
        {cerberus}
\bibfield{author}{\bibinfo{person}{Jelle Hellings}, \bibinfo{person}{Daniel~P.
  Hughes}, \bibinfo{person}{Joshua Primero}, {and} \bibinfo{person}{Mohammad
  Sadoghi}.} \bibinfo{year}{2020}\natexlab{}.
\newblock \bibinfo{title}{Cerberus: Minimalistic Multi-shard
  Byzantine-resilient Transaction Processing}.
\newblock
\newblock
\urldef\tempurl%
\url{https://arxiv.org/abs/2008.04450}
\showURL{%
\tempurl}


\bibitem[\protect\citeauthoryear{Hellings and Sadoghi}{Hellings and
  Sadoghi}{2019}]%
        {csp}
\bibfield{author}{\bibinfo{person}{Jelle Hellings} {and}
  \bibinfo{person}{Mohammad Sadoghi}.} \bibinfo{year}{2019}\natexlab{}.
\newblock \bibinfo{title}{The fault-tolerant cluster-sending problem}.
\newblock
\newblock
\urldef\tempurl%
\url{https://arxiv.org/abs/1908.01455}
\showURL{%
\tempurl}


\bibitem[\protect\citeauthoryear{Hellings and Sadoghi}{Hellings and
  Sadoghi}{2021}]%
        {byshard}
\bibfield{author}{\bibinfo{person}{Jelle Hellings} {and}
  \bibinfo{person}{Mohammad Sadoghi}.} \bibinfo{year}{2021}\natexlab{}.
\newblock \showarticletitle{{ByShard: Sharding in a Byzantine Environment}}.
\newblock \bibinfo{journal}{\emph{Proc. {VLDB} Endow.}} \bibinfo{volume}{14},
  \bibinfo{number}{11} (\bibinfo{year}{2021}), \bibinfo{pages}{2230--2243}.
\newblock


\bibitem[\protect\citeauthoryear{Katz and Lindell}{Katz and Lindell}{2014}]%
        {cryptobook}
\bibfield{author}{\bibinfo{person}{Jonathan Katz} {and} \bibinfo{person}{Yehuda
  Lindell}.} \bibinfo{year}{2014}\natexlab{}.
\newblock \bibinfo{booktitle}{\emph{Introduction to Modern Cryptography}
  (\bibinfo{edition}{2nd} ed.)}.
\newblock


\bibitem[\protect\citeauthoryear{Kokoris-Kogias, Jovanovic, Gasser, Gailly,
  Syta, and Ford}{Kokoris-Kogias et~al\mbox{.}}{2018}]%
        {omniledger}
\bibfield{author}{\bibinfo{person}{Eleftherios Kokoris-Kogias},
  \bibinfo{person}{Philipp Jovanovic}, \bibinfo{person}{Linus Gasser},
  \bibinfo{person}{Nicolas Gailly}, \bibinfo{person}{Ewa Syta}, {and}
  \bibinfo{person}{Bryan Ford}.} \bibinfo{year}{2018}\natexlab{}.
\newblock \showarticletitle{{OmniLedger}: A Secure, Scale-Out, Decentralized
  Ledger via Sharding}. In \bibinfo{booktitle}{\emph{2018 IEEE Symposium on
  Security and Privacy (SP)}}. \bibinfo{pages}{583--598}.
\newblock
\urldef\tempurl%
\url{https://doi.org/10.1109/SP.2018.000-5}
\showDOI{\tempurl}


\bibitem[\protect\citeauthoryear{Kotla, Alvisi, Dahlin, Clement, and
  Wong}{Kotla et~al\mbox{.}}{2007}]%
        {zyzzyva}
\bibfield{author}{\bibinfo{person}{Ramakrishna Kotla}, \bibinfo{person}{Lorenzo
  Alvisi}, \bibinfo{person}{Mike Dahlin}, \bibinfo{person}{Allen Clement},
  {and} \bibinfo{person}{Edmund Wong}.} \bibinfo{year}{2007}\natexlab{}.
\newblock \showarticletitle{Zyzzyva: Speculative Byzantine Fault Tolerance}.
\newblock \bibinfo{journal}{\emph{SIGOPS Oper. Syst. Rev.}}
  \bibinfo{volume}{41}, \bibinfo{number}{6} (\bibinfo{year}{2007}),
  \bibinfo{pages}{45--58}.
\newblock
\urldef\tempurl%
\url{https://doi.org/10.1145/1323293.1294267}
\showDOI{\tempurl}


\bibitem[\protect\citeauthoryear{Kotla, Alvisi, Dahlin, Clement, and
  Wong}{Kotla et~al\mbox{.}}{2010}]%
        {zyzzyvaj}
\bibfield{author}{\bibinfo{person}{Ramakrishna Kotla}, \bibinfo{person}{Lorenzo
  Alvisi}, \bibinfo{person}{Mike Dahlin}, \bibinfo{person}{Allen Clement},
  {and} \bibinfo{person}{Edmund Wong}.} \bibinfo{year}{2010}\natexlab{}.
\newblock \showarticletitle{Zyzzyva: Speculative Byzantine Fault Tolerance}.
\newblock \bibinfo{journal}{\emph{ACM Trans. Comput. Syst.}}
  \bibinfo{volume}{27}, \bibinfo{number}{4}, Article \bibinfo{articleno}{7}
  (\bibinfo{year}{2010}), \bibinfo{numpages}{39}~pages.
\newblock
\urldef\tempurl%
\url{https://doi.org/10.1145/1658357.1658358}
\showDOI{\tempurl}


\bibitem[\protect\citeauthoryear{Lamport}{Lamport}{1998}]%
        {paxos}
\bibfield{author}{\bibinfo{person}{Leslie Lamport}.}
  \bibinfo{year}{1998}\natexlab{}.
\newblock \showarticletitle{The Part-time Parliament}.
\newblock  (\bibinfo{year}{1998}).
\newblock


\bibitem[\protect\citeauthoryear{Li, Li, Zhou, Xu, Long, and Yao}{Li
  et~al\mbox{.}}{2018}]%
        {conflux}
\bibfield{author}{\bibinfo{person}{Chenxing Li}, \bibinfo{person}{Peilun Li},
  \bibinfo{person}{Dong Zhou}, \bibinfo{person}{Wei Xu}, \bibinfo{person}{Fan
  Long}, {and} \bibinfo{person}{Andrew Yao}.} \bibinfo{year}{2018}\natexlab{}.
\newblock \bibinfo{title}{Scaling Nakamoto Consensus to Thousands of
  Transactions per Second}.
\newblock
\newblock
\urldef\tempurl%
\url{https://arxiv.org/abs/1805.03870}
\showURL{%
\tempurl}


\bibitem[\protect\citeauthoryear{Luu, Narayanan, Zheng, Baweja, Gilbert, and
  Saxena}{Luu et~al\mbox{.}}{2016}]%
        {elastico}
\bibfield{author}{\bibinfo{person}{Loi Luu}, \bibinfo{person}{Viswesh
  Narayanan}, \bibinfo{person}{Chaodong Zheng}, \bibinfo{person}{Kunal Baweja},
  \bibinfo{person}{Seth Gilbert}, {and} \bibinfo{person}{Prateek Saxena}.}
  \bibinfo{year}{2016}\natexlab{}.
\newblock \showarticletitle{A Secure Sharding Protocol For Open Blockchains}.
  In \bibinfo{booktitle}{\emph{Proceedings of the 2016 ACM SIGSAC Conference on
  Computer and Communications Security}}. \bibinfo{publisher}{ACM},
  \bibinfo{pages}{17--30}.
\newblock
\urldef\tempurl%
\url{https://doi.org/10.1145/2976749.2978389}
\showDOI{\tempurl}


\bibitem[\protect\citeauthoryear{Malkhi and Reiter}{Malkhi and Reiter}{1998a}]%
        {byzq}
\bibfield{author}{\bibinfo{person}{Dahlia Malkhi} {and}
  \bibinfo{person}{Michael Reiter}.} \bibinfo{year}{1998}\natexlab{a}.
\newblock \showarticletitle{Byzantine quorum systems}.
\newblock \bibinfo{journal}{\emph{Distributed Computing}} \bibinfo{volume}{11},
  \bibinfo{number}{4} (\bibinfo{year}{1998}), \bibinfo{pages}{203--213}.
\newblock
\urldef\tempurl%
\url{https://doi.org/10.1007/s004460050050}
\showDOI{\tempurl}


\bibitem[\protect\citeauthoryear{Malkhi and Reiter}{Malkhi and Reiter}{1998b}]%
        {phalanx}
\bibfield{author}{\bibinfo{person}{Dahlia Malkhi} {and}
  \bibinfo{person}{Michael Reiter}.} \bibinfo{year}{1998}\natexlab{b}.
\newblock \showarticletitle{Secure and scalable replication in {Phalanx}}. In
  \bibinfo{booktitle}{\emph{Proceedings Seventeenth IEEE Symposium on Reliable
  Distributed Systems}}. \bibinfo{publisher}{IEEE}, \bibinfo{pages}{51--58}.
\newblock
\urldef\tempurl%
\url{https://doi.org/10.1109/RELDIS.1998.740474}
\showDOI{\tempurl}


\bibitem[\protect\citeauthoryear{Merkle}{Merkle}{1988}]%
        {merkle}
\bibfield{author}{\bibinfo{person}{Ralph~C. Merkle}.}
  \bibinfo{year}{1988}\natexlab{}.
\newblock \showarticletitle{A Digital Signature Based on a Conventional
  Encryption Function}. In \bibinfo{booktitle}{\emph{Advances in Cryptology ---
  CRYPTO '87}}. \bibinfo{publisher}{Springer}, \bibinfo{pages}{369--378}.
\newblock
\urldef\tempurl%
\url{https://doi.org/10.1007/3-540-48184-2_32}
\showDOI{\tempurl}


\bibitem[\protect\citeauthoryear{Miller, Xia, Croman, Shi, and Song}{Miller
  et~al\mbox{.}}{2016}]%
        {honeybadger}
\bibfield{author}{\bibinfo{person}{Andrew Miller}, \bibinfo{person}{Yu Xia},
  \bibinfo{person}{Kyle Croman}, \bibinfo{person}{Elaine Shi}, {and}
  \bibinfo{person}{Dawn Song}.} \bibinfo{year}{2016}\natexlab{}.
\newblock \showarticletitle{{The Honey Badger of BFT Protocols}}. In
  \bibinfo{booktitle}{\emph{Proceedings of the 2016 ACM SIGSAC Conference on
  Computer and Communications Security}} \emph{(\bibinfo{series}{CCS '16})}.
  \bibinfo{publisher}{ACM}, \bibinfo{pages}{31–42}.
\newblock
\urldef\tempurl%
\url{https://doi.org/10.1145/2976749.2978399}
\showDOI{\tempurl}


\bibitem[\protect\citeauthoryear{of~Economic~Advisers}{of~Economic~Advisers}{2018}]%
        {ecodam}
\bibfield{author}{\bibinfo{person}{The~Council of Economic~Advisers}.}
  \bibinfo{year}{2018}\natexlab{}.
\newblock \bibinfo{booktitle}{\emph{The Cost of Malicious Cyber Activity to the
  {U.S.}\ Economy}}.
\newblock \bibinfo{type}{{T}echnical {R}eport}. \bibinfo{institution}{Executive
  Office of the President of the United States}.
\newblock
\urldef\tempurl%
\url{https://www.whitehouse.gov/wp-content/uploads/2018/03/The-Cost-of-Malicious-Cyber-Activity-to-the-U.S.-Economy.pdf}
\showURL{%
\tempurl}


\bibitem[\protect\citeauthoryear{Ongaro and Ousterhout}{Ongaro and
  Ousterhout}{2014}]%
        {raft}
\bibfield{author}{\bibinfo{person}{Diego Ongaro} {and} \bibinfo{person}{John
  Ousterhout}.} \bibinfo{year}{2014}\natexlab{}.
\newblock \showarticletitle{In Search of an Understandable Consensus
  Algorithm}. In \bibinfo{booktitle}{\emph{ATC}}.
\newblock


\bibitem[\protect\citeauthoryear{{\"O}zsu and Valduriez}{{\"O}zsu and
  Valduriez}{2020}]%
        {distdb}
\bibfield{author}{\bibinfo{person}{M.~Tamer {\"O}zsu} {and}
  \bibinfo{person}{Patrick Valduriez}.} \bibinfo{year}{2020}\natexlab{}.
\newblock \bibinfo{booktitle}{\emph{Principles of Distributed Database
  Systems}}.
\newblock \bibinfo{publisher}{Springer}.
\newblock
\urldef\tempurl%
\url{https://doi.org/10.1007/978-3-030-26253-2}
\showDOI{\tempurl}


\bibitem[\protect\citeauthoryear{Qadah, Gupta, and Sadoghi}{Qadah
  et~al\mbox{.}}{2020}]%
        {qstore}
\bibfield{author}{\bibinfo{person}{Thamir Qadah}, \bibinfo{person}{Suyash
  Gupta}, {and} \bibinfo{person}{Mohammad Sadoghi}.}
  \bibinfo{year}{2020}\natexlab{}.
\newblock \showarticletitle{{Q-Store: Distributed, Multi-partition Transactions
  via Queue-oriented Execution and Communication}}. In
  \bibinfo{booktitle}{\emph{Proceedings of the 23rd International Conference on
  Extending Database Technology, {EDBT}}}.
  \bibinfo{publisher}{OpenProceedings.org}, \bibinfo{pages}{73--84}.
\newblock
\urldef\tempurl%
\url{https://doi.org/10.5441/002/edbt.2020.08}
\showDOI{\tempurl}


\bibitem[\protect\citeauthoryear{Rahnama, Gupta, Qadah, Hellings, and
  Sadoghi}{Rahnama et~al\mbox{.}}{2020}]%
        {resilientdb-demo}
\bibfield{author}{\bibinfo{person}{Sajjad Rahnama}, \bibinfo{person}{Suyash
  Gupta}, \bibinfo{person}{Thamir Qadah}, \bibinfo{person}{Jelle Hellings},
  {and} \bibinfo{person}{Mohammad Sadoghi}.} \bibinfo{year}{2020}\natexlab{}.
\newblock \showarticletitle{Scalable, Resilient and Configurable Permissioned
  Blockchain Fabric}.
\newblock \bibinfo{journal}{\emph{Proc. VLDB Endow.}} \bibinfo{volume}{13},
  \bibinfo{number}{12} (\bibinfo{year}{2020}), \bibinfo{pages}{2893--2896}.
\newblock
\urldef\tempurl%
\url{https://doi.org/doi.org/10.14778/3415478.3415502}
\showDOI{\tempurl}


\bibitem[\protect\citeauthoryear{Sadoghi and Blanas}{Sadoghi and
  Blanas}{2019}]%
        {tpbook}
\bibfield{author}{\bibinfo{person}{Mohammad Sadoghi} {and}
  \bibinfo{person}{Spyros Blanas}.} \bibinfo{year}{2019}\natexlab{}.
\newblock \bibinfo{booktitle}{\emph{{Transaction Processing on Modern
  Hardware}}}.
\newblock \bibinfo{publisher}{Morgan {\&} Claypool Publishers}.
\newblock
\urldef\tempurl%
\url{https://doi.org/10.2200/S00896ED1V01Y201901DTM058}
\showDOI{\tempurl}


\bibitem[\protect\citeauthoryear{Sheth and Larson}{Sheth and Larson}{1990}]%
        {federated-databases}
\bibfield{author}{\bibinfo{person}{Amit~P. Sheth} {and}
  \bibinfo{person}{James~A. Larson}.} \bibinfo{year}{1990}\natexlab{}.
\newblock \showarticletitle{{Federated Database Systems for Managing
  Distributed, Heterogeneous, and Autonomous Databases}}.
\newblock \bibinfo{journal}{\emph{ACM Comput. Surv.}} \bibinfo{volume}{22},
  \bibinfo{number}{3} (\bibinfo{date}{Sept.} \bibinfo{year}{1990}),
  \bibinfo{pages}{183–236}.
\newblock
\showISSN{0360-0300}
\urldef\tempurl%
\url{https://doi.org/10.1145/96602.96604}
\showDOI{\tempurl}


\bibitem[\protect\citeauthoryear{Skeen}{Skeen}{1982}]%
        {3pc}
\bibfield{author}{\bibinfo{person}{Dale Skeen}.}
  \bibinfo{year}{1982}\natexlab{}.
\newblock \bibinfo{booktitle}{\emph{A Quorum-Based Commit Protocol}}.
\newblock \bibinfo{type}{{T}echnical {R}eport}. \bibinfo{institution}{Cornell
  University}.
\newblock


\bibitem[\protect\citeauthoryear{Sompolinsky, Lewenberg, and Zohar}{Sompolinsky
  et~al\mbox{.}}{2016}]%
        {spectre}
\bibfield{author}{\bibinfo{person}{Yonatan Sompolinsky}, \bibinfo{person}{Yoad
  Lewenberg}, {and} \bibinfo{person}{Aviv Zohar}.}
  \bibinfo{year}{2016}\natexlab{}.
\newblock \bibinfo{title}{{SPECTRE}: A Fast and Scalable Cryptocurrency
  Protocol}.
\newblock
\newblock
\newblock
\shownote{https://eprint.iacr.org/2016/1159.}


\bibitem[\protect\citeauthoryear{Taubenfeld and Moran}{Taubenfeld and
  Moran}{1996}]%
        {possibleasync}
\bibfield{author}{\bibinfo{person}{Gadi Taubenfeld} {and}
  \bibinfo{person}{Shlomo Moran}.} \bibinfo{year}{1996}\natexlab{}.
\newblock \showarticletitle{Possibility and impossibility results in a shared
  memory environment}.
\newblock \bibinfo{journal}{\emph{Acta Informatica}} \bibinfo{volume}{33},
  \bibinfo{number}{1} (\bibinfo{year}{1996}), \bibinfo{pages}{1--20}.
\newblock
\urldef\tempurl%
\url{https://doi.org/10.1007/s002360050034}
\showDOI{\tempurl}


\bibitem[\protect\citeauthoryear{Tel}{Tel}{2001}]%
        {distalgo}
\bibfield{author}{\bibinfo{person}{Gerard Tel}.}
  \bibinfo{year}{2001}\natexlab{}.
\newblock \bibinfo{booktitle}{\emph{Introduction to Distributed Algorithms}
  (\bibinfo{edition}{2nd} ed.)}.
\newblock \bibinfo{publisher}{Cambridge University Press}.
\newblock


\bibitem[\protect\citeauthoryear{Thomson, Diamond, Weng, Ren, Shao, and
  Abadi}{Thomson et~al\mbox{.}}{2012}]%
        {calvin}
\bibfield{author}{\bibinfo{person}{Alexander Thomson},
  \bibinfo{person}{Thaddeus Diamond}, \bibinfo{person}{Shu-Chun Weng},
  \bibinfo{person}{Kun Ren}, \bibinfo{person}{Philip Shao}, {and}
  \bibinfo{person}{Daniel~J. Abadi}.} \bibinfo{year}{2012}\natexlab{}.
\newblock \showarticletitle{{Calvin: Fast Distributed Transactions for
  Partitioned Database Systems}}. In \bibinfo{booktitle}{\emph{Proceedings of
  the 2012 ACM SIGMOD International Conference on Management of Data}}
  \emph{(\bibinfo{series}{SIGMOD})}. \bibinfo{publisher}{ACM},
  \bibinfo{pages}{1--12}.
\newblock
\urldef\tempurl%
\url{https://doi.org/10.1145/2213836.2213838}
\showDOI{\tempurl}


\end{thebibliography}

\end{document}